\documentclass[a4paper,12pt,twoside]{memoir}
\usepackage{graphicx}
\usepackage{amsmath}
\usepackage{amssymb}
\pdfoutput=1

\chapterstyle{section}
\OnehalfSpacing
\makeheadrule{headings}{\textwidth}{0.5pt}

\setlrmargins{1.25in}{*}{*}
\checkandfixthelayout

\begin{document}

\begin{titlingpage}
\calccentering{\unitlength} 
\begin{adjustwidth*}{\unitlength}{-\unitlength}
\begin{center}
\begin{figure}
\calccentering{\unitlength} 
\begin{adjustwidth*}{\unitlength}{-\unitlength}
\begin{center}
\includegraphics[width=2.5cm]{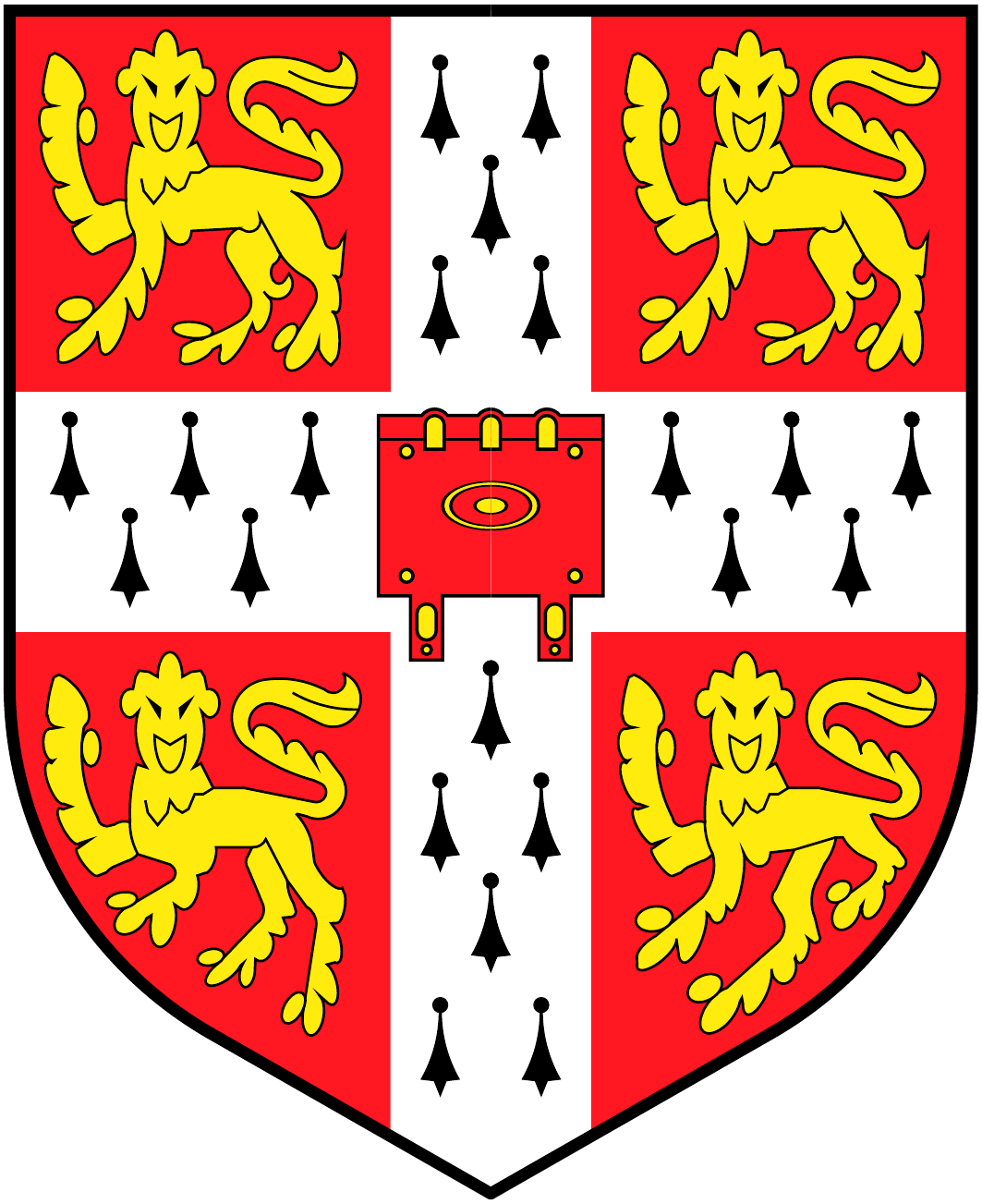}
\end{center}
\end{adjustwidth*}
\end{figure}

\vspace{0.05in}
\small{Cavendish Laboratory}\\
\small{University of Cambridge}\\
\vspace{0.50in}

{\Huge{Gaussian Approximation Potential: an interatomic potential derived from first principles
Quantum Mechanics}\\}
\vspace{0.90in}
{\Large{Albert Bart\'ok-P\'artay}\\}
\vspace{0.10in}
{\Large{Pembroke College} \\}

\vspace{0.50in}
\normalsize{This dissertation is submitted to the University of Cambridge for the degree of Doctor of Philosophy}\\
\vspace{1.0in}

November 2009\\
\end{center}
\end{adjustwidth*}
\end{titlingpage}

\cleardoublepage
\pagenumbering{roman}

\clearpage

\chapter*{Declaration\markboth{Declaration}{Declaration}}

The work described in this dissertation was carried out by the author in the Cavendish
Laboratory at the University of Cambridge between October 2006 and
November 2009. The contents are the original work of the author except where otherwise
indicated, and have not previously been submitted for any degree or qualification at
this or another institution.

The number of words in this thesis does not exceed 60,000 words.\\

\begin{flushright}
Albert Bart\'ok-P\'artay\\
November 2009
\end{flushright}

\clearpage

\chapter*{Acknowledgements\markboth{Acknowledgements}{Acknowledgements}}

I am most grateful to my supervisor, G\'abor Cs\'anyi, for his advice, help and discussions,
which often happened outside office hours. I would also like to thank Mike Payne for
giving the opportunity to carry out my research in the Theory of Condensed Matter Group. Thanks are
due to Risi Kondor, whose advice on Group Theory was invaluable. I am especially thankful to 
Edward Snelson, who gave useful advice on the details of machine learning algorithms. Further thanks
are due to my second supervisor, Mark Warner.

I am indebted to the members of G\'abor Cs\'anyi's research group: L\'ivia Bart\'ok-P\'artay, Noam
Bernstein, James Kermode, Anthony Leung, Wojciech Szlachta, Csilla V\'arnai and Steve Winfield for
useful discussions and inspiration at our group meetings. I am grateful to my peers in TCM, in
particular, Hatem Helal and Mikhail Kibalchenko, from whom I received support and company.

Many thanks to Michael Rutter for his advice on computer-related matters. Thanks also to Tracey Ingham
for helping with the administration.

I would like to thank my family for their patience and my wife, L\'ivia Bart\'ok-P\'artay for her love
and encouragement.

\clearpage

\tableofcontents
\cleardoublepage

\pagenumbering{arabic}

\newpage

\chapter*{Summary}

Simulation of materials at the atomistic level is an important tool in studying microscopic
structure and processes. The atomic interactions necessary for the simulation are correctly
described by Quantum Mechanics.  However, the computational resources required to solve the quantum
mechanical equations limits the use of Quantum Mechanics at most to a few hundreds of atoms and only
to a small fraction of the available configurational space.  This thesis presents the results of my
research on the development of  a new interatomic potential generation scheme, which we refer to as
Gaussian Approximation Potentials. In our framework, the quantum  mechanical potential energy
surface is interpolated between a set of  predetermined values at different points in atomic
configurational  space by a non-linear, non-parametric regression method, the Gaussian  Process. To
perform the fitting, we represent the atomic environments  by the bispectrum, which is invariant to
permutations of the atoms in  the neighbourhood and to global rotations. The result is a general
scheme, that allows one to generate interatomic potentials based on arbitrary quantum mechanical
data. We built a series of Gaussian Approximation Potentials using data obtained from Density
Functional Theory and tested the capabilities of the method. We showed that our models reproduce the
quantum  mechanical potential energy surface remarkably well for the group IV semiconductors, iron
and gallium nitride. Our potentials, while maintaining quantum mechanical accuracy, are several
orders of magnitude faster than Quantum Mechanical methods.

\chapter{Introduction}

Understanding the behaviour of materials at the atomic scale is fundamental to modern science and
technology. As many properties and phenomena are ultimately controlled by the details of the
atomic interactions, simulations of atomic systems provide useful information, which is often not
accessible by experiment alone. Observing materials on a microscopic level can help to
interpret physical phenomena and to predict the properties of previously unknown molecules and
materials. To perform such atomistic simulations, we have to use models to describe the atomic interactions, whose accuracy
has to be validated in order to ensure that the simulations are realistic.

Quantum Mechanics provides a description of matter, which, according to our current knowledge, is
ultimately correct, a conclusion which is strongly corroborated by experimental evidence. However, the solution of the
Schr\"odinger equation---apart from a few very simple examples---has to be performed numerically
using
computers. A series of approximations and sophisticated numerical techniques has led to various
implementations of the originally exact quantum mechanical theory, which can be now routinely used in studies of atomic
systems. In the last few decades, as computational speed capacities grew exponentially, the
description of more and more atoms has become tractable. In most practical applications, the electrons
and the nuclei are treated separately, and the quantum mechanical description of the nuclei is
dropped altogether.
This simplification, namely, that the nuclei move on a potential energy surface determined by the
interaction of the electrons, already makes quantum mechanical calculations several order of magnitudes faster.
However, determining macroscopic thermodynamical quantities of atomic systems requires a large number of samples
of different arrangements of atoms, and the number of atoms has to be large enough to minimise 
finite-size effects. In fact, the computational costs associated with the solution of the
Schr\"odinger equation are so large that the use of Quantum Mechanics is limited at most to a
hundred of
atoms and only a small fraction of the available configurational space.

The demand for faster calculations to allow calculations of larger systems or the exploration of
configurational space leads to the realm of analytical potentials, which are based on
substituting the solution of the electronic Schr\"odinger equation with evaluation using an analytic
function. Whereas the quantum mechanical description does not need validation---apart from ensuring
that the errors introduced by the approximations are minimised---, analytic potentials have to be
checked to determine whether the description remains valid. This is often done by comparing macroscopic
quantities computed by the model to experimental values. There is a high degree of arbitrariness in the
creation and validation of such potentials, and in practice it is found that they are significantly less
accurate than Quantum Mechanics.

As quantum mechanical calculations are becoming more widely available, we have access to a large
number of microscopic observables. The approach we present in this thesis is to create interatomic potentials based directly
on quantum mechanical data which are fast and have an accuracy close to the original method. To
achieve this, we have used a Gaussian Process to interpolate the quantum mechanical potential energy
surface. The Gaussian Process is routinely used by the machine-learning community for regression,
but it has never previously been adapted to represent the atomic potential energy surface. 

We describe the environment of the atoms by a vector, called the bispectrum, which is invariant to
rotations, translations and permutation of atoms in the neighbourhood. Within the bispectrum
representation, we regard the potential energy surface as a sum of atomic energy functions, whose
variables are the elements of the bispectrum. Our approach for generating interatomic potentials,
which we collectively refer to as Gaussian Approximation Potentials, has the favourable scaling and
speed of analytic potentials, while the accuracy is comparable with the underlying quantum
mechanical method. With Gaussian Approximation Potentials atomistic simulations can be taken to an
entirely new level.

\section{Outline of the thesis}

The thesis is organised as follows. In chapter~\ref{chapter:represent} I discuss the representation
of atomic environments by the bispectrum. I show how the rotational invariance of the  bispectrum
can be proved using Representation Theory and how the bispectrum is related to the widely used
bond-order parameters.
I summarise the Gaussian Process non-linear regression method we used in chapter~\ref{chapter:gp},
where I
show the derivation of the formulae based on the Bayes' Theorem and the extensions which allowed us
to use Gaussian Process for the regression of atomic potential energy surfaces.
I describe a number of interatomic potentials and the Gaussian Approximation Potential in detail in
chapter~\ref{chapter:interatomic_potentials}.
Details of the computational methods, which we used to test our model, are given in
chapter~\ref{chapter:methods}.
Finally, I present our results on generating Gaussian Approximation Potentials for several systems
and the validation of the models in chapter~\ref{chapter:results}.

\chapter{Representation of atomic environments}\label{chapter:represent}

\section{Introduction}
The quantitative representation of atomic environments is an important tool in modern
computational chemistry and condensed matter physics. For example, in structure search
applications\cite{pickard_nh3_01}, each configuration that is found during the procedure depends
numerically on the precise initial conditions and the path of the search, so it is important to be
able to identify equivalent structures or detect similarities. In other applications, such as
molecular dynamics simulation of phase transitions\cite{walesbook}, one needs good order parameters
capable of detecting changes in the local order around the atoms. In constructing interatomic
potentials\cite{neur01}, the functional forms depend on elements of a carefully chosen
representation of atomic neighbourhoods, e.g. bond lengths, bond angles, etc. 

Although the Cartesian coordinate system provides a simple and unequivocal description of atomic
systems, comparisons of structures based on it are difficult: the list of coordinates can be ordered
arbitrarily, or two structures might be mapped to each other by a rotation, reflection or
translation. Hence, two different lists of atomic coordinates can in fact represent the same or very
similar structures. In a good representation, permutational, rotational and translational symmetries
are built in explicitly, i.e. the representation is {\em invariant} with respect to these
symmetries, while retaining the faithfulness of the Cartesian coordinates. If a representation is
{\em complete}, a one-to-one mapping is obtained between the genuinely different atomic environments
and the set of invariants comprising the representation. 

The most well known invariants describing atomic neighbourhoods are the set of bond-order parameters
proposed by Steinhardt et al.\cite{QW02}. These have been successfully used as order parameters in
studies of nucleation\cite{QW01}, phase transitions\cite{melting_na_01} and
glasses\cite{blaaderen_colloid01}. In the following sections we show that the bond-order parameters actually form a
subset of a more general set of invariants called the bispectrum. We prove that the bispectrum
components indeed form a rotational and permutational invariant representation of atomic environments.
The formally infinite array of bispectral invariants provide an almost complete set, and by truncating it one
obtains representations whose sensitivity can be refined at will.  

\section{Translational invariants}

The concept of the power spectrum and the bispectrum was originally introduced by the signal
processing community. In the analysis of periodic signals the absolute phase is often irrelevant and
a hindering factor, for example, when comparing signals. The problem of eliminating the phase of a
periodic function is very similar to the problem of creating a rotationally invariant representation
of spatial functions. We show how the bispectrum of periodic functions can be defined and discuss
its possible uses in atomistic simulations.

\subsection{Spectra of signals}\label{sec:power_signal}
A periodic signal $f(t)$ (or a function defined on the circumference of a circle) where
$t\in[0,2\pi)$, can be represented by its Fourier series:
\begin{equation}
f(t) = \sum_n f_n \exp(i \omega_n t) \textrm{,}
\end{equation}
where the coefficients, $f_n$, can be obtained as follows:
\begin{equation}
f_n = \frac{1}{2\pi} \int_0^{2\pi} f(t) \exp(-i \omega_n t) \mathrm{d}t \textrm{.}
\end{equation}
A phase shift of the signal (or rotation of the function) by $t_0$ transforms the original signal
according to
\begin{equation}
f(t) \to f(t+t_0) \textrm{,}
\end{equation}
and the coefficients become
\begin{equation}
f_n \to \exp(i \omega_n t_0) f_n \textrm{.}
\end{equation}
It follows that the power spectrum of the signal defined as
\begin{equation}
p_n = f_n^* f_n
\end{equation}
is invariant to such phase shifts:
\begin{equation}
p_n = f_n^* f_n \to \left( f_n \exp(i \omega_n t_0) \right)^* \left(f_n \exp(i \omega_n t_0) \right) =
f_n^* f_n \textrm{,}
\end{equation}
but the information content of different channels becomes decoupled.
\begin{figure}
\begin{center}
\includegraphics[width=8cm]{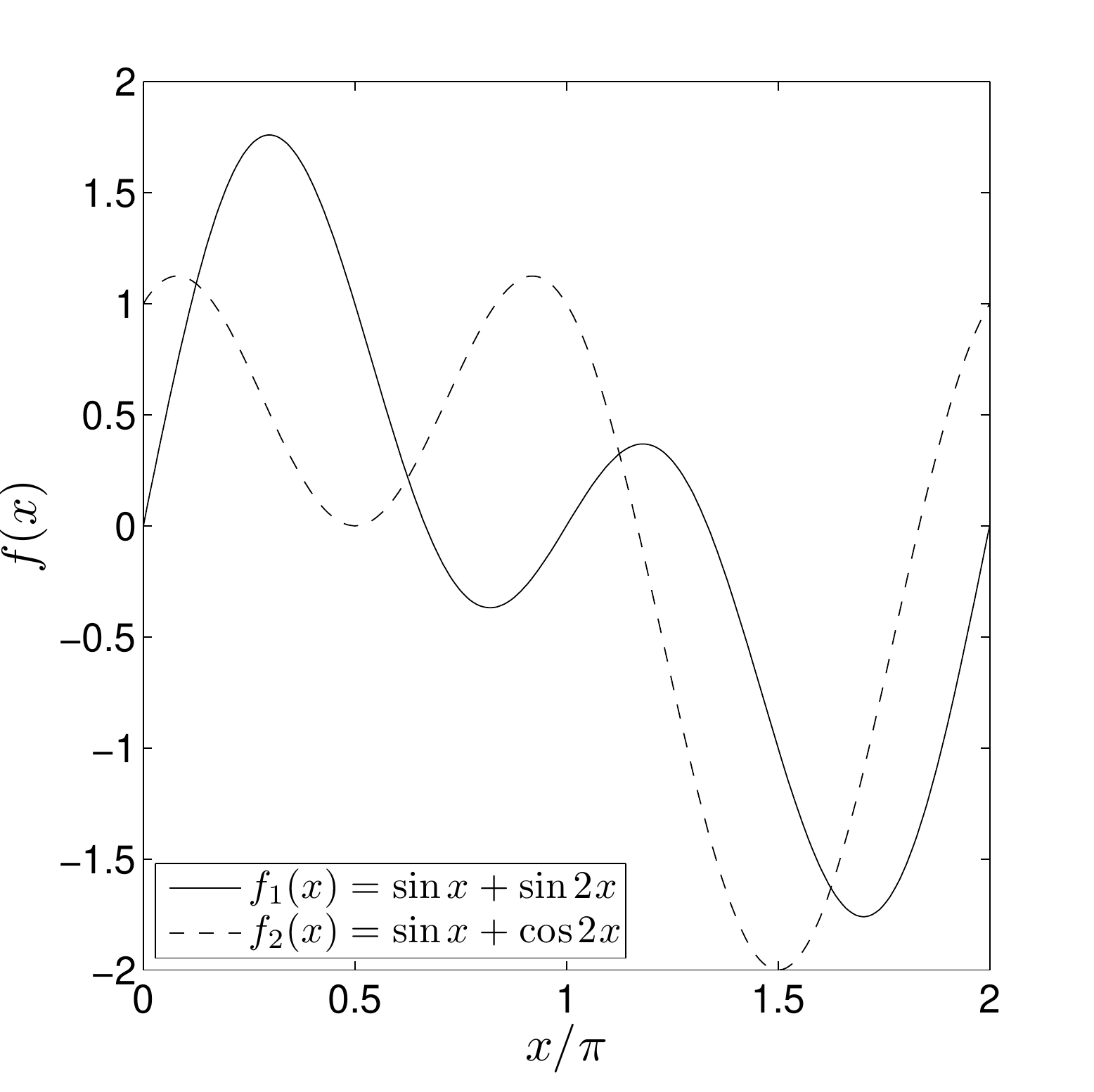}
\end{center}
\caption{\label{fig:1d_power} Two different periodic functions that share the same power spectrum
coefficients.}
\end{figure}
Figure~\ref{fig:1d_power} and table~\ref{tab:1d_power} demonstrate two functions, $f_1=\sin(t) +
\sin(2t)$ and $f_2=\sin(t) + \cos(2t)$, that can both be represented by the same power spectrum.
\begin{table}
\begin{center}
\begin{tabular}{c|r|r|r|r|r}
\hline
\hline
$\omega$       &$-2$ &$ -1$ &$ 0$ &$ 1$ &$ 2$ \\
\hline
\hline
$\mathbf{f}_1$ &$-i$ &$ -i$ &$ 0$ &$ i$ &$ i$ \\
\hline
$\mathbf{f}_2$ &$ 1$ &$ -i$ &$ 0$ &$ i$ &$ 1$ \\
\hline
$\mathbf{p}_1=\mathbf{p}_2$   &$ 1$ &$  1$ &$ 0$ &$ 1$ &$ 1$ \\
\hline
\hline
\end{tabular}
\end{center}
\caption{\label{tab:1d_power} Fourier and power spectrum coefficients of $f_1$ and $f_2$.}
\end{table}

\subsection{Bispectrum}

As the power spectrum is not complete, i.e. the original function cannot be reconstructed from it,
there is a need for an invariant representation from which the original function can (at least in
theory) be restored. The bispectrum contains the relative phase of the different channels, moreover,
it has been proven to be complete\cite{KakaralaPhD}.

A periodic function $f: \mathbb{R}^n \to \mathbb{C}$, whose period is $L_i$ in the $i$-th
direction, can be expressed in terms of a Fourier series:
\begin{equation}
   f(\mathbf{r}) = \sum_{\boldsymbol{\omega}} f(\boldsymbol{\omega}) \exp (i \boldsymbol{\omega} \mathbf{r} ) \textrm{,}
\end{equation}
where the Fourier-components can be obtained from
\begin{equation}
f(\boldsymbol{\omega}) = \prod_{i=1}^n \frac{1}{L_i} \int_V f(\mathbf{r}) \exp(i \boldsymbol{\omega} \mathbf{r} ) \mathrm{d}\mathbf{r} 
\end{equation}
and $\boldsymbol{\omega} = (\omega_1,\omega_2,\ldots,\omega_n)$.
An arbitrary translation $\hat{T}(\mathbf{r}_0)$ transforms $f$ as $f(\mathbf{r}) \to
f(\mathbf{r}-\mathbf{r}_0)$, thus the Fourier-coefficients change as $f(\boldsymbol{\omega}) \to
\exp(-i \boldsymbol{\omega} \mathbf{r}_0) f(\boldsymbol{\omega})$. The bispectrum of $f$ is defined
as the triple-correlation of the Fourier coefficients:
\begin{equation}
b(\boldsymbol{\omega}_1,\boldsymbol{\omega}_2) = f(\boldsymbol{\omega}_1) f(\boldsymbol{\omega}_2) f(\boldsymbol{\omega}_1 + \boldsymbol{\omega}_2)^* \textrm{.}
\end{equation}
The bispectrum is invariant to translations:
\begin{multline}
b(\boldsymbol{\omega}_1,\boldsymbol{\omega}_2) \to f(\boldsymbol{\omega}_1) \exp(i -\boldsymbol{\omega}_1 \mathbf{r}_0) 
f(\boldsymbol{\omega}_2) \exp(i -\boldsymbol{\omega}_2 \mathbf{r}_0) \\
\times f(\boldsymbol{\omega}_1 + \boldsymbol{\omega}_2)^* 
\exp \left( i (\boldsymbol{\omega}_1 + \boldsymbol{\omega}_2) \mathbf{r}_0 \right) = b(\boldsymbol{\omega}_1,\boldsymbol{\omega}_2)
\textrm{.}
\end{multline}

The bispectrum has been shown to be complete\cite{KakaralaPhD}. The proof, which is highly technical
and would be too long to reproduce here is based on Group Theory. Further, Dianat and Raghuveer
proved that in case of one- and two-dimensional functions the original function can be restored
using only the diagonal elements of the bispectrum, i.e. only the components for which $\boldsymbol{\omega}_1 =
\boldsymbol{\omega}_2$\cite{bis01}. 

\subsection{Bispectrum of crystals}

Crystals are periodic repetitions of a unit cell in space in each of the three directions defined by the
lattice vectors. A unit cell can be described as a parallelepiped (the description used by the
conventional Bravais system of lattices) containing some number of atoms at given positions. The
three independent edges of the parallelepiped are the lattice vectors, whereas the positions of the
atoms in the unit cell form the basis. Defining crystals in this way is not unique, as any subset of
a crystal which generates it by translations can be defined as a unit cell, for example, a
Wigner-Seitz cell, which is not even necessarily a parallelepiped.

Thus a crystal can be described by the coordinates of the basis atoms $\mathbf{r}_i$, where
$i=1,\dots,N$ and the three lattice vectors $\mathbf{a}_\alpha$, $\alpha=1,2,3$. The position of the
basis can be given in terms of the fractional coordinates $\mathbf{x}_i$, such that
\begin{equation}
\mathbf{r}_i = \sum_{\alpha=1}^3 x_{i \alpha} \mathbf{a}_\alpha \textrm{,}
\end{equation}
where $0 < x_{i \alpha} < 1$.

In the same way as in the case of atomic environments, the order of the atoms in the basis is
arbitrary. We introduce the permutational invariance through the atomic density:
\begin{equation}
\rho (\mathbf{x}) = \sum_i \delta(\mathbf{x}-\mathbf{x}_{i}) \textrm{.}
\end{equation}
$\rho$ is a periodic function in the unit cube, therefore we can expand it in a Fourier series and
calculate invariant features such as the power spectrum and bispectrum. It can be noted that the power
spectrum of $\rho$ is equivalent to the structure factor used in X-ray and neutron diffraction, and
it is clear from Section~\ref{sec:power_signal} why the structure factor is not sufficient to
determine the exact structure of a crystal. In contrast, the bispectrum of the atomic density
function could be used as a unique fingerprint of the crystal that is invariant to the permutation
and translation of the basis.

We note that permuting the lattice vectors of the crystal permutes the reciprocal lattice vectors
which therefore, mixes the elements of the bispectrum. This problem can be eliminated by first matching the
lattice vectors of the two structures which are being compared. The rotation of the entire lattice
does not change the fractional coordinates, hence the bispectrum is invariant to global rotations.

\section{Rotationally invariant features}\label{sec:rotinv}

Invariant features of atomic environments can be constructed by several methods, of which we list
a few here . In interatomic potentials, a set
of geometric parameters are used, such as bond lengths, bond angles and tetrahedral angles. These
are rotationally invariant by construction, but the size of a complete set of such parameters grows
as $\exp(N)$, where $N$ is the number of neighbours. The complete set is vastly redundant, but there
is no systematic way of reducing the number of parameters without losing completeness.

A more compact rotationally invariant representation of the atomic environment can be built in the form
of a matrix by using the bond vectors $\mathbf{r}_i$, $i=1,\ldots,N$ between the central atom and
its $N$ neighbours. The elements of the matrix are given by the dot product
\begin{equation}
M_{ij} = \mathbf{r}_i \cdot \mathbf{r}_j
\textrm{.}
\end{equation}
Matrix $\mathbf{M}$ contains the bond lengths on its diagonal, whereas the off-diagonal elements are
related to the bond angles. It can be shown that $\mathbf{M}$ is a complete
representation\cite{weylbook}. However, permuting the neighbouring atoms shuffles the columns and
rows of $\mathbf{M}$, thus $\mathbf{M}$ is not a suitable invariant representation.

Permutational invariance can be achieved by using the symmetric polynomials\cite{symm_pol}. These
are defined by
\begin{equation}
\Pi_k(x_1, x_2,\ldots,x_N) = \Pi_k(x_{\pi_1},x_{\pi_2},\ldots,x_{\pi_N})
\end{equation}
for every $\boldsymbol{\pi}$, where $\boldsymbol{\pi}$ is an arbitrary permutation of the vector
$(1,2,\ldots,N)$. The first three symmetric polynomials are
\begin{align}
\Pi_1(x_1,x_2,\ldots,x_N) &= \sum_{i}^N x_i \\
\Pi_2(x_1,x_2,\ldots,x_N) &= \sum_{i < j}^N x_i x_j \\
\Pi_3(x_1,x_2,\ldots,x_N) &= \sum_{i < j < k}^N x_i x_j x_k 
\textrm{.}
\end{align}
The series of polynomials form a complete representation, however, this set is not rotationally invariant.

\subsection{Bond-order parameters}\label{sec:bond_order}

As a first step to derive a more general invariant representation of atomic environments, we define
the local atomic density as
\begin{equation}\label{eq:atomic_density}
\rho_i (\mathbf{r}) = \sum_j \delta(\mathbf{r}-\mathbf{r}_{ij}) \textrm{,}
\end{equation}
where the index $j$ runs over the neighbours of atom $i$. The local atomic density is already
invariant to permuting neighbours, as changing the order of the atoms in the neighbour list only
affects the order of the summation. This function could be expanded in terms of spherical harmonics
(dropping the atomic index $i$ for clarity):
\begin{equation}
\label{eq:rho_expand}
\rho(\mathbf{r}) = \sum_{l=0} \sum_{m=-l}^l c_{lm}
Y_{lm}\left(\theta(\mathbf{r}),\phi(\mathbf{r})\right) \textrm{.}
\end{equation}
However, we should note that this representation does not contain information about the distances of
neighbours. In fact, $\rho(\mathbf{r})$ represented this way is the projection of the positions of
neighbouring atoms onto the unit sphere. The properties of functions defined on the unit sphere are
described by the group theory of SO(3), the group of rotations about the origin.

The spherical harmonics functions form an orthonormal basis set for $L_2$:
\begin{equation}
\langle Y_{lm} | Y_{l'm'} \rangle = \delta_{ll'} \delta_{mm'} \textrm{,}
\end{equation}
where the inner product of functions $f$ and $g$ is defined as
\begin{equation}
\langle f | g \rangle = \int f^*(\mathbf{r}) g(\mathbf{r}) \mathrm{d}\mathbf{r}
\textrm{.}
\end{equation}
The coefficients $c_{lm}$ can be determined as
\begin{equation}
c_{lm} = \langle \rho | Y_{lm} \rangle = \sum_j
Y_{lm}\left(\theta(\mathbf{r}_{ij}),\phi(\mathbf{r}_{ij})\right) \textrm{.}
\end{equation}

We note that the order parameters $Q_{lm}$ introduced by Steinhardt et al\cite{QW02} are
proportional to the coefficients $c_{lm}$. In their work, they defined the bonds in the system as
vectors joining neighbouring atoms. Defining which atoms are the neighbours of a particular atom can
be done by using a simple distance cutoff or via the Voronoi analysis. Once the set of neighbours
has been defined, each bond $\mathbf{r}_{ij}$ connecting neighbour atoms $i$ and $j$ is represented
by a set of spherical harmonics coefficients
\begin{equation}
Y_{lm}(\hat{\mathbf{r}}_{ij}) = Y_{lm}(\theta(\mathbf{r}_{ij}),\phi(\mathbf{r}_{ij})) \textrm{.}
\end{equation}
Averaging the coefficients for atom $i$ provides the atomic order parameters for that atom
\begin{equation}
Q_{lm}^i = \frac{1}{N_i} \sum_j Y_{lm} (\hat{\mathbf{r}}_{ij}) \textrm{,}
\end{equation}
where $N_i$ is the number of neighbours of atom $i$. Similarly, averaging over all bonds in the
system gives a set of global order parameters
\begin{equation}
\bar{Q}_{lm} = \frac{1}{N_b}\sum_{ij} Y_{lm}(\hat{\mathbf{r}}_{ij}) \textrm{,}
\end{equation}
where $N_b$ is the total number of bonds. Both of these order parameters are invariant to
permutations of atoms and to translations, but they still depend on the orientation of the reference
frame. However, rotationally invariant combinations of these order parameters can be constructed as
follows
\begin{align}
Q_l^i &= \left( \frac{4 \pi}{2l+1} \sum_{m=-l}^l (Q_{lm}^i)^* Q_{lm}^i \right)^{1/2}
\textrm{and} \\
W_l^i &= \sum_{m_1, m_2, m_3 = -l}^l \left(
\begin{array}{ccc}
l   & l   & l \\
m_1 & m_2 & m_3
\end{array} \right) Q_{lm_1}^i Q_{lm_2}^i Q_{lm_2}^i
\end{align}
for atoms and
\begin{align}
\bar{Q}_l &= \left( \frac{4 \pi}{2l+1} \sum_{m=-l}^l \bar{Q}_{lm}^* \bar{Q}_{lm} \right)^{1/2} \\
\bar{W}_l &= \sum_{m_1, m_2, m_3 = -l}^l \left(
\begin{array}{ccc}
l   & l   & l \\
m_1 & m_2 & m_3
\end{array} \right) \bar{Q}_{lm_1} \bar{Q}_{lm_2} \bar{Q}_{lm_2}
\end{align}
for global structures. The factor in parentheses is the Wigner-3jm symbol, which is nonzero only for
$m_1+m_2+m_3=0$.

$Q_l^i$ and $W_l^i$ are called second-order and third-order bond-order parameters, respectively. It
is possible to normalise $W_l^i$ such that it does not depend strongly on the number of neighbours
as follows:
\begin{equation}
\hat{W}_l^i = W_l^i / \left( \sum_{m=-l}^l (Q_{lm}^i)^* Q_{lm}^i \right)^{3/2}
\textrm{.}
\end{equation}

Bond-order parameters were originally introduced by Steinhardt et al\cite{QW02} for studying the order in liquids and glasses, but their approach was adopted soon for a wide range of applications. For example, the bond-order parameters, when averaged over all bonds in the system, can be used as reaction coordinates in phase transitions\cite{lj_melting_qw01}.

For symmetry reasons, bond order parameters with $l \ge 4$ have non-zero values in clusters with
cubic symmetry and $l \ge 6$ for clusters with icosahedral symmetry. The most widely calculated bond
order parameters are $l=4$ and $l=6$. Different values correspond to crystalline materials with different
symmetry, while the global values vanish in disordered phases, such as in liquids. This feature made
the $Q$ and $W$ invariants attractive for use as bond order parameters in many applications.

\subsection{Power spectrum}

Using some basic concepts from representation theory, we can now prove that the second-order invariants are rotationally invariant, then we show a more general form of invariants, a superset consisting of third-order invariants\cite{Kondor07}. An arbitrary rotation $\hat{R}$ operating on a spherical harmonic function $Y_{lm}$ transforms it into a linear combination of spherical harmonics with the same $l$ index:
\begin{equation}
\hat{R} Y_{lm} = \sum_{m'=-l}^l D^{(l)}_{mm'}(R) Y_{lm'} \textrm{,}
\end{equation}
where the matrices $\mathbf{D}^{(l)}(R)$ are also known as the Wigner-matrices. The elements of the
Wigner matrices can be generated by
\begin{equation}
D^{(l)}_{mm'}(R) = \langle Y_{lm} | \hat{R} | Y_{lm'} \rangle \textrm{.}
\end{equation}
It follows that the rotation operator $\hat{R}$ acts on the function $\rho$ as
\begin{multline}
\hat{R} \rho = \hat{R} \sum_{l=0} \sum_{m=-l}^l c_{lm} Y_{lm} = \sum_{l=0} \sum_{m=-l}^l c_{lm} \hat{R} Y_{lm} \\
             = \sum_{l=0} \sum_{m=-l}^l \sum_{m'=-l}^l c_{lm} D^{(l)}_{mm'}(R) Y_{lm'} = \\
             = \sum_{l=0} \sum_{m'=-l}^l c'_{lm} Y_{lm'} \textrm{,}
\end{multline}
thus the vector of coefficients $\mathbf{c}_l$ transform under rotation as
\begin{equation}
\mathbf{c}_l \to \mathbf{D}^{(l)}(R) \mathbf{c}_l \textrm{.}
\end{equation}

Making use of the fact that rotations are unitary operations, it is possible to show that the
matrices $\mathbf{D}^{(l)}$ are unitary, i.e.
\begin{equation}
\left( \mathbf{D}^{(l)} \right)^\dagger \mathbf{D}^{(l)} = \mathbf{I} \textrm{,}
\end{equation}
leading us to a set of rotationally invariant coefficients, the rotational power spectrum:
\begin{equation}
p_l = \mathbf{c}_l^\dagger \mathbf{c}_l \textrm{.}
\end{equation}
The coefficients of the power spectrum remain invariant under rotations:
\begin{equation}
p_l = \mathbf{c}_l^\dagger \mathbf{c}_l \to \left( \mathbf{c}_l ^\dagger \left( \mathbf{D}^{(l)}
\right) ^\dagger \right) \left( \mathbf{D}^{(l)} \mathbf{c}_l \right) = \mathbf{c}_l^\dagger \mathbf{c}_l
\textrm{.}
\end{equation}
It can be directly seen that the second-order bond-order parameters are related to the power spectrum via the simple equation
\begin{equation}
Q_l = \left( \frac{4 \pi}{2l+1} p_l \right)^{1/2} \textrm{.}
\end{equation}
The power spectrum is a very impoverished representation of the original function $\rho$, because
all $p_l$ coefficients are rotationally invariant \emph{independently}, i.e. different $l$ channels
are decoupled. This representation, although rotationally invariant, is, in turn, severely
incomplete.

\begin{figure}
\begin{center}
\includegraphics[width=6cm]{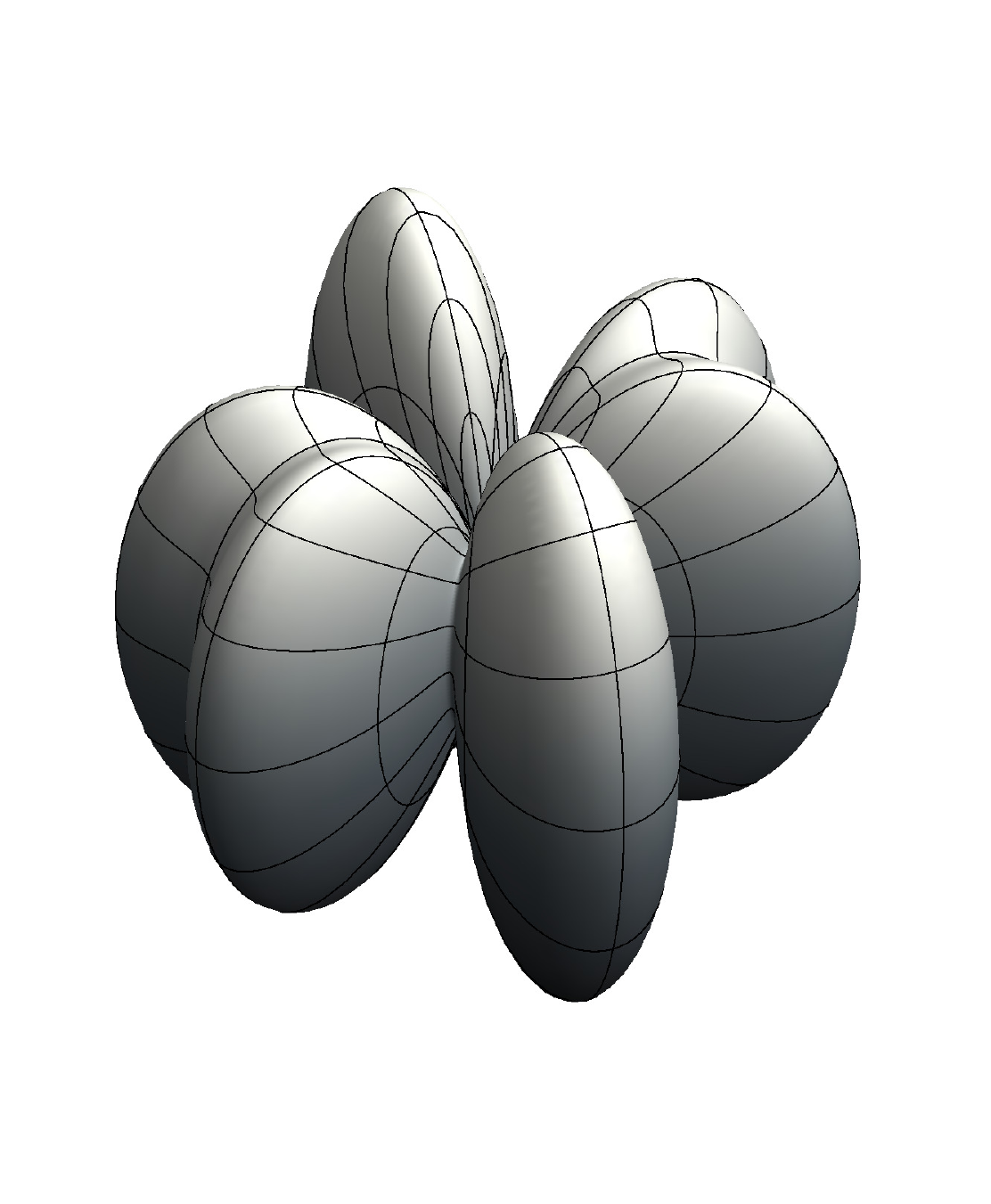}
\includegraphics[width=6cm]{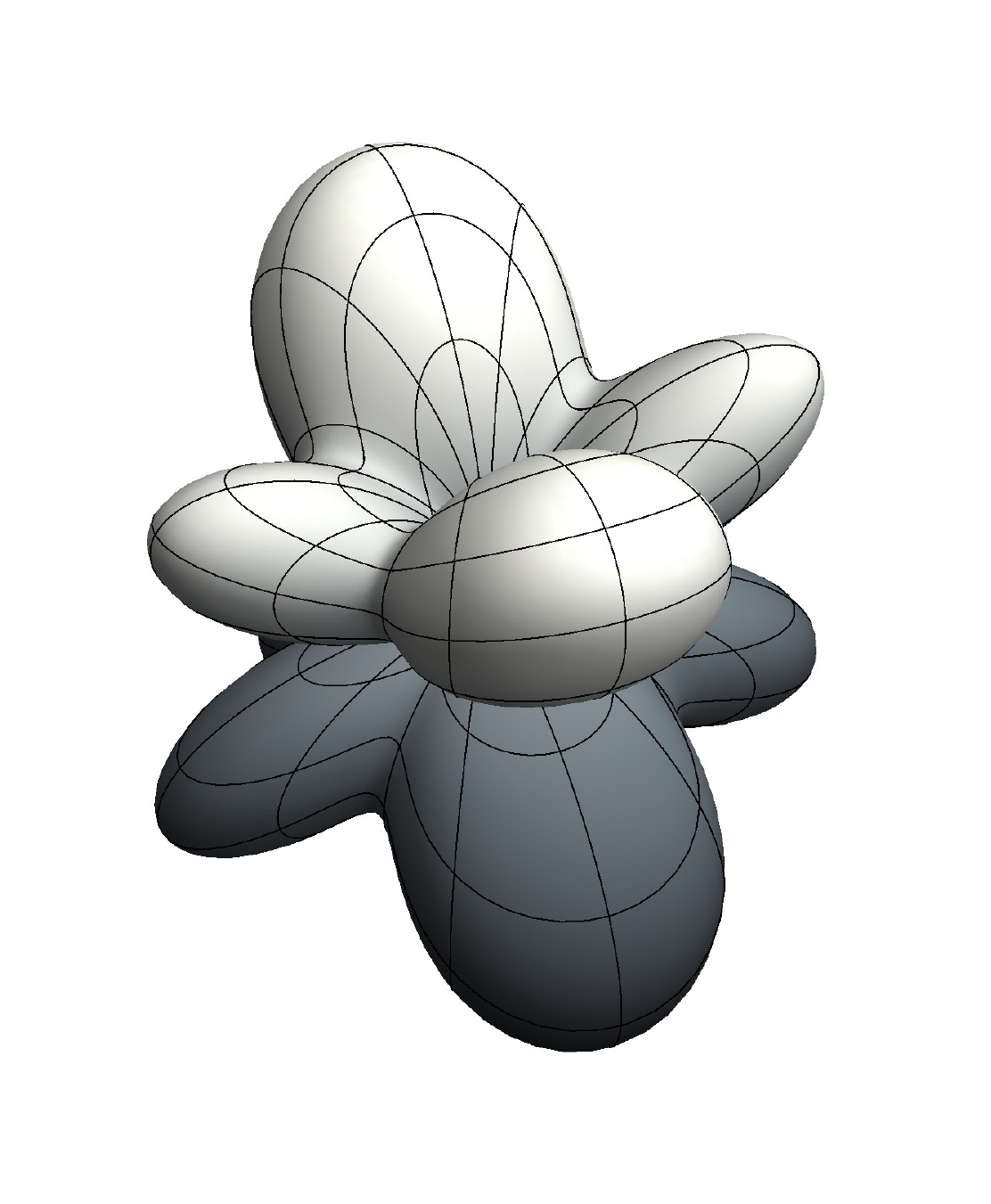}
\end{center}
\caption{\label{fig:so3_power} Two different angular functions that share the same power spectrum coefficients.}
\end{figure}

The incompleteness of the power spectrum can be demonstrated by the following example. Assuming a function $f$ in the form
\begin{equation}
f(\hat{\mathbf{r}}) = \sum_{m=-l_1}^{l_1} \alpha_m Y_{l_1m}(\hat{\mathbf{r}}) + \sum_{m=-l_2}^{l_2}
\beta_m Y_{l_2 m}(\hat{\mathbf{r}}) \textrm{,}
\end{equation}
its power spectrum elements are $p_{l_1} = |\boldsymbol{\alpha}|^2 $ and $p_{l_1} =
|\boldsymbol{\beta}|^2$. Thus only the \emph{length} of the vectors $\boldsymbol{\alpha}$ and
$\boldsymbol{\beta}$ are constrained by the power spectrum, their relative orientation is lost,
i.e. the information content of channels $l_1$ and $l_2$ becomes decoupled. Figure~\ref{fig:so3_power} shows two different angular functions, $f_1 = Y_{22} + Y_{2-2} + Y_{33} + Y_{3-3}$ and $f_2 = Y_{21} + Y_{2-1} + Y_{32} + Y_{3-2}$ that have the same power spectrum $p_2=2$ and $p_3=2$.

\subsection{Bispectrum}

We will now generalise the concept of the power spectrum in order to obtain a more complete set of
invariants via the coupling of the different angular momentum channels\cite{Kondor07}. Let us
consider the direct product $\mathbf{c}_{l_1} \otimes
\mathbf{c}_{l_2}$, which transforms under a rotation as
\begin{equation}
\mathbf{c}_{l_1} \otimes \mathbf{c}_{l_2} \to \left( \mathbf{D}^{(l_l)} \otimes \mathbf{D}^{(l_2)}
\right) \left( \mathbf{c}_{l_1} \otimes \mathbf{c}_{l_2} \right) \textrm{.}
\end{equation}
It follows from the representation theory of groups that the direct product of two irreducible
representations can be decomposed into direct sum of irreducible representations of the same group. In case of the SO(3) group, the direct product of two Wigner-matrices can be decomposed into a direct sum of Wigner-matrices in the form
\begin{equation}
\mathbf{D}^{(l_l)} \otimes \mathbf{D}^{(l_2)} = \left( \mathbf{C}^{l_1,l_2} \right)^\dagger
\left[ \bigoplus_{l=|l_1-l_2|}^{l_1+l_2} \mathbf{D}^{(l)} \right] \mathbf{C}^{l_1,l_2} \textrm{,}
\end{equation}
where $\mathbf{C}^{l_1,l_2}$ denote the Clebsch-Gordan coefficients. The matrices of Clebsch-Gordan coefficients are themselves unitary, hence the vector $\mathbf{C}^{l_1,l_2} \left( \mathbf{c}_{l_1}
\otimes \mathbf{c}_{l_2} \right) $ transforms as
\begin{equation}
\mathbf{C}^{l_1,l_2} \left( \mathbf{c}_{l_1} \otimes \mathbf{c}_{l_2} \right) \to
\left[ \bigoplus_{l=|l_1-l_2|}^{l_1+l_2} \mathbf{D}^{(l)} \right] \mathbf{C}^{l_1,l_2} 
\left( \mathbf{c}_{l_1} \otimes \mathbf{c}_{l_2} \right) \textrm{.}
\end{equation}
We define $\mathbf{g}_{l_1,l_2,l}$ as
\begin{equation}
\bigoplus_{l=|l_1-l_2|}^{l_1+l_2} \mathbf{g}_{l_1,l_2,l} 
\equiv \mathbf{C}^{l_1,l_2} \left( \mathbf{c}_{l_1} \otimes \mathbf{c}_{l_2} \right),
\end{equation}
i.e. the $\mathbf{g}_{l_1,l_2,l}$ is that part of the RHS which transforms under rotation as
\begin{equation}
\mathbf{g}_{l_1,l_2,l} \to \mathbf{D}^{(l)} \mathbf{g}_{l_1,l_2,l} \textrm{.}
\end{equation}

Analogously to the power spectrum, the bispectrum components or cubic
invariants, can be written as
\begin{equation}
b_{l_1,l_2,l} = \mathbf{c}_l^\dagger \mathbf{g}_{l_1,l_2,l} \textrm{,} 
\end{equation}
which are invariant to rotations:
\begin{equation}
b_{l_1,l_2,l} =  \mathbf{c}_l^\dagger \mathbf{g}_{l_1,l_2,l} \to 
\left( \mathbf{c}_l \mathbf{D}^{(l)} \right)^\dagger \mathbf{D}^{(l)} \mathbf{g}_{l_1,l_2,l}
= \mathbf{c}_l^\dagger \mathbf{g}_{l_1,l_2,l}
\end{equation}

Kondor showed that the bispectrum of the SO(3) space is not complete, i.e. the bispectrum does not
determine uniquely the original function. This is a deficiency due to the fact that the unit sphere,
$S_2$ is a homogeneous space. However, he states that the bispectrum is still a remarkably rich
invariant representation of the function.

Rewriting the bispectrum formula as
\begin{equation}
\label{eq:bispectrum_so3}
b_{l_1,l_2,l} = \sum_{m=-l}^l \sum_{m_1=-l_1}^{l_1} \sum_{m_2=-l_2}^{l_2} c_{lm}^*
C^{lm}_{l_1 m_1 l_2 m_2} c_{l_1 m_1} c_{l_2 m_2} \textrm{,}
\end{equation}
the similarity to the third-order bond-order parameters becomes apparent. Indeed, the Wigner
3jm-symbols are related to the Clebsch-Gordan coefficients through
\begin{equation}
\left(
\begin{array}{ccc}
l_1 & l_2 & l_3 \\
m_1 & m_2 & m_3
\end{array} \right) = \frac{(-1)^{l_1-l_2-m_3}}{\sqrt{2l_3+1}} C^{lm}_{l_1 m_1 l_2 m_2}
\textrm{.}
\end{equation}
For the spherical harmonics $Y_{lm} = (-1)^m Y^*_{l-m}$, thus the third-order parameters $W_l$ are
simply the diagonal elements of the bispectrum $b_{l,l,l}$ up to a scalar factor, and thus, the
bispectrum is a superset of the third-order bond-order parameters. Further, considering that $Y_{00}
\equiv 1$, therefore the coefficient $c_{00}$ is simply the number of neighbours $N$, and
$C_{m,0,m_2}^{l,0,l_2} = \delta_{l,l_2} \delta_{m,m_2}$, we notice that the bispectrum elements
$l_1=0$, $l=l_2$ are the power spectrum components, previously introduced:
\begin{equation}
b_{l,0,l} = N_i \sum_{m=-l}^l \sum_{m_2=-l}^{l} c_{lm}^* \delta_{m,m_2} c_{l m_2} = N_i \sum_{m=-l}^l
c_{lm}^* c_{lm} = N_i p_l \textrm{.}
\end{equation}

Finally, the relationship between the bond-order parameters and the bispectrum can be summarised as
\begin{align}
Q_l &\propto \sqrt{p_l} \propto \sqrt{b_{l,0,l}}\\
W_l & \propto b_{l, l, l}
\textrm{.}
\end{align}

\subsubsection{Radial dependence}

The bispectrum is still a very incomplete representation, as it uses the unit-sphere projection of
the atomic environment, i.e. the distance of the atoms from the centre is not represented. One way to
improve this shortcoming---namely, the lack of radial information---is to introduce radial basis
functions\cite{qw_meam01}, completing the basis for three-dimensional space. In
equation~\ref{eq:rho_expand}, we use the product of spherical harmonics and a linearly independent
set of radial functions $g_n$:
\begin{equation}
\rho(\mathbf{r}) = \sum_n \sum_{l=0} \sum_{m=-l}^l c_{nlm} g_n(r)
Y_{lm}\left(\theta(\mathbf{r}),\phi(\mathbf{r})\right) \textrm{.}
\end{equation}
If the set of radial basis functions is not orthonormal, i.e. $\langle g_n | g_m \rangle = S_{nm} \ne \delta_{nm}$, after obtaining the coefficients $c'_{nlm}$ with
\begin{equation}
c'_{nlm} = \langle g_n Y_{lm} | \rho \rangle \textrm{,}
\end{equation}
the elements $c_{nlm}$ are given as
\begin{equation}
c_{nlm} = \sum_{n'} \left(S^{-1}\right)_{n'n} c'_{n'lm} \textrm{.}
\end{equation}
In practice, when constructing the invariants, both $c'_{nlm}$ and $c_{nlm}$ can be used.

Rotational invariance only applies globally, therefore the different angular momentum channels
corresponding to various radial basis functions need to be coupled. Simply extending
equation~\ref{eq:bispectrum_so3} to the form
\begin{equation}
b_{n,l_1,l_2,l} = \sum_{m=-l}^l \sum_{m_1=-l_1}^{l_1} \sum_{m_2=-l_2}^{l_2} c_{nlm}^*
C^{lm}_{l_1 m_1 l_2 m_2} c_{nl_1 m_1} c_{nl_2 m_2} \textrm{,}
\end{equation}
provides a set of invariants describing the three-dimensional neighbourhood of the atom. In fact,
this formula can easily lead to a poor representation, if the radial functions have little overlap
with each other, as the coefficients belonging to different $n$ channels become decoupled. To avoid
this, it is necessary to choose wide, overlapping radial functions, although this greatly reduces the
sensitivity of each channel. The fine-tuning of the basis set is rather arbitrary, and there does
not necessarily exist an optimum for all systems.
\begin{figure}
\begin{center}
\includegraphics[width=6cm]{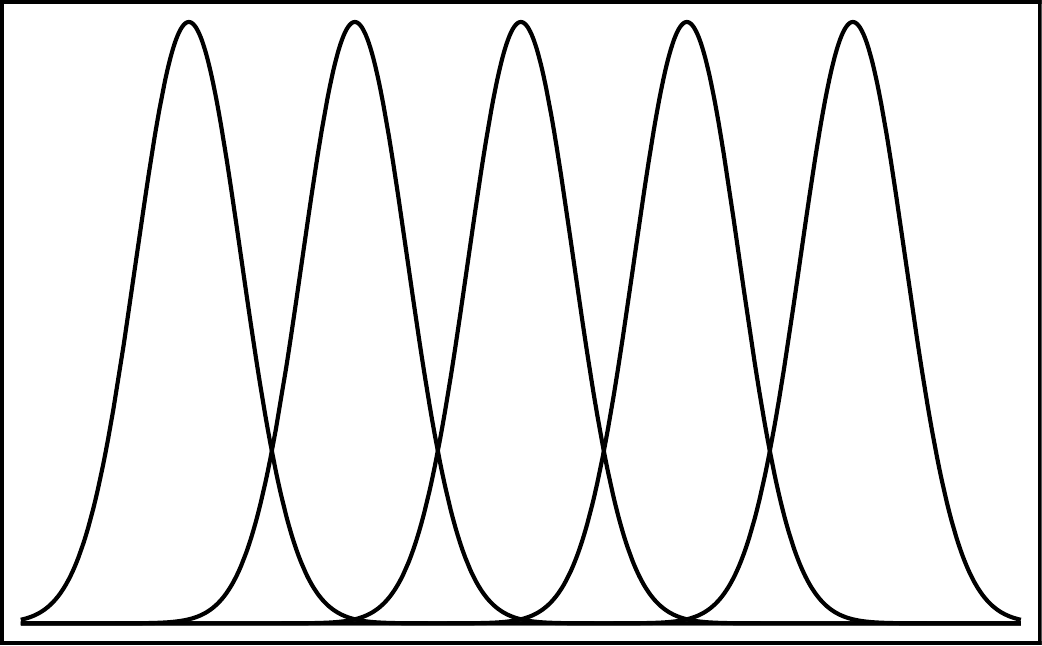}
\includegraphics[width=6cm]{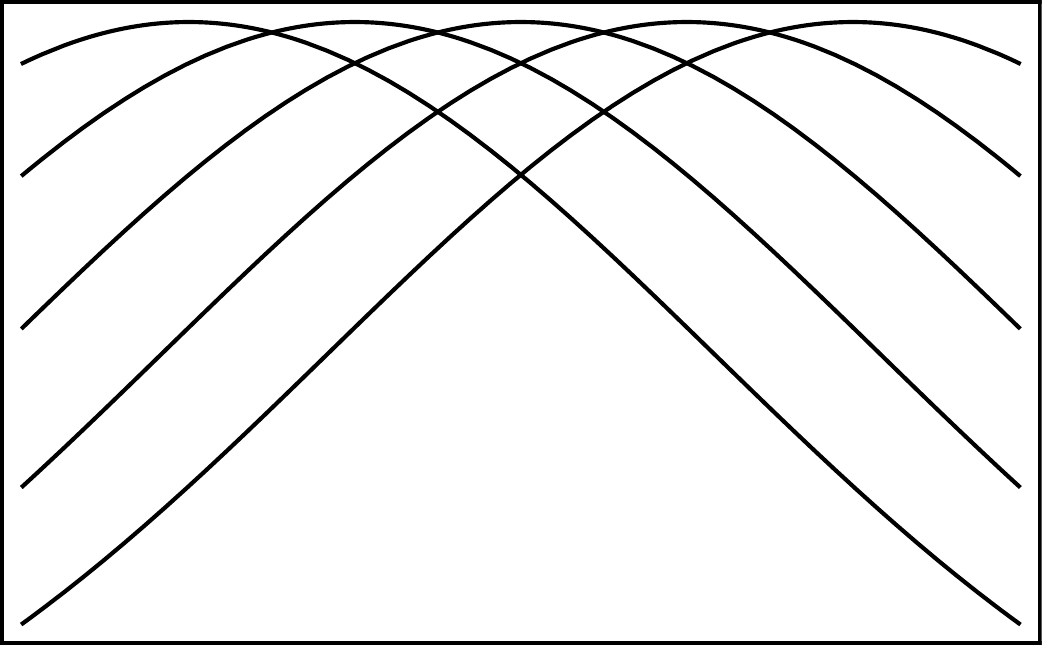}
\end{center}
\caption{\label{fig:radial_basis} Two possible sets of radial basis functions, Gaussians centred at different radii. The narrow Gaussians are more sensitive to changes in radial positions, but the coupling between them is weaker.}
\end{figure}
An alternative way to construct invariants from $\mathbf{c}$ is to couple different radial channels,
for example, as
\begin{equation}
b_{n_1,n_2,l_1,l_2,l} = \sum_{m=-l}^l \sum_{m_1=-l_1}^{l_1} \sum_{m_2=-l_2}^{l_2} c_{n_1 lm}^*
C^{lm}_{l_1 m_1 l_2 m_2} c_{n_2 l_1 m_1} c_{n_2 l_2 m_2} \textrm{.}
\end{equation}
Now we ensure that radial channels cannot become decoupled, but at the price of increasing the
number of invariants quadratically. Although adding a suitable set of radial functions allows one to
construct a complete representation, we found this approach overly complicated. A high degree of
arbitrariness is introduced by having to choose a radial basis. 

\subsection{4-dimensional bispectrum}

\begin{figure}
\begin{center}
\includegraphics[width=6cm]{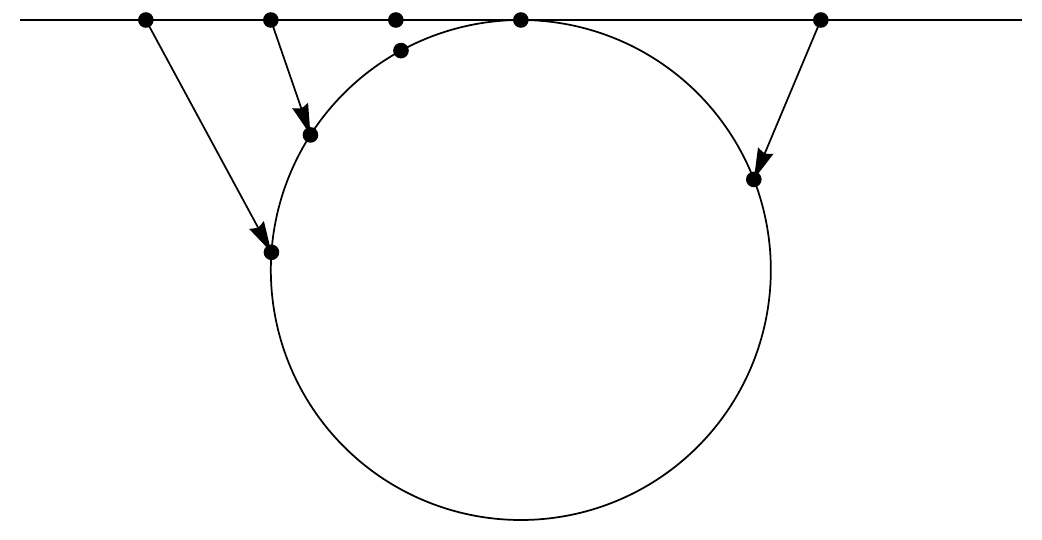}
\includegraphics[width=6cm]{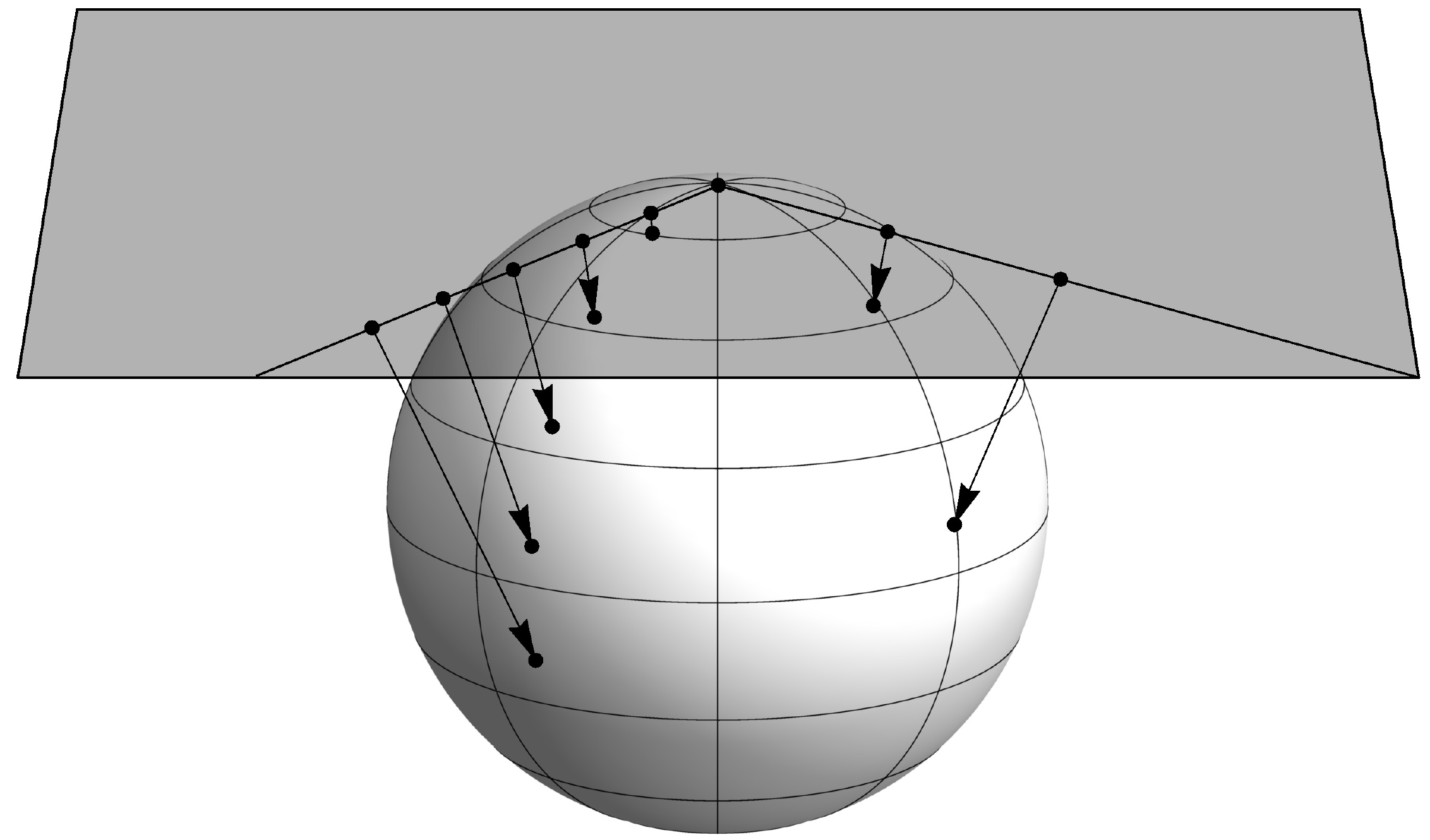}
\end{center}
\caption{\label{fig:riemann} Projection of a line to a circle (left panel), projection the two-dimensional plane onto the three-dimensional sphere (right panel). The projection we use is in equation~\ref{eq:tfm} the generalisation to one more dimension.}
\end{figure}

Instead of using a rather arbitrary radial basis set, we propose a generalisation of the power
spectrum and bispectrum that does not require the explicit introduction of a radial basis set, yet
still forms a complete basis of three-dimensional space.  We start by projecting the atomic
neighbourhood density onto the surface of the four-dimensional unit sphere, in a similar fashion to
the Riemann-construction:
\begin{equation}\label{eq:tfm}
\mathbf{r}\equiv\left(\begin{matrix}x\\y\\z\end{matrix}\right) \rightarrow
\begin{array}{lcl}\phi &=& \arctan(y/x) \\ \theta &=& \arccos(z/|\mathbf{r}|) \\ \theta_0 &=&
|\mathbf{r}|/r_0 \end{array} \textrm{,}
\end{equation}
where $r_0 > r_\textrm{cut}/\pi$. Using this projection, rotations in the three-dimensional space
correspond to rotations in the four-dimensional space. Figure~\ref{fig:riemann} shows such 
projections for 1 and 2 dimensions, which can be more easily drawn than the three-dimensional case
that we use here.

An arbitrary function $\rho$ defined on the surface of a 4D sphere can be numerically represented
using the hyperspherical harmonics functions $U^j_{m'm}(\phi, \theta, \theta_0)$:
\begin{equation}\label{eq:4d_exp}
\rho = \sum _{j=0} ^\infty \sum _{m,m'=-j}^{j} c^j_{m'm} U^j_{m'm} \textrm{.}
\end{equation}
The hyperspherical harmonics form an orthonormal basis set, thus the expansion coefficients
$c^j_{m'm}$ can be calculated via
\begin{equation}
c^j_{m'm} = \langle U^j_{m'm} | \rho \rangle \textrm{,}
\end{equation}
where $\langle . | . \rangle$ denotes the inner product in 4-dimensional space. Although the
coefficients $c^j_{m'm}$ have two indices for each $j$, they are vectors and, for clarity, we denote
them as $\mathbf{c}^j$. Similarly to the three-dimensional case, a unitary operation $\hat{R}$, such
as a rotation, acts on the hyperspherical harmonics functions as
\begin{equation}
\hat{R} U^j_{m_1' m_1} = \sum_{m_2' m_2} R^j_{m_1' m_1 m_2' m_2} U^j_{m_2' m_2} \textrm{,}
\end{equation}
where the matrix elements $R^j_{m_1' m_1 m_2' m_2}$ are given by
\begin{equation}
R^j_{m_1' m_1 m_2' m_2} = \langle U^j_{m_1' m_1} | \hat{R} | U^j_{m_2' m_2} \rangle \textrm{.}
\end{equation}
Hence the rotation $\hat{R}$ acting on $\rho$ transforms the coefficient vectors $\mathbf{c}^j$
according to
\begin{equation}
\mathbf{c}^{j} \to \mathbf{R}^j \mathbf{c}^j \textrm{.}
\end{equation}
$\mathbf{R}^j$ are unitary matrices, i.e. $\left( \mathbf{R}^j \right) ^\dagger \mathbf{R}^j = \mathbf{I}$.

The product of two hyperspherical harmonics functions can be expressed as the linear combination of hyperspherical harmonics \cite{angmombook}:
\begin{equation} \label{eq:4d_cg}
U^{l_1}_{m'_1 m_1} U^{l_2}_{m'_2 m_2} = \sum _{l = |l_1-l_2|} ^{l_1+l_2} C^{lm}_{l_1 m_1 l_2 m_2}
C^{lm}_{l_1 m_1 l_2 m_2} U^l_{m'm} \textrm{,}
\end{equation}
where $C^{lm}_{l_1 m_1 l_2 m_2}$ are the well-known Clebsch-Gordan coefficients.
We can recognise in equation~\ref{eq:4d_cg} the four dimensional analogues of the Clebsch-Gordan expansion coefficients, defined as $H^{l m m'}_{l_1 m_1 m_1',l_2 m_2 m_2'} \equiv
C^{lm}_{l_1 m_1 l_2 m_2} C^{lm'}_{l_1 m_1' l_2 m_2'}$. Using the matrix notation of the expansion coefficients, it can be shown that the direct product of the four-dimensional rotation matrices decompose according to
\begin{equation}
\mathbf{R}^{j_1} \otimes \mathbf{R}^{j_2} = \left( \mathbf{H}^{j_1,j_2} \right)^\dagger
\left[ \bigoplus_{j=|j_1-j_2|}^{j_1+j_2} \mathbf{R}^{j} \right] \mathbf{H}^{j_1,j_2} \textrm{.}
\end{equation}
The remainder of the derivation continues analogously to the 3D case. Finally, we arrive at the
expression for the bispectrum elements, given by
\begin{equation}
B_{j_1,j_2,j} = \sum_{m'_1, m_1 = -j_1}^{j_1} \sum_{m'_2, m_2 = -j_2}^{j_2}
\sum_{m', m = -j}^{j}
\bigl( c^j_{m'm} \bigr)^* \\C^{jm}_{j_1 m_1 j_2 m_2} C^{jm'}_{j_1 m'_1 j_2 m'_2}
c^{j_1}_{m'_1 m_1} c^{j_2}_{m'_2 m_2} \textrm{.}
\end{equation}
Note that the 4D power spectrum can be constructed as
\begin{equation}
P_j = \sum_{m', m = -j}^{j} \left( c^j_{m'm} \right)^* c^j_{m'm} \textrm{.}
\end{equation}

The 4D bispectrum is invariant with respect to rotations of four-dimensional space, which
include three-dimensional rotations. However, there are
additional rotations, associated with the third polar angle $\theta_0$, which, in our case,
represents the radial information. In order to eliminate the invariance with respect to the third
polar angle, we modified the atomic density as follows
\begin{equation}
\rho_i (\mathbf{r}) = \delta(\mathbf{0}) + \sum_j \delta(\mathbf{r}-\mathbf{r}_{ij}) \textrm{,}
\end{equation}
i.e. by adding the central atom as a reference point.

The magnitude of the elements of the bispectrum scale as the cube of the number of neighbours, so we
take the cube-root of the coefficients in order to make the comparison of different spectra easier.

\subsection{Results}

\begin{figure}[htb]
\begin{center}
\includegraphics[width=12cm]{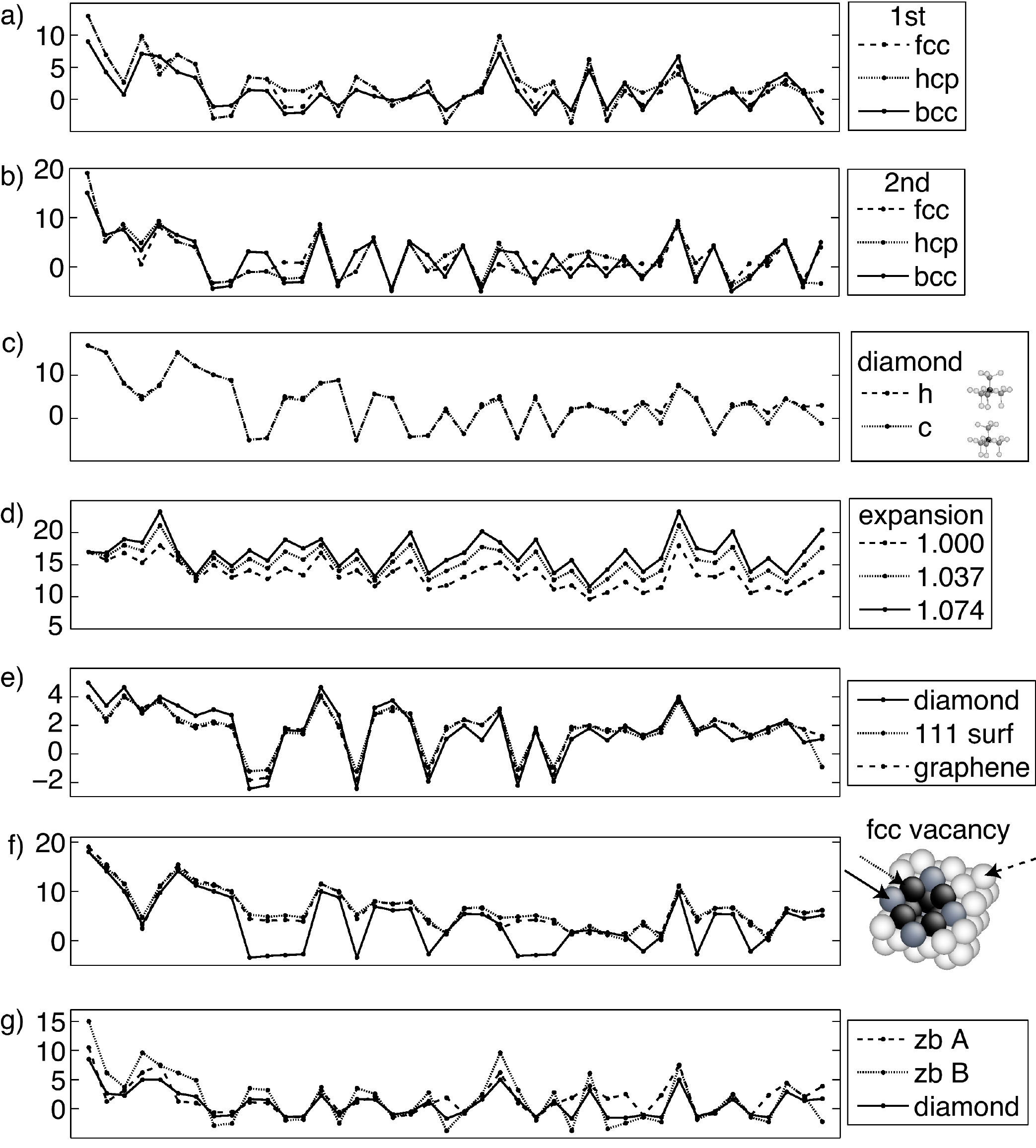}
\end{center}
\caption{\label{fig:bispectra} 4D bispectra of atoms in various structures: a) fcc/hcp/bcc lattices with a first neighbour cutoff; b) fcc/hcp/bcc lattices with a second neighbour cutoff; c) hexagonal and cubic diamond lattice; d) expansion of a diamond lattice; e) bulk diamond, $(111)$ surface of diamond and graphene; f) fcc vacancy; g) the A and B atoms in a zincblende structure, compared with diamond.}
\end{figure}

In practice, the infinite spherical harmonic expansion of the atomic neighbourhood is truncated to
obtain a finite array of bispectral invariants. In Figure~\ref{fig:bispectra} we show the 4D
bispectra of atoms in a variety of environments, truncated to $j \leq 4$, which gives 42 bispectrum
coefficients. In each case the $r_0$ parameter was set to highlight differences between the
bispectral elements. 

\begin{figure}[htb]
\begin{center}
\includegraphics[width=4cm]{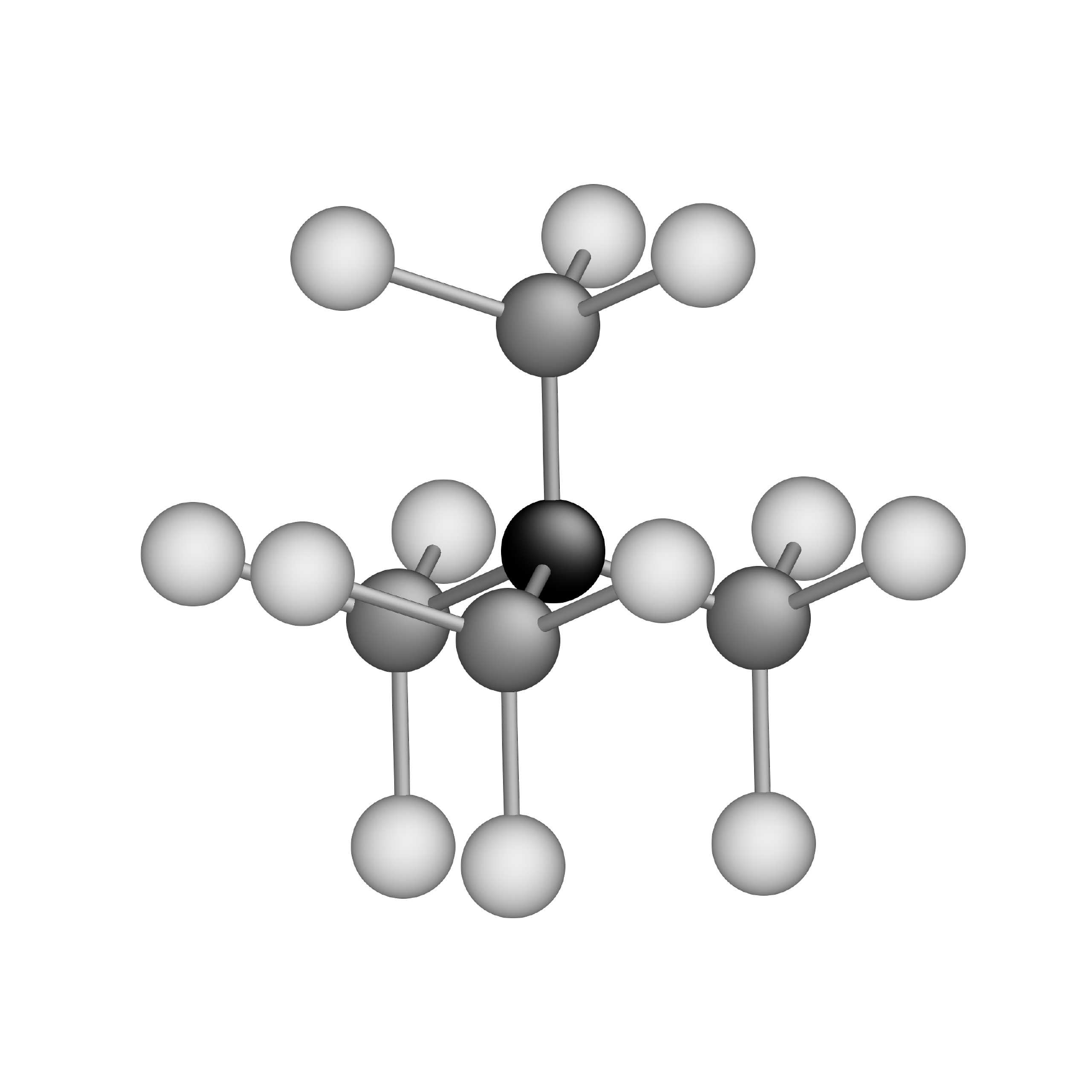}
\includegraphics[width=4cm]{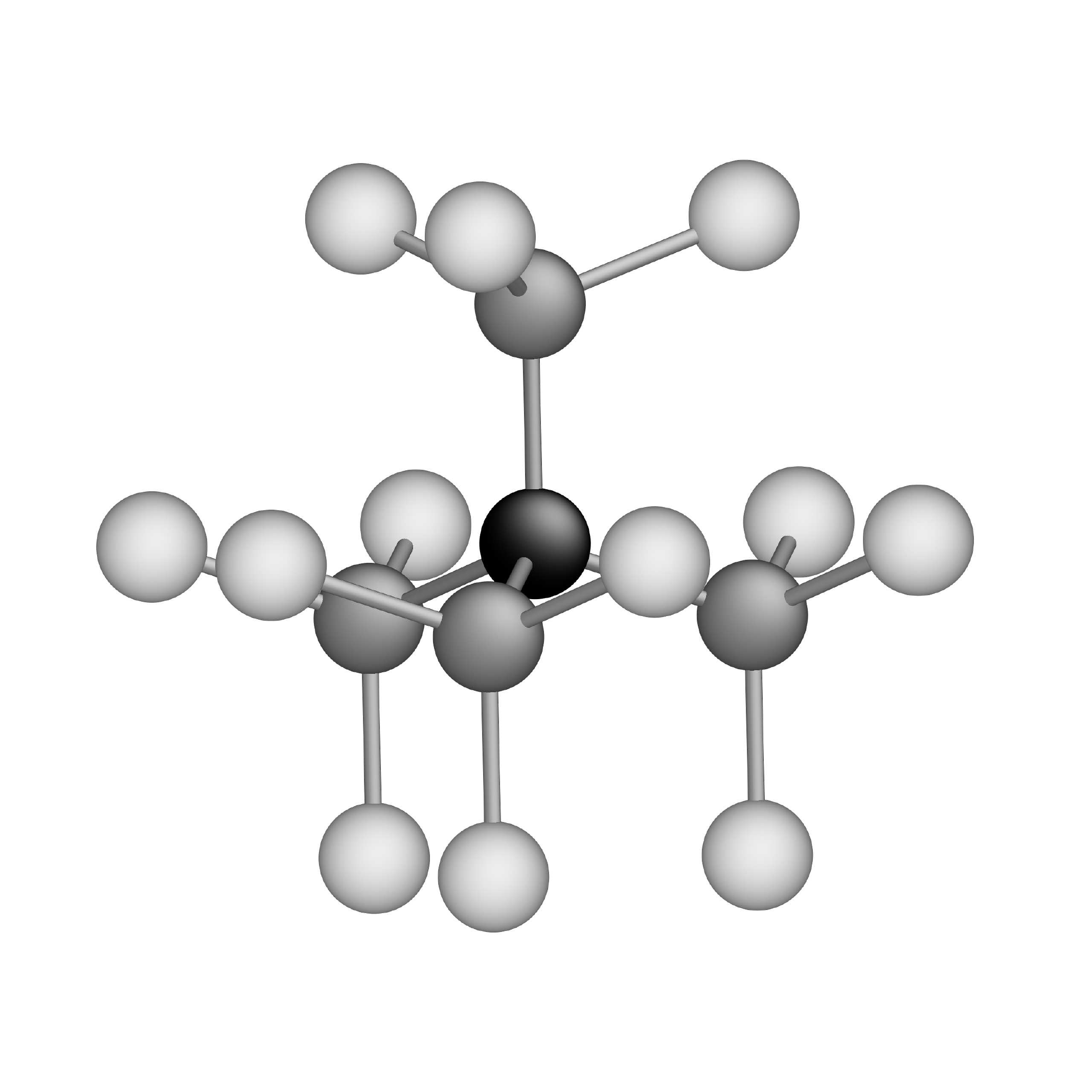}
\end{center}
\caption{\label{fig:cubhex} Cubic and hexagonal diamond. Cubic diamond is shown in the left panel.}
\end{figure}

It can be seen from figure~\ref{fig:bispectra} that the bispectrum is capable of distinguishing very
subtle differences in atomic neighbourhood environments. Some points of particular interest are 
the following. The difference between the face-centred cubic (fcc) and the hexagonal close-packed (hcp)  structures is very small
within the first neighbour shell, as is the difference between the corresponding bispectra
(panel a). However, the difference is much more pronounced once second neighbours are included (panel b). The
difference between the cubic and hexagonal diamond lattices is the stacking order of the $(111)$
sheets. The positions of the four nearest neighbours and nine atoms of the second-nearest neighbour
shell are the same and, only the positions of the remaining three neighbours are different, as shown in
figure~\ref{fig:cubhex}.
The curves in figure~\ref{fig:bispectra}c reflect the similarity of these two structures: most of
the bispectrum coefficients are equal, except a few, which can be used for distinguishing the
structures.  Figure~\ref{fig:bispectra}d shows the bispectra of three atoms in perfect diamond
lattices, which differ in the lattice constants. This plot illustrates the sensitivity of the
bispectrum in the radial dimension because the expansion of a lattice leaves all angular coordinates
the same. It can be seen that the first element of the bispectrum array remains the same, because
this is proportional only to the number of neighbours. 

\begin{figure}[htb]
\begin{center}
\includegraphics[width=12cm]{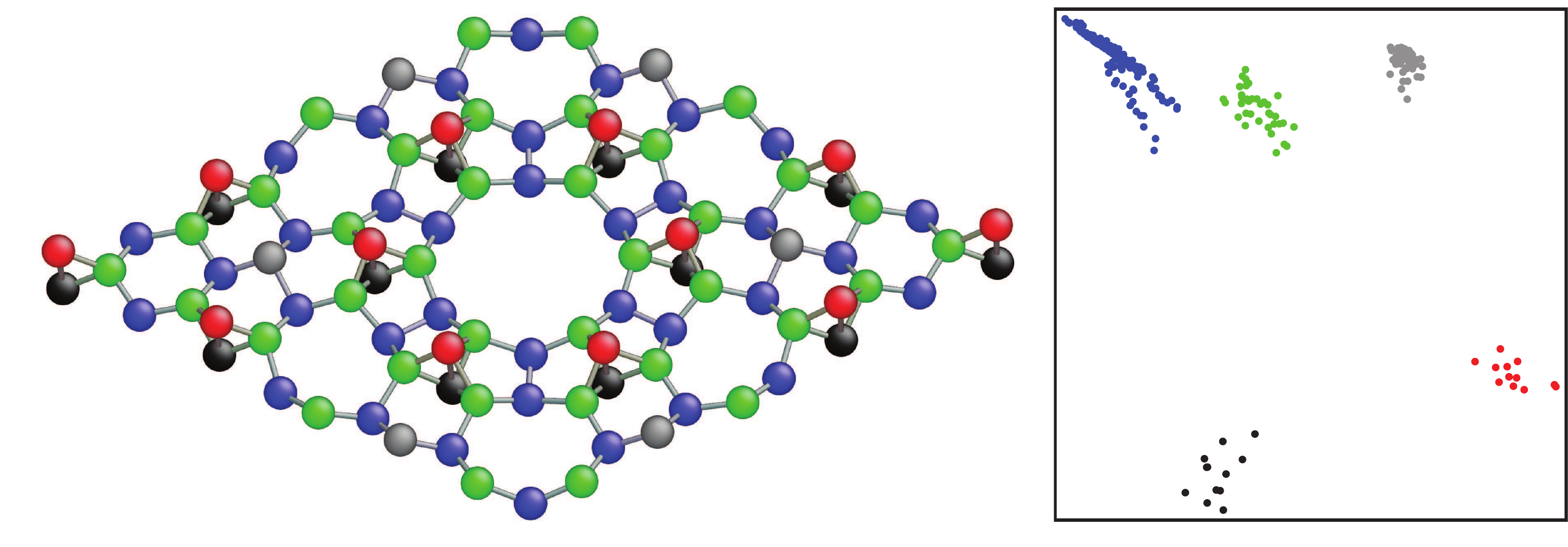}
\end{center}
\caption{\label{fig:surf_111} Principle component analysis of the bispectrum of atoms on the
$7\times7$ reconstruction of the $(111)$ surface of silicon.}
\end{figure}

We performed the principle component analysis\cite{pca01} on the bispectra of atoms in a slab of
silicon. On the surface of the slab, the atoms were arranged according to the $7\times7$
reconstruction\cite{pandey77}. The position of the atoms were randomised by 0.3~{\AA}. We projected
the 42-dimensional space of the bispectrum---which corresponds to $j \le 4$---to the two-dimensional
plane and clustered the points using the k-means algorithm\cite{datacluster01}. In
figure~\ref{fig:surf_111}, we show the result of the principle component analysis. Different colours
are assigned to each cluster identified by the k-means method, and we coloured the atoms with
respect to the cluster they belong. This example demonstrates that the bispectrum can be used to
identify atomic environments in an automatic way.

It is straightforward to describe multi-species atomic environments using the bispectrum. We modify
the atomic density function defined in equation~\ref{eq:atomic_density} as
\begin{equation}\label{eq:atomic_density_multi}
\rho_i (\mathbf{r}) = s_i \delta(\mathbf{0}) + \sum_j s_j \delta(\mathbf{r}-\mathbf{r}_{ij}) \textrm{,}
\end{equation}
where $\mathbf{s}$ contains an arbitrary set of coefficients, different for each species, which are thus distinguished. Figure~\ref{fig:bispectra}g shows the resulting bispectra for the
two different atoms in the zincblende lattice, as well as the diamond lattice for comparison. It can
be seen that the bispectrum successfully distinguishes between the different species.

\chapter{Gaussian Process}\label{chapter:gp}
\section{Introduction}

Regression methods are important tools in data analysis. Parametric models can be expressed in
functional forms that contain free parameters that are fitted such that the models reproduce
observations. The model can often be formulated in a way that the functional form is a linear
combination of the parameters. The fitting procedure in such cases is called linear regression.
Non-linear regression is needed if the functional form cannot be expressed as a simple linear
combination of the parameters, but this case does not differ conceptually from the linear case. However,
there is often no theory or model describing a particular process---or it is just too complicated to
write the model in a closed functional form---, but it is still important to make predictions of the
outcome of the process. Non-parametric approaches, such as neural networks or Gaussian Processes,
can be used to approximate the underlying function given a set of previously collected data. As
neural network methods form a subset of Gaussian Processes\cite{RW0}, we decided to use the latter
approach in our work.

\section{Function inference}

Gaussian Processes predict the values of a function whose form is not explicitly known by using function
observations as evidence. If $\mathbf{t}=\{t_i\}_{i=1}^N$ are values of a function $f:\mathbb{R}^n
\to \mathbb{R}$ measured at the points $\mathbf{X} = \{\mathbf{x}_i\}_{i=1}^N$ with some error,
predicting the value $t_{N+1}$ at $\mathbf{x}_{N+1}$ can be formulated as a Bayesian inference
problem. Bayes' theorem states that
\begin{equation}
   P(t_{N+1} | \mathbf{t}) = \frac{ P( \mathbf{t}|t_{N+1} ) P(t_{N+1}) }{ P( \mathbf{t} )} \propto 
P( \mathbf{t} | t_{N+1} ) P( t_{N+1} ) \textrm{,}
\end{equation}
where $P(t_{N+1})$ is a Gaussian prior on the function space. It is possible to introduce a Gaussian
prior on function $f$ as
\begin{equation}\label{eq:function_basis}
f(\mathbf{x}) = \sum_h w_h \phi_h (\mathbf{x})
\end{equation}
where $\{\phi_h\}_{h=1}^H$ form a complete basis set and the distribution of $\mathbf{w}$ is a
Gaussian with zero mean and variance $\sigma_h$: $w_h \sim N(0,\sigma_h) $. Each function value
$f_n$ is a linear combination of the basis functions:
\begin{equation}
f_n = \sum_{h=1}^H w_h \phi_h(\mathbf{x}_n) = \sum_{h=1}^H w_h R_{nh} \textrm{,}
\end{equation}
where $R_{nh} \equiv \phi_h(\mathbf{x}_n)$. The covariance matrix of the function values
$\mathbf{f}$ is the matrix of expectation values
\begin{equation}\label{eq:cov_matrix}
\mathbf{Q} = \langle \mathbf{f} \, \mathbf{f}^T \rangle = \langle \mathbf{R} \mathbf{w} \mathbf{w}^T \mathbf{R}^T \rangle 
= \mathbf{R} \langle \mathbf{w} \mathbf{w}^T \rangle \mathbf{R}^T = \sigma_h^2 \mathbf{R} \mathbf{R}^T
\end{equation}
Thus the prior distribution of $\mathbf{f}$ is $N(\mathbf{0},\mathbf{Q}) = N(\mathbf{0},\sigma_h^2
\mathbf{R} \mathbf{R}^T)$. However, each measurement contains noise, which we assume to be Gaussian
with zero mean and variance $\sigma_{\nu}$. The vector of data points also has Gaussian
distribution: $P( \mathbf{t} ) \sim N(0, \mathbf{Q} + \sigma_{\nu}^2)$. We denote the covariance
matrix of $\mathbf{t}$ by $\mathbf{C} \equiv \mathbf{Q} + \sigma_{\nu}^2\mathbf{I} $.

The distribution of the joint probability of observing $t_{N+1}$ having previously observed $\mathbf{t}$
can be written as
\begin{equation}
P(t_{N+1} | \mathbf{t} ) \propto P( [\mathbf{t} \, t_{N+1}] ) \textrm{,}
\end{equation}
where $ P( [\mathbf{t} \, t_{N+1}] ) \sim N(0, \mathbf{C}_{N+1})$, or explicitly
\begin{equation}\label{eq:t_dist}
P( [\mathbf{t} \, t_{N+1}] ) \propto \exp \left( -\frac{1}{2} [\mathbf{t} \, t_{N+1}]^T C_{N+1}^{-1}
[\mathbf{t} \, t_{N+1}] \right) \textrm{.}
\end{equation}
The covariance matrix $\mathbf{C}_{N+1}$ and its inverse can be written as
\begin{equation}
\mathbf{C}_{N+1} = \left[
\begin{array}{cc}
\mathbf{C}_N & \mathbf{k} \\
\mathbf{k}^T & \kappa
\end{array}
\right] 
\end{equation}
and
\begin{equation}
\mathbf{C}_{N+1}^{-1} = \left[
\begin{array}{cc}
\mathbf{M} & \mathbf{m} \\
\mathbf{m}^T & m
\end{array}
\right] \textrm{.}
\end{equation}
The submatrices of $\mathbf{C}_{N+1}^{-1}$ can be calculated via
\begin{equation}
\mathbf{C}_{N+1} \mathbf{C}_{N+1}^{-1} = \left[
\begin{array}{cc}
\mathbf{C}_N \mathbf{M} + \mathbf{k} \mathbf{m}^\top & \mathbf{C}_N \mathbf{m} + m \mathbf{k} \\
\mathbf{k}^\top \mathbf{M} + \kappa \mathbf{m}^\top & \mathbf{k}^\top \mathbf{m} + \kappa m
\end{array}
\right] = \mathbf{I} \textrm{,}
\end{equation}
which leads to
\begin{align}
m &= \left( \kappa - \mathbf{k}^T \mathbf{C}_N^{-1} \mathbf{k} \right)^{-1} \\
\mathbf{m} &= - m \mathbf{C}_N^{-1} \mathbf{k} \\
\mathbf{M} &= \mathbf{C}_N^{-1} + \frac{1}{m} \mathbf{m} \mathbf{m}^T
\textrm{.}
\end{align}
Substituting these into equation~\ref{eq:t_dist}, we obtain
\begin{equation}\label{eq:training}
P(t_{N+1} | \mathbf{t} ) \propto \exp \left( - \frac{ \left( t_{N+1} - \hat{t}_{N+1} \right) ^2 }{2 \sigma ^2 _{\hat{t}_{N+1}}} \right) \textrm{,}
\end{equation}
where the new variables $\hat{t}_{N+1}$ and $\sigma_{\hat{t}_{N+1}}$ are defined as
\begin{equation}\label{eq:gp_predict}
\hat{t}_{N+1} \equiv \mathbf{k}^T \mathbf{C}_N^{-1} \mathbf{t} 
\end{equation}
and
\begin{equation}
\sigma ^2 _{ \hat{t}_{N+1} } \equiv \kappa - \mathbf{k}^T \mathbf{C}_N^{-1} \mathbf{k} \textrm{,}
\end{equation}
i.e. $t_{N+1}$ has Gaussian distribution with mean $\hat{t}_{N+1}$ and variance $\sigma ^2
_{\hat{t}_{N+1} }$. We use this formula to predict function values and error bars.

\begin{figure}
\begin{center}
\includegraphics[width=8cm]{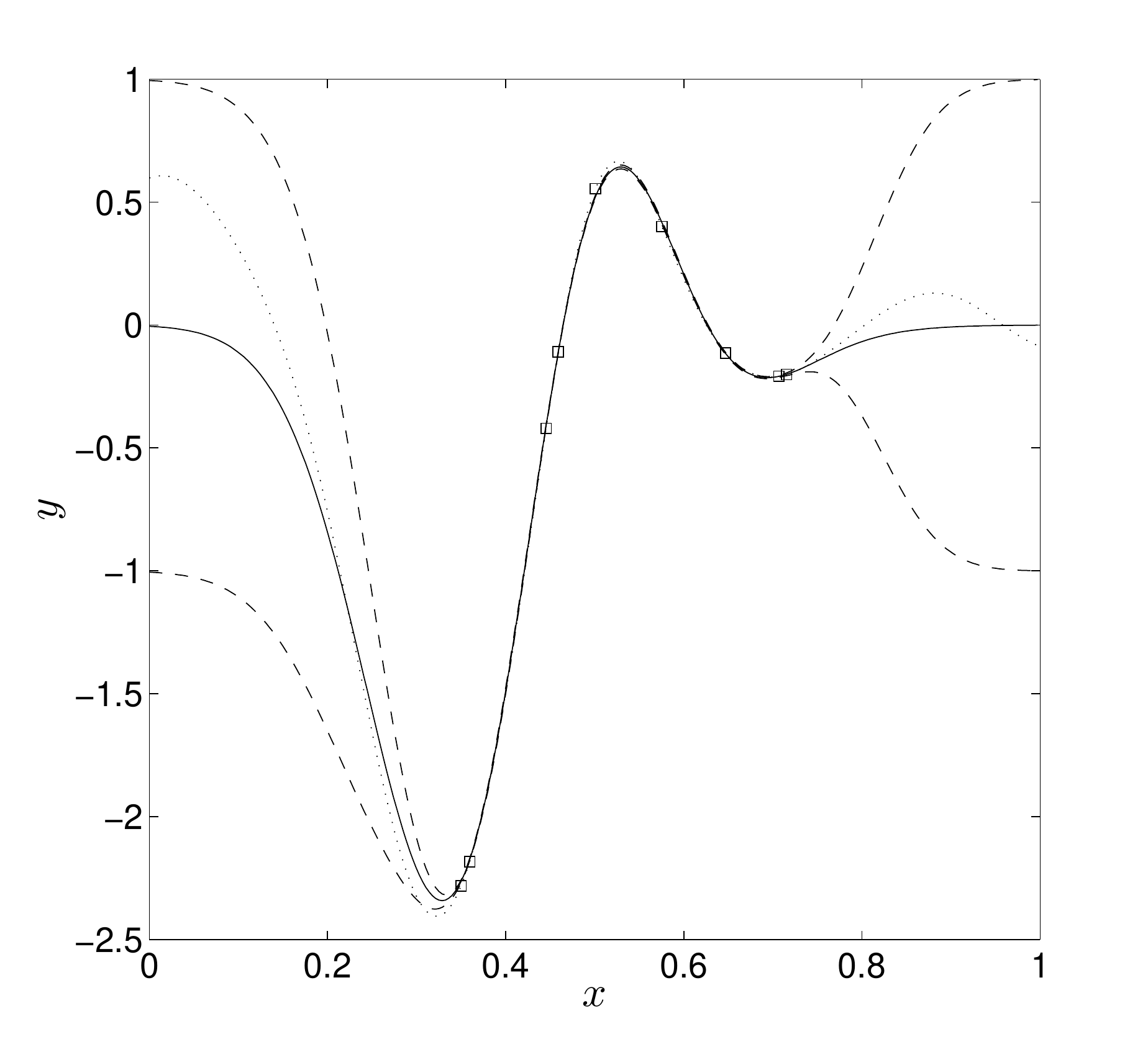}
\end{center}
\caption{\label{fig:gp_example} Gaussian Process regression in one dimension. The original function
(dotted line) was sampled at ten random points (open squares). The predicted function values (solid
line) and the errors (dashed line) are shown.}

\end{figure}

Figure~\ref{fig:gp_example} shows a one-dimensional example of the Gaussian Process regression. We
sampled an arbitrary function at ten random points between the interval $(\frac{1}{4},\frac{3}{4})$ and used
these samples as the training points. We present the predicted values and the predicted errors in
the entire interval $(0,1)$. It can be seen that inside the fitting region, the predicted values are
very close to the original functions, and the predicted variance is also small. Outside the fitting
region, the prediction is meaningless, and this is indicated by the large variance.

\subsection{Covariance functions}

The elements of the covariance matrix $\mathbf{Q}$ defined in equation~\ref{eq:cov_matrix} can be determined as
\begin{equation}\label{eq:cov_matrix_sum}
Q_{nn'} = \sigma_h^2 \sum_h R_{nh} R_{n'h} = \sigma_h^2 \sum_h \phi_h(\mathbf{x}_n) \phi_h(\mathbf{x}_{n'}) \textrm{.}
\end{equation}
In our work, we used Gaussians centred at different points as basis functions. In one dimension,
these would have the form
\begin{equation}
\phi_h (x) = \exp \left( -\frac{(x-x_h)^2}{2 r^2} \right)
\end{equation}
If the basis set consists of infinitely many basis functions which are distributed uniformly, the
summation in equation~\ref{eq:cov_matrix_sum} can be replaced by an integration:
\begin{equation}
Q_{nn'} \propto \int \exp \left(-\frac{(x_n-x_h)^2}{2 r^2} \right) \exp \left(-\frac{(x_{n'}-x_h)^2}{2 r^2} \right) \mathrm{d}r \textrm{.}
\end{equation}
The integral of the product of two Gaussian is also a Gaussian, leading to the final
expression---also known as the \emph{kernel}---of the covariance matrix elements
\begin{equation}
Q_{nn'} = \delta^2 \exp \left(-\frac{(x_n - x_{n'})^2}{2 \theta^2} \right) \textrm{,}
\end{equation}
where $\delta$ and $\theta$ are usually referred to as hyperparameters. This finding demonstrates
that the Gaussian Process method is, in fact, an example of non-parametric regression with
infinitely many basis functions, but where it is not necessary to determine the coefficients of the
basis functions explicitly.  We note that using Gaussians as basis functions is a convenient choice,
as the elements of the covariance matrix can be calculated analytically using a simple Gaussian
kernel, but depending on the nature of the target function, there is a large variety of alternative
basis functions and kernels.

In the case of multidimensional input data, the Gaussian kernel could be
modified such that different length scales are associated with different directions:
\begin{equation}\label{eq:cov_mat_element}
Q_{nn'} = \delta^2 \exp \left( -\frac{1}{2} \sum_i \frac{(x_{ni} - x_{n'i})^2}{\theta_i^2} \right) \textrm{,}
\end{equation}
where the vector $\boldsymbol{\theta} \equiv \{\theta_i\}_{i=1}^N$ contains the typical
decorrelation length of the function in each dimension $i$. If we assume that the initial Gaussian
basis functions are not aligned in the directions of the original input vectors, the kernel can be
written in the form
\begin{equation}
Q_{nn'} = \delta^2 \exp \left( -\frac{1}{2} \mathbf{x}_n^T \boldsymbol{\Theta}^T \boldsymbol{\Theta} \mathbf{x}_n \right) \textrm{,}
\end{equation}
where $\boldsymbol{\Theta}$ is the matrix of hyperparameters.

\subsection{Hyperparameters}

The choice of hyperparameters $\delta$, $\boldsymbol{\theta}$ and $\sigma_{\nu}$ depends strongly on
the dataset.  $\boldsymbol{\theta}$ represents the width of the basis functions, i.e. it characterises
the typical length scale over which the function values become uncorrelated.
$\delta$ places a prior on the variance of the parameter vector $\mathbf{w}$, describing the typical
variance of the function, while $\sigma_{\nu}$ is the assumed noise in the measured data values.
Ideally, a prediction for $t_{N+1}$ would be made by evaluating the integral
\begin{equation}
P(t_{N+1} | x_{N+1}, \mathbf{t},\mathbf{X} )  = \int P(t_{N+1} | x_{N+1},\mathbf{t}, \mathbf{X},\mathbf{h} )
P(\mathbf{h}|\mathbf{t},\mathbf{X}) \mathrm{d}\mathbf{h}
\textrm{,}
\end{equation}
but depending on the model, the analytic form of the integral may or may not be known. Although it is
always possible to carry out the integration numerically, for example, by Markov chain Monte Carlo or Nested
Sampling\cite{nest03}, a computationally less demanding method is to approximate the integral at the most
probable value of $\mathbf{h}$.
It is often possible to choose good hyperparameters based on known features of the function, but the
hyperparameters can also be optimised if needed.  If we consider the probability distribution of a
hyperparameter set $\mathbf{h}$ given a dataset $D$:
\begin{equation}\label{eq:likelihood_general}
P(\mathbf{h}|D) \propto P(D|\mathbf{h}) P(\mathbf{h}) \textrm{,}
\end{equation}
optimal hyperparameters can be obtained by maximising this probability, known as the marginal
likelihood.
Assuming a uniform prior on the hyperparameters and using the result found in
equation~\ref{eq:cov_matrix}, i.e. $P( \mathbf{t} | \mathbf{X} ) \sim N(0, \mathbf{C})$, the
logarithm of the likelihood is
\begin{equation}
\ln P(\mathbf{t} | \mathbf{X}, \mathbf{h}) = -\frac{1}{2} \mathbf{t}^T \mathbf{C}^{-1} \mathbf{t} -\frac{1}{2} \ln \det \mathbf{C} -\frac{N}{2} \ln 2 \pi \textrm{.}
\end{equation}
Maximising the logarithm of the likelihood with respect to the hyperparameters can be performed by
gradient-based methods such as Conjugate Gradients\cite{bib:cg1}, where that gradients can be calculated
as
\begin{equation}
\frac{\partial  \ln P}{\partial h_i} = \frac{1}{2} \mathbf{t}^T  \mathbf{C}^{-1} \frac{\partial \mathbf{C}}{\partial h_i} \mathbf{C}^{-1}
\mathbf{t} - \frac{1}{2} \mathrm{tr} \left( \mathbf{C}^{-1} \frac{\partial \mathbf{C}}{\partial h_i} \right)
\textrm{.} 
\end{equation}

\subsection{Predicting derivatives and using derivative observations}\label{sec:gp_derivatives}

Predicting the values of derivatives using a Gaussian Process can be performed by simply
differentiating the expectation value $\hat{t}$ in equation~\ref{eq:gp_predict}:
\begin{equation}
\frac{\partial \hat{t}}{\partial x_i} = \frac{\partial \mathbf{k}^T}{\partial x_i} \mathbf{C}_N^{-1} \mathbf{t}
\textrm{.}
\end{equation}
The elements of $\mathbf{k}$ are given by the covariance function, hence we need to differentiate
the covariance function,
\begin{equation}
\frac{\partial k_n}{\partial x_i} = \frac{\partial C(\mathbf{x}_n,\mathbf{x})}{\partial x_i}
\end{equation}
which gives
\begin{equation}
\frac{\partial k_n}{\partial x_i} = \frac{x_{ni} - x_i}{\theta_i^2}  \delta^2
\exp \left( -\frac{1}{2} \sum_i \frac{(x_{ni} - x_i)^2}{\theta_i^2} \right)
\end{equation}
in the case of Gaussian kernels.

It is also possible that values of derivatives have been measured and these are also available. In
order to use this data, we differentiate equation~\ref{eq:function_basis}
\begin{equation}
\frac{\partial f}{\partial x_i} \bigg| _{\mathbf{x}_n} 
= \sum_h w_h \frac{\partial \phi_h}{\partial x_i} \bigg| _{\mathbf{x}_n} \textrm{,}
\end{equation}
thus we need to substitute $\frac{\partial \phi_h}{\partial x_i}$ for the basis functions in
equation~\ref{eq:cov_matrix_sum} to give
\begin{equation}
Q_{nn'} = \sigma_h^2 \sum_h R_{nh} R_{n'h} = \sigma_h^2 \sum_h \frac{\partial \phi_h}{\partial x_i} \bigg| _{\mathbf{x}_n} 
\phi_h(\mathbf{x}_{n'}) \textrm{.}
\end{equation}
\begin{equation}
Q_{nn'} = \sigma_h^2 \sum_h R_{nh} R_{n'h} = \sigma_h^2 \sum_h \frac{\partial \phi_h}{\partial x_i} \bigg| _{\mathbf{x}_n} 
\frac{\partial \phi_h}{\partial x_j} \bigg| _{\mathbf{x}_{n'}}
\end{equation}
For Gaussian kernels, the covariance between a derivative and a function value observation is
\begin{equation}
Q_{nn'} = \frac{x_{ni} - x_{n'i}}{\theta_i^2}  \delta^2
\exp \left( -\frac{1}{2} \sum_k \frac{(x_{nk} - x_{n'k})^2}{\theta_i^2} \right) \textrm{,}
\end{equation}
or between two derivative observations the covariance is
\begin{equation}
Q_{nn'} = \left( \frac{1}{\theta_i \theta_j} - \frac{1}{2}\frac{x_{ni} - x_{n'i}}{\theta_i^2}
\frac{x_{nj} - x_{n'j}}{\theta_j^2} \right) \delta^2
\exp \left( -\frac{1}{2} \sum_k \frac{(x_{nk} - x_{n'k})^2}{\theta_k^2} \right) \textrm{.}
\end{equation}
Finally, if the function is a composite function of the form $f(\mathbf{x}) \equiv
f(\mathbf{y}(\mathbf{x}))$ and the derivatives $\frac{\partial f}{\partial x_i}$ are available, the
Gaussian covariance function between a derivative ($n$-th) and function value ($n'$-th) observation
is
\begin{equation}
Q_{nn'} = \sum_k \frac{y_{nk} - y_{n'k}}{\theta_k^2} \frac{\partial y_{nk}}{\partial x_i} \delta^2
\exp \left( -\frac{1}{2} \sum_k \frac{(y_{nk} - y_{n'k})^2}{\theta_k^2} \right) \textrm{,}
\end{equation}
and between two derivative observations $\frac{\partial f}{\partial x_i}$ and $\frac{\partial
f}{\partial x_j}$ is
\begin{equation}
Q_{nn'} = \left( \sum_k \frac{1}{\theta_k^2} \frac{\partial y_{nk}}{\partial x_i}
\frac{\partial y_{n'k}}{\partial x_j} - D_{ij} \right) \delta^2
\exp \left( -\frac{1}{2} \sum_k \frac{(y_{nk} - y_{n'k})^2}{\theta_k^2} \right) \textrm{,}
\end{equation}
with
\begin{equation}
D_{ij} = \frac{1}{2} \left( \sum_k \frac{y_{nk} - y_{n'k}}{\theta_k^2}
\frac{\partial y_{nk}}{\partial x_i} \right)
\left( \sum_k \frac{y_{nk} - y_{n'k}}{\theta_k^2} \frac{\partial y_{n'k}}{\partial x_j} \right)
\textrm{.}
\end{equation}

Using the same model for observations of function values and their derivatives enables us to
incorporate the available information into a single regression allowing us to infer both function
values and derivatives.

Since there is no reason to assume that the noise is the same in case of both the function value and
derivative observations, we use two distinct noise hyperparameters.

\subsection{Linear combination of function values}\label{sec:gp_linear_combinations}

It is possible that linear combinations of function values can be observed during the data collection process:
\begin{equation}
f'_m = \sum_n L_{mn} f(\mathbf{x}_n) = \sum_{n,h} L_{mn} R_{nh} w_h \textrm{.}
\end{equation}
If this is the case, equation~\ref{eq:cov_matrix} is thereby modified, so the covariance matrix of
the observed values can be obtained as
\begin{equation}\label{eq:cov_matrix_lin}
\mathbf{Q'} = \langle \mathbf{f'} \mathbf{f'}^T \rangle =
\langle \mathbf{L} \mathbf{R} \mathbf{w} \mathbf{w}^T \mathbf{R}^T \mathbf{L}^T \rangle 
= \sigma_h^2 \mathbf{L} \mathbf{R} \mathbf{R}^T \mathbf{L}^T = \mathbf{L} \mathbf{Q} \mathbf{L}^T \textrm{.}
\end{equation}
In our work, equation~\ref{eq:cov_matrix_lin} proved to be very useful, as only the total energy of
an atomic system can be obtained using quantum mechanical calculations. However, we view
the energy as arising from the sum of atomic contributions.
Thus, in this case, the matrix $\mathbf{L}$ describing the relationship of the observations (total
energy) to the unknown function values (atomic energies) consists of zeros and ones.

\subsection{Sparsification}\label{sec:sparse}

Snelson and Ghahramani\cite{spgp01} introduced a modification to the standard Gaussian Process
regression model for large, correlated data sets.  The computational cost of the training process
described in equation~\ref{eq:training} scales as the cube of the number of data points, due to the
computational cost of inverting the covariance matrix. In case of large data sets, the training
process can become computationally expensive.  Although the computational cost of predicting
function values scales linearly with the number of teaching points, this cost can also be
computationally demanding.  If the data set is highly correlated, i.e. observations are made at
closely spaced points, it is feasible to use a sparse approximation of the full Gaussian Process,
which has significantly reduced computational requirements but only a little less accuracy.

We used the sparsification procedure described in \cite{spgp01}.  In the sparsification procedure, a
set of $M$ pseudo-inputs $\{\overline{\mathbf{x}}_m\}_{m=1}^M$ are chosen from the full dataset of
$N$ input values $\{ \mathbf{x}_n\}_{n=1}^N$, and the covariance matrices $\mathbf{C}_{NM}$ and
$\mathbf{C}_M$ are calculated as
\begin{equation}
\left[\mathbf{C}_M \right]_{mm'} = C(\overline{\mathbf{x}}_m,\overline{\mathbf{x}}_{m'}) \\
\end{equation}
and
\begin{equation}
\left[\mathbf{C}_{NM}\right]_{nm} = [\mathbf{k}_n]_m = C(\mathbf{x}_n,\overline{\mathbf{x}}_m) \textrm{.}
\end{equation}
In order to simulate the full covariance matrix, the matrix
\begin{equation}
\boldsymbol{\Lambda} = \mathrm{Diag}( \mathrm{diag}(\mathbf{C}_{N} - \mathbf{C}_{NM} \mathbf{C}_M^{-1} \mathbf{C}_{MN}))
\end{equation}
is also needed, where $\mathbf{C}_{N}$ is the full $N \times N$ covariance matrix, although only the
diagonal elements are calculated.  The elements of the covariance vector $\mathbf{k}$ are calculated
from the coordinates of the pseudo-inputs and the test point $\mathbf{x}_*$:
\begin{equation}
k_m = C(\overline{\mathbf{x}}_m,\mathbf{x}_*) \textrm{.}
\end{equation}
The pseudo-covariance matrix of the sparsified data set is
\begin{equation}
\mathbf{Q}_M = \mathbf{C}_M + \mathbf{C}_{MN} ( \mathbf{\Lambda} + \sigma^2 \mathbf{I})^{-1} \mathbf{C}_{NM} \textrm{,}
\end{equation}
which can now be used to predict the function value and the error estimate at the test point as
\begin{align}
\hat{t} &= \mathbf{k}^T \mathbf{Q}_M^{-1} \mathbf{C}_{MN} ( \boldsymbol{\Lambda} + \sigma^2 \mathbf{I})^{-1} \mathbf{t} \\
\sigma^2_{\hat{t}} &= C(\mathbf{x}_*,\mathbf{x}_*) - \mathbf{k}_*^T (\mathbf{C}_M^{-1}-\mathbf{Q}_M^{-1}) \mathbf{k} + \sigma^2
\textrm{.}
\end{align}
In order to obtain an optimal set of hyperparameters and pseudo-inputs, the likelihood function
\begin{multline}\label{eq:sparse_likelihood}
\log L= -\frac{1}{2}\mathbf{t}^T (\mathbf{C}_{NM} \mathbf{C}_M^{-1} \mathbf{C}_{MN} + \mathbf{\Lambda}
+ \sigma^2 \mathbf{I})^{-1} \mathbf{t} \\
-\frac{1}{2} \log |\mathbf{C}_{NM} \mathbf{C}_M^{-1}
\mathbf{C}_{MN} + \mathbf{\Lambda}+ \sigma^2 \mathbf{I}| -\frac{n}{2} \log 2 \pi 
\end{multline}
is maximised in the space of hyperparameters and pseudo-inputs.

In our work, observation of single function values is not possible, i.e. only total energies (sum of
atomic energies) and forces (sum of derivatives of local energies) are accessible.  Depending on the
number of atoms in the cell, in the case of total energy observations, and the number of atoms within the
chosen cutoff radius, in the case of force observations, a large number of input values has to be added
to the training set, regardless of whether the neighbourhood of a particular atom is different from
the ones previously encountered.  Thus in our case, the sparsification process is crucial in order to
develop a tractable computational scheme.

\chapter{Interatomic potentials}\label{chapter:interatomic_potentials}

\section{Introduction}

A wide variety of models have been developed to describe atomic interactions, ranging from the very
accurate and extremely expensive to the fast but very approximate. Quantum Mechanics ultimately
provides a true description of matter via solving the Schr\"odinger equation, but even in its
crudest approximation, the use of Quantum Mechanics is limited to a few hundreds of atoms or a few
hundreds of different configurations, which is inadequate to sample the entire phase space of a
system. A series of further simplifications leads to the realm of analytic potentials that can be
used to describe larger systems or more configurations. The so-called empirical potentials are based
on fixed functional forms, which are equally based on theoretical considerations and intuition,
making the creation of new potentials a combination of ``art and science''\cite{brenner02}. Analytic
potentials can be described as non-linear parametric regression from the statistical point of view,
where the fitting process is based on experimental or quantum mechanical data. Further, the
parametric formula that is chosen to describe the behaviour of the real system is often fitted to
reproduce some well-known equilibrium properties, such as the lattice constant and elastic constants
of the bulk material or the structure of a liquid, and it is assumed that the same function will
perform well in very different configurations. This clearly implies that analytic potentials are
expected to be able to extrapolate to very different environments on the basis of the physical
insight used when the particular functional form was chosen. Even if there exists such a functional
form, it follows from the overly complicated nature of regression in such high dimensions that
finding the right form and fitting it to each new interesting material is extremely difficult. Our
work focuses on the development of a potential based on non-linear, non-parametric regression
methods that infers the interactions directly from quantum mechanical data, though the approach can
be adopted irrespective of the origin of the data.

\section{Quantum Mechanics}

In the general case, the Schr\"odinger equation takes the form
\begin{equation}
i \hbar \frac{\partial \Psi(\mathbf{r},t)}{\partial t} = \hat{H} \Psi(\mathbf{r},t) \textrm{,}
\end{equation}
where $\Psi$ is the time-dependent wave-function, $\mathbf{r}$ contains the coordinates of all the 
particles in the system and $\hat{H}$ is the Hamiltonian operator.  The Hamiltonian can be written
as
\begin{equation}\label{eq:hamiltonian_general}
\hat{H} = - \sum_i \frac{\hbar^2}{2m_i} \nabla^2_i + V(\mathbf{r}) \textrm{,}
\end{equation}
where $V(\mathbf{r})$ is the potential energy. The standing wave solution of the time dependent
Schr\"odinger equation is
\begin{equation}
\Psi(\mathbf{r},t) = \psi(\mathbf{r}) \exp \left( -\frac{i E t}{\hbar} \right) \textrm{,}
\end{equation}
which leads to the time-independent form of the Schr\"odinger equation
\begin{equation}\label{eq:schrodinger}
\hat{H} \psi(\mathbf{r}) = E \psi(\mathbf{r}) \textrm{.}
\end{equation}
Atomic systems consist of electrons and nuclei, hence equation~\ref{eq:hamiltonian_general} becomes
\begin{multline}\label{eq:hamiltonian_atomic}
\hat{H} = - \sum_i^{\textrm{elec.}} \frac{\hbar^2}{2m_e} \nabla^2_i \, + \sum_{i<j}^{\textrm{elec.}} \frac{q_e^2}{r_{ij}}
\, - \sum_{i}^{\textrm{elec.}} \sum_{A}^{\textrm{nuclei}} Z_A \frac{q_e^2}{r_{iA}} \\
- \sum_A^{\textrm{nuclei}} \frac{\hbar^2}{2m_A} \nabla^2_A \, 
+ \sum_{A<B}^{\textrm{nuclei}} Z_A Z_B \frac{q_e^2}{r_{AB}}
\end{multline}
where $m_e$ and $q_e$ are the mass and the charge of an electron, $m_A$ and $Z_A$ are the mass and
atomic number of the nucleus $A$.
The Born-Oppenheimer approximation further simplifies the solution of equation~\ref{eq:schrodinger}
by assuming that the coupling of the electrons and nuclei is negligible. The basis of this
assumption is that the mass of the nuclei is at least three order of magnitudes larger than the mass
of the electrons, thus the electrons adapt to the nuclei adiabatically.  The Born-Oppenheimer
approximation can be expressed as
\begin{align}\label{bo_approximation}
\begin{split}
\left( - \sum_i^{\textrm{elec.}} \frac{\hbar^2}{2m_e} \nabla^2_i + \sum_{i<j}^{\textrm{elec.}} \frac{q_e^2}{r_{ij}}
- \sum_{i}^{\textrm{elec.}} \sum_{A}^{\textrm{nuclei}} Z_A \frac{q_e^2}{r_{iA}} \right)
\psi(\mathbf{r},\mathbf{R}) \\
+ \sum_{A<B}^{\textrm{nuclei}} Z_A Z_B \frac{q_e^2}{r_{AB}}
&= E_e(\mathbf{R}) \psi(\mathbf{r},\mathbf{R}) 
\end{split} \\
\left( - \sum_A^{\textrm{nuclei}} \frac{\hbar^2}{2m_A} \nabla^2_A + E_e(\mathbf{R}) \right) \chi(\mathbf{R}) 
&= E \chi(\mathbf{R}) \textrm{,}
\end{align}
where the electronic wavefunction $\psi(\mathbf{r},\mathbf{R})$ only depends on the coordinates of
the electrons $\mathbf{r}$ and the coordinates of the nuclei $\mathbf{R}$ are regarded as
parameters.  The solutions of equation~\ref{bo_approximation}, the so-called electronic
Schr\"odinger equation provides the \emph{potential energy surface} (PES) $E_e(\mathbf{R})$, which
describes the interactions of the nuclei.  The nuclear Schr\"odinger equation is often replaced by
the classical equations of motion.

\subsection{Density Functional Theory}

The analytic solution of the electronic Schr\"odinger equation is impossible for systems more
complicated than a hydrogen molecular-ion $\textrm{H}_2^+$.  There exists a wide range of methods
that are concerned with determining the electronic structure, ranging from the very approximate
tight-binding\cite{finnisbook} approach to the essentially exact full configuration
interaction\cite{szabobook} method.  In our work, we used Density Functional Theory as the underlying quantum mechanical method.

Density Functional Theory aims to find the ground state electron density rather than the wavefunction.
\begin{equation}
\rho(\mathbf{r}) = \int \! |\psi(\mathbf{r},\mathbf{r}_2,\ldots,\mathbf{r}_N)|^2
\mathrm{d}\mathbf{r}_2 \, \ldots \, \mathrm{d}\mathbf{r}_N
\end{equation}
The density depends only on three spatial coordinates instead of $3N$, reducing the complexity of
the task enormously.  The Hohenberg-Kohn principles prove that the electron density is the most
central quantity determining the electronic interactions and forms the basis of an exact expression
of the electronic ground state.

\subsubsection{The Hohenberg-Kohn principles}

The basic lemma of Hohenberg and Kohn\cite{dft05} states that the ground state electron density of a system of
interacting electrons in an arbitrary external potential determines this potential uniquely.  The
proof is given by the variational principle.  If we consider a Hamiltonian $ \hat{H}_1 $ of an
external potential $ V_1 $ as
\begin{equation}
\hat{H}_1 = \hat{T} + \hat{U} + \hat{V}_1
\textrm{,}
\end{equation}
where $\hat{T}$ is the kinetic energy operator and $\hat{U}$ is the electron-electron interaction
operator. The solution of the Schr\"odinger equation
\begin{equation}
\hat{H}_1 \psi = E \psi
\end{equation}
is the ground state wavefunction $ \psi_1 $, which corresponds to the electron density $ \rho_1 $.
The ground state energy is then
\begin{equation}
E_1 = \langle \psi_1 | \hat{H}_1 | \psi_1 \rangle =
\int \! V_1(\mathbf{r}) \rho(\mathbf{r}) + \langle \psi_1 | \hat{T} + \hat{U} | \psi_1 \rangle
\textrm{.}
\end{equation}
Considering another potential $ V_2 $, which cannot be obtained as $ V_1 + \mathrm{constant} $, with
a ground state wavefunction $ \psi_2 $, which generates the same electron density, the ground state energy is
\begin{equation}
E_2 = \int \! V_2(\mathbf{r}) \rho(\mathbf{r}) + \langle \psi_2 | \hat{T} + \hat{U} | \psi_2 \rangle
\textrm{.}
\end{equation}
According to the variational principle,
\begin{multline}
E_1 < \langle \psi_2 | \hat{H}_1 | \psi_2 \rangle = 
\int \! V_1(\mathbf{r}) \rho(\mathbf{r}) + \langle \Psi_2 | \hat{T} + \hat{U} | \Psi_2 \rangle  \\
= E_2 + \int \! \left[ V_1(\mathbf{r}) - V_2(\mathbf{r}) \right] \rho(\mathbf{r})
\end{multline}
and
\begin{multline}
E_2 < \langle \psi_1 | \hat{H}_2 | \psi_1 \rangle = 
\int \! V_2(\mathbf{r}) \rho(\mathbf{r}) + \langle \psi_1 | \hat{T} + \hat{U} | \psi_1
\rangle  \\
= E_1 + \int \! \left[ V_2(\mathbf{r}) - V_1(\mathbf{r}) \right] \rho(\mathbf{r})
\textrm{.}
\end{multline}
By adding the two inequalities together, we find the contradiction
\begin{equation}
E_1 + E_2 < E_1 + E_2 \textrm{.}
\end{equation}
This is the indirect proof that no two different external potentials can generate the same electron
density.

The second Hohenberg-Kohn theorem establishes a link between the total energy and the electron density,
namely that there exists a universal energy functional, which is valid for every external potential, and
its global minimum corresponds to the ground state of the system and the ground state electron
density.
To prove this theorem, we write the total energy functional as a universal functional
\begin{equation}
E[\rho] = F_{\mathrm{HK}}[\rho] + \int \! V(\mathbf{r}) \rho(\mathbf{r}) + E_{ZZ} \textrm{,}
\end{equation}
where $F_{\mathrm{HK}}$ applies to every electronic system. It determines the entire electronic
energy
except the energy due to the external potential $V(\mathbf{r})$. $E_{ZZ}$ is the interaction
between the nuclei.
The ground state energy is given by
\begin{equation}
E = \langle \psi | \hat{H} | \psi \rangle = E[\rho] \textrm{.}
\end{equation}
According to the variational principle, changing the wavefunction to a different $\psi'$, which in turn corresponds to a different electron density $\rho'$, the resulting energy 
\begin{equation}
E < E' = \langle \psi' | \hat{H} | \psi' \rangle \textrm{,}
\end{equation}
is greater than $E$, thus $\rho$ cannot correspond to the exact ground state.

We note that the ground state wavefunction can be found from the variational principle
\begin{equation}
E = \mathrm{min}_{\tilde{\psi}} \langle \tilde{\psi} | \hat{T} + \hat{U} + \hat{V} | \tilde{\psi} \rangle \textrm{,}
\end{equation}
where $ \tilde{\psi} $ is a trial wavefunction. The variational principle can be reformulated in terms of trial densities, $ \tilde{\rho} $:
\begin{equation}
E = \mathrm{min}_{\tilde{\rho}} E[\tilde{\rho}]
\end{equation}

\subsubsection{The self-consistent Kohn-Sham equations}

The Hohenberg-Kohn principles provide the theoretical basis of Density Functional Theory,
specifically that the total energy of a quantum mechanical system is determined by the electron
density through the Kohn-Sham functional.  In order to make use of this very important theoretical
finding, Kohn-Sham equations are derived, and these can be used to determine the electronic ground state of
atomic systems.

The total energy of a system of interacting electrons in the external potential of the classic nuclei can be written as
\begin{equation}
E[\rho] = T[\rho] + E_\mathrm{H}[\rho] + E_\mathrm{xc}[\rho] + E_{Ze}[\rho] + E_{ZZ} \textrm{,}
\end{equation}
where $ T[\rho] $ is the kinetic energy functional, $ E_\mathrm{xc} $ is the exchange-correlation functional,
$ E_\mathrm{H} $ is the  Hartree interaction between electrons, $ E_{Ze} $ is the interaction between the electrons and
the nuclei and $ E_{ZZ} $ is the nuclei-nuclei interaction. The latter three energies have the forms
\begin{align}
E_\mathrm{H}[\rho] &= \int \frac{\rho(\mathbf{r}) \rho(\mathbf{r}')}{|\mathbf{r}-\mathbf{r}'|}
\mathrm{d}\mathbf{r} \mathrm{d}\mathbf{r}' \\
E_{Ze}[\rho] &= \sum_A^{\textrm{nuclei}} \int Z_A \frac{\rho(\mathbf{r})}{|\mathbf{r}-\mathbf{r}_A|}
\mathrm{d}\mathbf{r} \\
E_{ZZ} &= \sum_{A<B}^{\textrm{nuclei}} \frac{Z_A Z_B}{|\mathbf{r}_A-\mathbf{r}_B|} \textrm{,}
\end{align}
whereas the exact form of functionals $ T[\rho] $ and $ E_\mathrm{xc}[\rho] $ is not specified by
the theory.  However, according to the Hohenberg-Kohn principle, any system of interacting electrons
can be described as a system of independent electrons moving in an effective potential, meaning that
the kinetic energy functional can represented by the kinetic energy of non-interacting electrons,
$T_\mathrm{S}$.  The difference between the true kinetic energy functional and $ T_\mathrm{S} $
\begin{equation}
\Delta T = T[\rho] - T_\mathrm{S} 
\end{equation}
is included in the exchange-correlation functional, which still needs to be determined.  The
non-interacting kinetic energy operator $ T_\mathrm{S} $ is simply written as
\begin{equation}
T_\mathrm{S} = -\frac{\hbar^2}{2m_e} \sum_n^{\textrm{elec.}} \langle \psi_n | \nabla_n^2 | \psi_n \rangle \textrm{,}
\end{equation}
where $ \psi_n $ are the independent electron orbitals. The one-electron orbitals determine the charge density as
\begin{equation}
\rho (\mathbf{r}) = \sum_n \psi_n^*(\mathbf{r}) \psi_n (\mathbf{r}) \textrm{.}
\end{equation}
Hence the ground state will correspond to the electronic density at which the functional derivative
of the total energy with respect to $ \psi_n $ is zero, while maintaining the orthogonality constraints
\begin{equation}
\langle \psi_i | \psi_j \rangle = \delta_{ij}
\end{equation}
via the Lagrange multipliers $ \epsilon_{ij} $. Thus minimising the energy functional and the constraints 
\begin{equation}
\frac{\delta \left[ E - \sum_{ij} \epsilon_{ij} (\langle \psi_i | \psi_j \rangle - \delta_{ij}) \right] }
{\delta \psi_n} = 0 \textrm{, } \forall n
\end{equation}
leads to the Kohn-Sham equations,
\begin{multline}
0 = \frac{\delta T_\mathrm{S}}{\delta \psi_n} +
\frac{ E_\mathrm{H}[\rho] + E_\mathrm{xc}[\rho] + E_{Ze}[\rho] }{\delta \rho} \frac{\delta \rho}{\delta \psi_n} -
\frac{ \sum_{ij} \epsilon_{ij} (\left< \psi_i | \psi_j \right> }{\delta \psi_n} = \\
= \Delta_n \psi_n + \hat{V}_\mathrm{eff} \psi_n - \sum_j \epsilon_{nj}\psi_j \textrm{,}
\end{multline}
which can be solved as $n$ independent equations,
\begin{equation}
\Delta_n \psi'_n + \hat{V}_\mathrm{eff} \psi'_n = \epsilon'_n \psi'_n
\textrm{,}
\end{equation}
since there exists a basis set where the energy matrix is diagonal.  Although the minimisation can
be performed directly, as implemented in CASTEP as conjugate gradients for insulating systems or
EDFT\cite{castep01}, an iterative approach is more often used. The effective potential
$\hat{V}_\mathrm{eff}$ depends on the electronic density, thus it is calculated using some initial
guess for the density, then the Kohn-Sham equations are solved, resulting in a new density. This
process is repeated until the electron density becomes self-consistent.

\section{Empirical potentials}

The Born-Oppenheimer approximation, as given in equation~\ref{bo_approximation}, suggests that when
considering solely the interactions between the nuclei, the electrons do not have to be explicitly taken in
account.  The reason why the Schr\"odinger equation has to be solved in many
applications is the need for the accurate description of the Potential Energy Surface provided by
Quantum Mechanics.  If there were an
alternative way to determine the Potential Energy Surface felt by the nuclei $V(\mathbf{R}) \equiv
E(\mathbf{R})$, Quantum Mechanics could be bypassed entirely. Empirical potentials, as well as our
research, aim to achieve this.

\subsection{Hard-sphere potential}

The simplest interatomic potential is the hard-sphere potential, that can be characterised as
\begin{equation}
V(r) = \left\{ \begin{array}{ll}
0 & \textrm{if $r \leq r_0$} \\
\infty & \textrm{if $r>r_0$}
\end{array} \right. \textrm{,}
\end{equation}
where $r_0$ is the radius of the sphere. Even this simple functional form can describe the fact that
atoms repel each other due to the Pauli exclusion principle, albeit in a rather crude way. As this
potential completely lacks attractive terms, its use is usually limited to bulk phases. The
hard-sphere model is often used for testing purposes, as despite of its simplicity, a system of
hard-spheres shows a fluid-solid phase transition\cite{bib:HS1,bib:HS2}. More recently, systems of colloid
particles were also modelled as hard spheres\cite{bib:HScolloid1,bib:HScolloid2}, and the results of these
simulations have received strong experimental support.

\subsection{Lennard-Jones potential}

The Lennard-Jones potential 
\begin{equation}
V(r) = 4 \epsilon \left( \frac{\sigma^{12}}{r^{12}} - \frac{\sigma^{6}}{r^{6}} \right)
\end{equation}
was originally introduced to describe the interaction between argon atoms\cite{LJ02}.  The two terms
in the expression are the repulsion due to Pauli exclusion and the attraction which arises from
dispersion interactions.  The $r^{-6}$ variation is obtained by considering the interaction of two
induced dipoles on closed-shell atoms.  Although the $r^{-12}$ term has been introduced primarily
because it is the square of the other term---therefore its computation is very efficient---, and
has no theoretical justification, the Lennard-Jones potential reproduces the properties of argon
remarkably well\cite{bib:Allen_Tildesley}.  In the case of other noble gases, quantum effects (for He and Ne),
contribution from the interaction of higher order moments and relativistic effects (for Kr, Xe, Ra)
become more significant and so the Lennard-Jones model is not so successful.  The Lennard-Jones
potential has been applied to different types of systems, because of the ease of computation and the
strong physical basis.  Potentials for ions are often built as Lennard-Jones spheres and point
charges\cite{bib:LJcharge1,bib:LJcharge2}, the most successful water models are based on partial charges and
Lennard-Jones term(s)\cite{bib:model_SPCE,bib:model_TIP4P}, or even groups of atoms, such as methyl groups are modelled as
a single Lennard-Jones particle\cite{bib:LJ_unitedatom2}.  While being a relatively simple potential, systems
composed of Lennard-Jones particles show complex phase behaviour, which makes the use of this
potential attractive as test systems in such studies and method
development\cite{nest04,bib:LJ_phase1,bib:LJ_phase3}.

\subsection{The embedded-atom model}

The embedded-atom model was developed by Daw and Baskes\cite{EAM01} and was originally intended to
describe metallic systems.  In general, the potential takes the form
\begin{equation}\label{eq:eam}
E = \sum_i F(\rho_i) + \frac{1}{2} \sum_{j \ne i} \Phi(r_{ij}) \textrm{,}
\end{equation}
where $\rho_i$ is the electron density at the centre of atom $i$ due to the atoms at 
neighbouring sites
\begin{equation}\label{eq:eam_rho}
\rho_i = \sum_{j \ne i} \rho_j(r_{ij}) \textrm{,}
\end{equation}
where $\rho_j$ is the electronic density of atom $j$.  $F$ is the embedding functional and $\Phi$
represents the core-core repulsion.  This potential is derived from density functional theory, where
the electron density is approximated by a sum of atomic contributions and the
energy functional is substituted by a simple analytic function.  The parameters in the embedded atom potentials
used in the original applications were fitted to experimental observables, such as lattice constants
and elastic moduli.

More recently, a particularly interesting new formulation of the embedded-atom model, called the force-matching method has been
published by Ercolessi and Adams\cite{EAM05}.  In this work, no
prior assumptions were made on the actual functional forms in equations~\ref{eq:eam} and
\ref{eq:eam_rho}.  All functions were described by splines, and the splines were fitted such that
the difference between the forces predicted by the model and the forces determined by
first-principle calculations is minimal.  This method is an early example of using a flexible
regression for building interatomic potentials. The differences between the forces predicted by the 
Ercolessi-Adams potential and Density Functional Theory are remarkably small in bulk fcc aluminium,
although the description of surfaces is less accurate.

\subsection{The modified embedded-atom model}

Although the embedded atom model proved to be a good potential for metallic systems, it fails to
describe covalent materials, such as semiconductors.  The reason for this is that the electron
density in equation~\ref{eq:eam_rho} is assumed to be isotropic, which is a good approximation in
close packed systems, like fcc crystals, but in the case of covalent bonds, the electron density is
higher along the bonds.  In order to correct this, an angle-dependent density term was introduced by
Baskes\cite{MEAM01} for silicon
\begin{equation}
\rho_i = \sum_{j \ne i} \rho(r_{ij}) + \sum_{
j \ne i, \, k \ne i } \rho(r_{ij}) \rho(r_{ik}) g(\cos \theta_{jik})
\textrm{,}
\end{equation}
where $\theta_{jik}$ is the bond angle between the $ji$ and $ki$ bonds.
The original formulation used the fixed functional form 
\begin{equation}
g(\cos \theta_{jik}) = 1 - 3 \cos^2 \theta_{jik}
\end{equation}
for the angle-dependency, which biased the equilibrium bond angle preference to tetrahedral angles,
resulting in a poor description of liquid or non-tetrahedral phases of silicon.
Lenosky et al. adopted the force-matching method for the modified embedded-atom model\cite{MEAM02}.

Taylor showed an elegant generalisation of the modified embedded atom model in \cite{qw_meam01}.  In
this work, he formulated a Taylor-expansion of the total energy functional around the
ground-state density of atoms in terms of density variations, which led to a general expression for
the total energy of the system as a function of the atomic coordinates.  The energy of an atomic system
is determined as a functional of the atomic density as
\begin{equation}
E = \Phi[\rho(\mathbf{r})] \textrm{,}
\end{equation}
where
\begin{equation}
\rho(\mathbf{r}) = \sum_i \delta(\mathbf{r} - \mathbf{r}_i)
\end{equation}
and $\delta$ is the Dirac-delta function.
This form is, in fact, an alternative description of the total energy as given by Density Functional
Theory.  The atomic density determines, through Poisson's equation, the external potential through
which the electrons move as
\begin{equation}
\nabla^2 V_{\textrm{ext}} = -\frac{\rho}{\epsilon_0} \textrm{,}
\end{equation}
which in turn corresponds to a ground state electron density and a total energy.  If $E_0$ is the
minimum of the total energy with respect to the atomic density, the energy can be expressed in
a Taylor series in variations in the density $\rho = \rho_0 + \delta \rho$ as
\begin{equation}\label{eq:taylor_atomic_density}
E = E_0 + \int \frac{\delta E}{\delta \rho} \bigg|_{\mathbf{r}} \delta \rho(\mathbf{r}) \mathrm{d} \mathbf{r} +
\int\int \frac{\delta^2 E}{\delta \rho^2} \bigg|_{\mathbf{r},\mathbf{r}'} \delta \rho(\mathbf{r}) \delta \rho(\mathbf{r}') \mathrm{d} \mathbf{r} \mathrm{d} \mathbf{r}' + \ldots \textrm{.}
\end{equation}
The density variation $\delta \rho$ is given by
\begin{equation}\label{eq:atomic_density_var}
\delta \rho (\mathbf{r}) = \sum_i [ \delta(\mathbf{r} - \mathbf{r}_i) - \delta(\mathbf{r} - \mathbf{r}_i^0) ] \textrm{,} 
\end{equation}
where $\mathbf{r}_i^0$ are the equilibrium positions of the atoms, corresponding to the ground state
atomic density.  The first-order term in equation~\ref{eq:taylor_atomic_density} disappears because
the Taylor-expansion is performed around the minimum. Substituting \ref{eq:taylor_atomic_density} in 
equation~\ref{eq:atomic_density_var}, then integrating results in
\begin{equation}
E = E_0 = \sum_{i,j} \frac{ \delta^2 \Phi}{ \delta \rho^2} \bigg|_{\mathbf{r}_i,\mathbf{r}_j} - 
\frac{ \delta^2 \Phi}{ \delta \rho^2} \bigg|_{\mathbf{r}_i^0,\mathbf{r}_j} -
\frac{ \delta^2 \Phi}{ \delta \rho^2} \bigg|_{\mathbf{r}_i,\mathbf{r}_j^0} +
\frac{ \delta^2 \Phi}{ \delta \rho^2} \bigg|_{\mathbf{r}_i^0,\mathbf{r}_j^0} \textrm{.}
\end{equation}
Introducing the new functions
\begin{equation}
f(\mathbf{r}_i,\mathbf{r}_j) = \frac{ \delta^2 \Phi}{ \delta \rho^2} \bigg|_{\mathbf{r}_i,\mathbf{r}_j} \\
\end{equation}
and
\begin{equation}
g(\mathbf{r}_i) = \sum_j \frac{ \delta^2 \Phi}{ \delta \rho^2} \bigg|_{\mathbf{r}_i,\mathbf{r}_j^0} = 
\sum_j \frac{ \delta^2 \Phi}{ \delta \rho^2} \bigg|_{\mathbf{r}_j^0,\mathbf{r}_i}
\end{equation}
we can write the total energy as a sum of one- and two-body terms
\begin{equation}
E = E_0' + \sum_i g(\mathbf{r}_i) + \sum_{i,j} f(\mathbf{r}_i,\mathbf{r}_j) + \ldots \textrm{.}
\end{equation}

Similarly, if we consider the local atomic densities around atom $i$ 
\begin{equation}
\delta_i(\mathbf{r}) = \sum_{j \ne i} \delta(\mathbf{r}-\mathbf{r}_j) w(r_{ij}) \textrm{,}
\end{equation}
where $w$ is a screening function, we obtain the total energy expression up to second order
\begin{multline}
E = \sum_i E_{i,0} + \sum_i \sum_{j \ne i} g(\mathbf{r}_{ij}) w(r_{ij}) + \\
\sum_i \sum_{j \ne i} \sum_{k \ne i} f(\mathbf{r}_{ij},\mathbf{r}_{ik}) w(r_{ij}) w(r_{ik}) \textrm{.}
\end{multline}
This expression has the same form as the modified embedded atom model.
Taylor represented the local atomic density by bond-order parameters and different radial functions as discussed in section~\ref{sec:bond_order}.
By choosing appropriate radial functions, he obtained the original modified embedded-atom formula, but systematic improvement of the formula is also possible in his framework.

\subsection{Tersoff potential}

The form of interatomic potential suggested by Tersoff\cite{TER01} is an example of the wider family
of bond-order potentials\cite{bop06}.  The total energy is written as a sum of pair like terms,
\begin{align}
E &= \frac{1}{2} \sum_{i \ne j} V_{ij} \\
V_{ij} &= f_{\textrm{cut}} (r_{ij}) [ f_{\textrm{R}}(r_{ij}) + b_{ij} f_{\textrm{A}}(r_{ij}) ]
\end{align}
where $f_{\textrm{R}}$ and $f_{\textrm{A}}$ are repulsive and attractive terms, $f_{\textrm{cut}}$
is a cutoff function, and $b_{ij}$ is the bond-order term
\begin{align}
f_{\textrm{R}}(r_{ij}) &= A_{ij} \exp( -\lambda_{ij} r_{ij}) \\
f_{\textrm{A}}(r_{ij}) &= -B_{ij} \exp( -\mu_{ij} r_{ij}) \\
f_{\textrm{cut}} (r_{ij}) &= \left\{
\begin{array}{ll}
1 & \textrm{if $r_{ij} < R_{ij}$} \\
\frac{1}{2} + \frac{1}{2} \cos \left( \pi \frac{r_{ij}-R_{ij}}{S_{ij}-R_{ij}} \right) & \textrm{if $R_{ij} <r_{ij} < S_{ij}$} \\
0 & \textrm{if $r_{ij} > S_{ij}$} 
\end{array} \right. \\
b_{ij} &= \chi_{ij} (1+\beta_i^{n_i} \zeta_{ij}^{n_i})^{1/2n_i} \\
\zeta_{ij} &= \sum_{k \ne i,j} f_{\textrm{cut}} (r_{ik}) \omega_{ik} g(\theta_{ijk}) \\
g(\theta_{ijk}) &= 1 + \frac{c_i^2}{d_i^2} - \frac{c_i^2}{d_i^2 + (h_i - \cos \theta_{ijk})^2} \textrm{.}
\end{align}
The resulting potential is, in fact, a many-body potential, as the bond-order terms depend on the
local environment.  Bond-order potentials can also be derived from a quantum mechanical method,
tight-binding\cite{bop06} and can be regarded as an analytical approximation of the solutions of the
Schr\"odinger equation.

\section{Long-range interactions}

The electrostatic contribution to the total energy is often not negligible. If there is charge
transfer between atoms or polarisation effects are significant, the interaction between
charges, dipoles or even higher order multipoles needs to be calculated. There are well-established
methods to determine the electrostatic energy and forces, such as the Ewald-summation
technique\cite{ewald03}. The central question is the values of the electric charges and multipoles in a
particular model. In many cases fixed charges are used, for example, most water
potentials\cite{Horn04} and models of ionic crystals\cite{nacl01} have predetermined charges. Classical
water potentials describe the structure of bulk liquid water well, however, the representation of
solutions is often poor due to the fact that these models no longer describe the interactions
correctly in the modified environment and the resulting electric fields.

The electronegativity equalisation method\cite{EEM01} and the charge equilibration method\cite{QEq01}
were designed to introduce charges which depend on the atomic environment and the local electric
field. The atomic charges predicted by these methods agree well with the experimental values and with the ones
determined by quantum mechanical methods for ionic crystals and organic molecules.

Electrostatic models including multipoles have also been developed. The multipoles are often deduced from the
electronic structure determined by ab initio methods, for example, by using Wannier
functions\cite{Sagui04}. The dependence of the multipoles on the local electric field is accounted
for by including polarisability in the model. An example of a polarisable model is the shell model, where
a charge is attached to the atom by a spring, hence the dipole of the atom reacts to changes in the
local electric field. 

\section{Neural network potentials}\label{sec:neur}

Behler and Parrinello presented a new scheme for generating interatomic potentials using neural
networks that are trained to reproduce quantum mechanical data\cite{neur01}.  The main assumption of
the model is that the total energy of an atomic system can be described as a sum of atomic
contributions
\begin{equation}
E = \sum_i E_i \textrm{,}
\end{equation}
where each individual term $E_i$ depends only on the configuration of the neighbouring atoms within
a given cutoff distance. This local environment is represented using a set of symmetry functions
\begin{equation}
G^{1\alpha}_i = \sum_{j \ne i} \exp [ -\eta_{\alpha} (r_{ij} - r_{s\alpha})^2 ] f_{\textrm{cut}}(r_{ij}) \\
\end{equation}
\begin{multline}
G^{2\beta}_i = 2^{1-\zeta_{\beta}} \sum_{j \ne i} \sum_{k \ne i} (1 + \lambda_{\beta} \cos
\theta_{ijk})^{\zeta_{\beta}} \\
\exp [ -\eta_{\beta} (r_{ij}^2 + r_{ik}^2 + r_{jk}^2)] f_{\textrm{cut}}(r_{ij}) f_{\textrm{cut}}(r_{ik}) f_{\textrm{cut}}(r_{jk})
\textrm{,}
\end{multline}
where the cutoff function is
\begin{equation}
f_{\textrm{cut}} (r) = \left\{
\begin{array}{ll}
\frac{1}{2} + \frac{1}{2} \cos \left( \frac{\pi r}{r_{\textrm{cut}}} \right) & \textrm{if $r_{ij} \leq r_{\textrm{cut}}$} \\
0 & \textrm{if $r_{ij} > r_{\textrm{cut}} $} 
\end{array} \right. \textrm{.}
\end{equation}
Thus the atomic local energies $E_i$ depend on the set of symmetry variables $\{G^{1\alpha}_i,
G^{2\beta}_i\}_{\alpha,\beta}$ in an unknown way.  Instead of trying to find a parametric model for
this function, Behler and Parrinello used non-parametric regression via neural networks.  The input
data used to perform the regression is a set of total energies from reference calculations, in this
case these were Density Functional Theory calculations of different configurations of bulk silicon.
The parameters in the layers of the neural network were optimised such that the difference between
the reference energies and the energies predicted by the neural network is minimal.  The resulting potential
can then be used to describe an arbitrary number of silicon atoms.  For each atom, the symmetry
variables are first determined, then these are fed to the neural network and the neural network predicts
the atomic energies, which are added together to obtain the total energy.

\section{Gaussian Approximation Potentials}

Our aim is to formulate a generic interatomic potential, which can be reliably used in a wide
variety of applications. Arguably, Quantum Mechanics is such an interatomic potential, as it
provides \emph{ab initio} data that, to our current knowledge, is ultimately correct to the extent
that any inaccuracies are due to the limitation of the Born-Oppenheimer approximation or the
employed quantum mechanical model. The great advantage of quantum mechanical methods is that they
have true and proven predictive power, whereas classical potentials can be regarded as parametric
regression formulas that, in general, cannot be used outside their fitting regime, which usually
cannot be unambiguously classified. However, the solution of quantum mechanical equations is
computationally expensive, which limits the use of Quantum Mechanics to a modest number of atoms and a few
nanoseconds of simulation time---woefully inadequate for biomolecular and nanotechnological applications.

As in the case of other interatomic potentials, we base Gaussian Approximation Potentials on the
assumption that the total energy of the system can be written as a sum of two terms: the first
is a local, atomic contribution and the second is the long-range, electrostatic part
\begin{equation}\label{eq:total_energy}
E = \sum_i^{\textrm{atoms}} \varepsilon_i + \frac{1}{2} \sum_{i<j}^{\textrm{atoms}} \hat{L}_i \hat{L}_j
\frac{1}{r_{ij}} \textrm{,}
\end{equation}
where the operator $\hat{L}$ can be written as
\begin{equation}
\hat{L}_i = q_i + \mathbf{p}_i \cdot \nabla_i + \mathbf{Q}_i : \nabla_i
\nabla_i + \cdots \textrm{,} 
\end{equation}
and $q_i$, $\mathbf{p}_i$ and $\mathbf{Q}_i$ denote the charge, dipole and quadrupole of the $i$-th
atom, respectively. We formulate the locality of the atomic energy contributions as
\begin{equation}
\varepsilon_i \equiv \varepsilon(\{\mathbf{r}_{ij}\})
\textrm{,}
\end{equation}
where only the relative positions $\mathbf{r}_{ij}$ of the neighbouring $j$ atoms within a spherical
cutoff are considered. In atomic systems, for which charge transfer between atoms and
polarisation effects are negligible, we can simply drop the second term in
equation~\ref{eq:total_energy}. We note that short-range, well screened electrostatic effects can be
implicitly merged into the first term in equation~\ref{eq:total_energy} without great sacrifices in
accuracy.

The strict localization of $\varepsilon$ enables the independent computation of atomic energies.

The central challenge in the development of interatomic potentials is finding the form of
$\varepsilon(\{\mathbf{r}_{ij}\})$. In our approach, we do not make any prior assumptions about the
functional form of the potential. Instead, we use non-parametric, non-linear regression in the form
of a Gaussian Process to find the function values at arbitrary values. In the regression, quantum
mechanical data, such as total energies and atomic forces are used as evidence. Gaussian
Approximation Potentials can be regarded as interpolation of the quantum mechanical potential energy
surface. Moreover, the Gaussian Process framework allows us to to build into the model a strong bias,
namely, that the atomic energy function is smooth.

The advantage of Gaussian Approximation Potentials is that they are very flexible. In contrast to
analytic potentials, the accuracy of Gaussian Approximation Potentials can be improved by adding
more quantum mechanical data at various points in configurational space without changing the fit
globally. As the Gaussian Process predicts its own accuracy, it is possible to use it as a ``learn
on the fly'' method, i.e. if the predicted variance of the energy of the force in the case of a new configuration is higher than a
pre-set tolerance, the energy and forces for the new configuration can be calculated using Quantum Mechanics,
then the obtained data is added to the database in order to improve the fit. The flexibility of the
fit ensures that the best possible fit is achieved for any given data.

The Gaussian Approximation Potential scheme is similar to the Neural Network potentials introduced
by Behler and Parinello\cite{neur01}, as both uses non-linear, non-parametric regression instead of
fixed analytic forms. However, the representation of the atomic environments in GAP is complete and the
Gaussian Process uses energies and forces for regression. Moreover, the training of the neural
network involves the optimisation of the weights, whereas the training in the case of Gaussian
Process is a simple matrix inversion.

\subsection{Technical details}\label{sec:gap_formula}

The atomic energy function $\varepsilon$ depends on the atomic neighbourhood, but it is invariant
under rotation, translation and permutation of the atoms. One of the key ideas in the present work
is to represent atomic neighbourhoods in a transformed system of coordinates that accounts for these
symmetries. Ideally, this mapping should be one-to-one: mapping different neighbourhood
configurations to the same coordinates would introduce systematic errors into the model that cannot
be improved by adding more quantum mechanical data. In section~\ref{sec:rotinv} we described a
number of transformations that can be adapted to construct an invariant neighbourhood representation.
For our work, we have chosen the four dimensional bispectrum elements. In order to ensure that the
representation is continuous in space, we modified the atomic density in
equation~\ref{eq:atomic_density} to
\begin{equation}
\rho_i (\mathbf{r})= \delta(\mathbf{r}) + \sum_j \delta(\mathbf{r}-\mathbf{r}_{ij})
f_\textrm{cut}(r_{ij})
\textrm{,}
\end{equation}
where $f_\textrm{cut}$ is a cutoff function, in our case
\begin{equation}
f_\textrm{cut}(r) = \left\{ \begin{array}{ll}
0 & \textrm{if $r > r_\textrm{cut}$} \\
1/2 + \cos(\pi r / r_\textrm{cut})/2 & \textrm{if $r \leq r_\textrm{cut} $}
\end{array} \right.
\textrm{.}
\end{equation}

In Quantum Mechanics, atomic energies are not directly accessible, only the total energy of a
configuration and the forces on each atom can be determined. The forces contain cross-terms of the derivatives of
the local energies. The force on atom $i$ can be obtained by differentiating the total energy with
respect to the Cartesian coordinates of atom $i$, written as
\begin{equation}\label{eq:gap_force}
f_{i\alpha} = \frac{\partial E}{\partial r_{i\alpha}} = \sum_j^\textrm{atoms}
\frac{\partial \varepsilon_j}{\partial r_{i\alpha}}
\textrm{.}
\end{equation}
As $\varepsilon \equiv 0$ for any $r_{ij} > r_\textrm{cut}$, this summation only runs over the $N_i$
neighbours of atom $i$. The atomic energies depend directly on the bispectrum elements, which are
determined by the neighbourhood, thus the force becomes
\begin{equation}
f_{i\alpha} = \sum_j^{N_i} \sum_k \left( \frac{\partial \varepsilon}{\partial b_k}
\right)_{\mathbf{b}_j} \frac{\partial b_k}{\partial r_{i\alpha}}
\textrm{,}
\end{equation}
where $b_k$ is the $k$-th element of the bispectrum vector, and $\mathbf{b}_j$ is the bispectrum of
atom $j$. Therefore we can substitute total energy observations in the form of sums of atomic
energies, and forces, in the form of sums of derivatives of atomic energies, directly in the
formulae shown in sections~\ref{sec:gp_derivatives} and \ref{sec:gp_linear_combinations}.

If $N$ is the number of teaching points, the computational resources required for Gaussian Process
regression scales as $N^3$ for training and as $N$ for predicting values and as $N^2$ for predicting
variances. Due to the fact that we cannot add single atomic energy observations to the database,
only total energies or forces, the size of the training set and therefore the computational costs
would grow enormously. For example, if we intend to add configurations with defects to a database
that up to this point contains data for bulk atoms only, we have to add all the atomic neighbourhoods in the configuration that
contains the defect, despite of the fact that most of them are redundant because they incorporate
the bulk data that is already in the database. Similarly, a single configuration can contain many
correlated neighbourhoods.

A possible solution for this problem was given by Snelson and Ghahramani\cite{spgp01} and it was
described in section~\ref{sec:sparse}. By choosing $M$ sparse points from the complete training set,
the computational resources required for the training process scale as $NM^2$, while the cost of the
prediction of function values and variances scales as $M$ and $M^2$, respectively.

\subsection{Multispecies potentials}

It is possible to extend the scope of Gaussian Approximation Potentials to cases where there are more
than one atomic species present in the system. There are two main differences with respect to the method
described above for monoatomic
potentials. On the one hand, the different species have to be distinguished in the atomic neighbourhood
while retaining the rotational and permutational invariance, and, on the other hand, charge transfer
between different types of atoms might occur, in which case the long-range interactions have to be taken
in account. The latter is not necessary in every multispecies system, for example, in hydrocarbons or
metallic alloys there are no significant long-range interactions present\cite{brenner01}.

By modifying the atomic density function in equation~\ref{eq:atomic_density} as in
equation~\ref{eq:atomic_density_multi}:
\begin{equation}
\rho_i (\mathbf{r}) = s_i \delta(\mathbf{0}) + \sum_j s_j \delta(\mathbf{r}-\mathbf{r}_{ij})
\textrm{,}
\end{equation}
where the different species are distinguished by the different weights of the Dirac-delta functions.
The bispectrum of $\rho_i$ remains invariant to the global rotation of the atomic neighbourhood and
to permutations of atoms of the same species.

In this study, we have not developed any potentials that contain electrostatics explicitly, but
there is good evidence\cite{madden_8pole01}, that electrostatic parameters, such as charges and multipoles can
be obtained from electronic structure calculations. It is possible to fix these parameters, but
in general, the charges and multipoles will be determined by the local neighbourhood and the local
electric field, and so these effects must be incorporated any accurate potential. This branch of our
research awaits implementation.

\chapter{Computational methods}\label{chapter:methods}
\section{Lattice dynamics}

\subsection{Phonon dispersion}

Crystalline materials are composed of periodic replicas of unit cells. In our case, the unit cell is
a parallelepiped defined by the edge vectors $\mathbf{a}_1$, $\mathbf{a}_2$ and $\mathbf{a}_3$.  The
volume of the unit cell is the absolute value of determinant of the lattice matrix $\mathbf{A} =
[\mathbf{a}_1, \mathbf{a}_2, \mathbf{a}_3]$, which is nonzero, as the column vectors of the matrix
are linearly independent. The smallest unit cell is called the primitive cell. The positions $\mathbf{r}_j^0$ of the
atoms in the primitive cell form the basis of the crystal.

The crystal is built by translating the primitive cell by all the translation vectors
\begin{equation}
\mathbf{R}_{\mathbf{l}} = l_1 \mathbf{a}_1 + l_2 \mathbf{a}_2 + l_3 \mathbf{a}_3 \textrm{,}
\end{equation}
where $l_1$, $l_2$ and $l_3$ are integers. Hence the equilibrium position of the $i$-th atom in the
crystal can be written as
\begin{equation}
\mathbf{r}_i^0 = \mathbf{r}_{\mathbf{l}j}^0 = \mathbf{r}_j^0 + \mathbf{R}_{\mathbf{l}} \textrm{.}
\end{equation}
At finite temperature, atoms vibrate around their equilibrium positions, and their displacement can
be described by a small vector $\mathbf{u}$. The actual position of an atom is given by
\begin{equation}
\mathbf{r}_{\mathbf{l}j} = \mathbf{r}_{\mathbf{l}j}^0 + \mathbf{u}_{\mathbf{l}j} \textrm{.}
\end{equation}

The total potential energy $\phi$ of the crystal is a function of the positions of the atoms. The
Taylor-expansion of the potential energy is
\begin{equation}\label{eq:pot_taylor}
\phi = \phi_0 + \sum_{\mathbf{l},j,\alpha} \phi_{\mathbf{l}j\alpha} u_{\mathbf{l}j\alpha} +
\frac{1}{2} \sum_{\mathbf{l},j,\alpha} \sum_{\mathbf{l}',j',\alpha'} \phi_{\mathbf{l}j\alpha,\mathbf{l}'j'\alpha'} 
u_{\mathbf{l}j\alpha} u_{\mathbf{l}'j'\alpha'} + \ldots \textrm{,}
\end{equation}
where $\phi_0$ is the equilibrium energy. The first term in equation~\ref{eq:pot_taylor} is the
related to the force through
\begin{equation}
\phi_{\mathbf{l}j\alpha} = \frac{\partial \phi}{\partial u_{\mathbf{l}j\alpha}} = 
-f_{\mathbf{l}j\alpha} \textrm{.}
\end{equation}
This term is zero, because we perform the Taylor expansion around the minimum. The second term
contains the harmonic force constants, given by
\begin{equation}
\phi_{\mathbf{l}j\alpha,\mathbf{l}'j'\alpha'} = \frac{\partial^2 \phi} {\partial u_{\mathbf{l}j\alpha} \partial u_{\mathbf{l}j\alpha}} 
\textrm{.}
\end{equation}
In the harmonic approximation, higher order terms in the Taylor-expansion are neglected.
Newton's equations of motion are therefore written as
\begin{equation}\label{eq:harmonic_newton}
m_j \ddot{u}_{\mathbf{l}j\alpha} = \sum_{\mathbf{l}'j'\alpha'} \phi_{\mathbf{l}j\alpha,\mathbf{l}'j'\alpha'} u_{\mathbf{l}'j'\alpha'}
\textrm{,}
\end{equation}
which have wavelike solutions
\begin{equation}\label{eq:wave_sol}
\mathbf{u}_{\mathbf{l}j}(t) = \frac{1}{\sqrt{N m_j}} \sum_{\mathbf{k}\nu} A(\mathbf{k},\nu) \mathbf{e}(\mathbf{k}, \nu, j) 
\exp\left[ i (\mathbf{k} \mathbf{r}^0_{\mathbf{l}j} - \omega(\mathbf{k},\nu) t) \right] \textrm{.}
\end{equation}

Substituting \ref{eq:wave_sol} in \ref{eq:harmonic_newton}, we obtain the eigenvalue equation 
\begin{equation}\label{eq:eig_freq}
\omega^2(\mathbf{k}) e_{\alpha}(\mathbf{k}i, \nu, j) = \sum_{\alpha'j'\nu'} D_{\alpha \alpha'} \left( 
\begin{array}{c}
\mathbf{k} \\
j\nu, j'\nu'
\end{array}
  \right) e_{\alpha'}(\mathbf{k}, \nu', j') \textrm{,}
\end{equation}
where $\mathbf{D}$ is the dynamical matrix, the Fourier transform of the force constant matrix:
\begin{equation}
D_{\alpha \alpha'} \left( 
\begin{array}{c}
\mathbf{k} \\
j\nu, j'\nu'
\end{array}
  \right) = \frac{1}{m_j m_{j'}} \sum_{} \phi_{\mathbf{l}j\alpha,\mathbf{l}'j'\alpha'} 
\exp \left[ i \mathbf{k} ( \mathbf{r}^0_{\mathbf{l}j} - \mathbf{r}^0_{\mathbf{l}'j'}  ) \right]
\textrm{.}
\end{equation}
Non-trivial solutions of equation~\ref{eq:eig_freq} can be found by solving the secular determinant
\begin{equation}
|  \mathbf{D}(\mathbf{k}) - \omega^2 \mathbf{I} | = 0
\textrm{,}
\end{equation}
where the solutions are the  frequencies of different phonon modes at wavevector $\mathbf{k}$.
Substituting these solutions into \ref{eq:eig_freq}, the mode eigenvectors can also be obtained, and
these correspond to the normal modes of the vibrations. A more complete discussion of lattice
dynamics can be found, for example, in~\cite{dovebook}.

In our work, we first constructed a large supercell from the primitive cell, then perturbed each
atom in the original $\mathbf{l}=(0,0,0)$ cell by a small amount along the coordinate axes and
calculated the forces on the atoms in the perturbed supercell. We obtained an approximate force
constant matrix by the numerical differentiation of the forces, which we Fourier-transform to obtain
the dynamical matrix.  This procedure can be performed using any interatomic potential model,
although using Quantum Mechanics can be particularly expensive in the case of large supercells, i.e.
for small wavenumbers. However, this large computational cost in DFT can be avoided by calculating 
phonon dispersion relations using Density Functional Perturbation Theory, as described in
\cite{dfpt01}.

\subsection{Molecular Dynamics}

Alternatively, the phonon frequencies can also be obtained from molecular dynamics
runs\cite{mem01}. The
relative displacements $\mathbf{u}_{\mathbf{l} j}$ in equation~\ref{eq:wave_sol} can be
Fourier-transformed, leading to
\begin{equation}\label{eq:displacement_FT}
\epsilon_{\mathbf{k} j}(t) = \frac{1}{N_{\textrm{cell}}} \sum_j \sum_{\mathbf{l}} \exp( -i \,
\mathbf{k} \, \mathbf{R}_{\mathbf{l}}) \mathbf{u}_{\mathbf{l} j}
\propto \sum_{\nu} \exp(-i \omega(\mathbf{k},\nu) t)
\textrm{,}
\end{equation}
where $N_{\textrm{cell}}$ is the number of primitive cells in the supercell. Fourier-transforming
equation~\ref{eq:displacement_FT} to frequency space gives 
\begin{equation}
\epsilon_{\mathbf{k}}(\omega) \propto \sum_{\nu} \delta(\omega - \omega_{\mathbf{k},\nu})
\textrm{.}
\end{equation}
The spectral analysis of $\epsilon_{\mathbf{k}}(\omega)$, i.e. finding sharp peaks in the power
spectrum
\begin{equation}
P_{\mathbf{k}j} \equiv |\epsilon_{\mathbf{k}j}(\omega)|^2
\end{equation}
gives the phonon frequencies.

The advantage of this method is that it can be used for more complicated systems, where explicit calculation of
the full dynamical matrix would be extremely expensive. Furthermore, we can calculate the temperature
dependence of the phonon spectrum by simply performing molecular dynamics simulations at different
temperatures. The temperature dependence of the phonon spectrum is due to anharmonic effects, i.e.,
at larger displacements when terms higher than second order contribute to the potential energy in
equation~\ref{eq:pot_taylor}.

\subsection{Thermodynamics}\label{sec:tmd2tqm}

The quantum mechanical solution of a system of harmonic oscillators\cite{dovebook} states that the
allowed energies of a phonon mode labelled by $\mathbf{k}$ and $\nu$ are
\begin{equation}\label{eq:qm_harmonic}
E_{\mathbf{k} \nu} = \left( \frac{1}{2} + n \right) \omega(\mathbf{k},\nu)
\textrm{,}
\end{equation}
where $\hbar$ is the reduced Planck constant, and $n$ is a non-negative integer. The canonical
partition function of a system can be calculated as
\begin{equation}
Z = \sum_j \exp( - \beta E_j )
\textrm{,}
\end{equation}
where $E_j$ is the energy of the $j$-th state and $\beta = \frac{1}{k_B T}$.
Substituting \ref{eq:qm_harmonic} into this expression, we obtain
\begin{equation}
Z_{\textrm{vib.}} = \prod_{\mathbf{k},\nu} \left[ \sum_{n_{\mathbf{k} \nu}=0}^\infty \exp \left(
-\beta \left( \frac{1}{2} + n_{\mathbf{k} \nu} \right) \hbar \omega(\mathbf{k},\nu) \right) \right] 
\end{equation}
which can be simplified by using
\begin{equation}
\sum_{n=0}^\infty \exp(-nx) = \frac{1}{1-\exp(-x)}
\end{equation}
to
\begin{equation}
Z_{\textrm{vib.}} = \prod_{\mathbf{k},\nu} \frac{\exp( -\beta \hbar \omega(\mathbf{k},\nu) / 2)}
{1 - \exp( -\beta \hbar \omega(\mathbf{k},\nu) )}
\textrm{.}
\end{equation}
In the case of a crystal, the total partition function is
\begin{equation}
Z = \exp( -\beta \phi_0) Z_{\textrm{vib.}}
\textrm{.}
\end{equation}
The partition function can be used to obtain all thermodynamic quantities.  For example, the
free-energy can be obtained as
\begin{align}
F &= -k_B T \ln Z \\
  &= \phi_0 + k_B T \sum_{\mathbf{k},\nu} \ln \left[ 2 \sinh( \beta \hbar \omega(\mathbf{k},\nu) / 2)
\right]
\textrm{,}
\end{align}
and the internal energy is
\begin{align}
U &= \frac{1}{Z} \frac{\partial Z}{\partial \beta} \\
  &= \phi_0 + \sum_{\mathbf{k}, \nu} \hbar \omega(\mathbf{k}, \nu) \left( \frac{1}{2} + \frac{1}{\exp(-\beta
\hbar \omega(\mathbf{k},\nu)) -1} \right)
\textrm{.}
\end{align}
This result leads us to a rather crude method for approximating the real temperature in the case of a classical
molecular dynamics run\cite{phonon02}. We equate the kinetic energy $
E_{\textrm{kin.}}(T_{\textrm{MD}})$  to the quantum mechanical vibration energy
$U_{\textrm{vib.}}(T_{\textrm{QM}})$ and find the temperature $T_{\textrm{QM}}$ when
$U_{\textrm{vib.}}(T_{\textrm{QM}}) = E_{\textrm{kin.}}(T_{\textrm{MD}})$.  In the high temperature
limit $T_{\textrm{QM}} = T_{\textrm{MD}}$, but this expression allows us to relate results from
low-temperature molecular dynamics runs to experimental values.

The constant-volume heat capacity is defined as
\begin{equation}
c_V = \frac{\partial U}{\partial T}
\textrm{,}
\end{equation}
which, in the case of harmonic crystals, can be calculated as
\begin{equation}
C_V = \sum_{\mathbf{k},\nu} c_{\mathbf{k},\nu} 
\textrm{,}
\end{equation}
where $c_{\mathbf{k},\nu}$ is the contribution to the specific heat from mode $(\mathbf{k},\nu)$
\begin{equation}
c_{\mathbf{k},\nu} = k_B \left( \frac{\hbar \omega(\mathbf{k},\nu)}{k_B T} \right)^2
\frac{\exp(\beta \hbar \omega(\mathbf{k},\nu))}{\left[\exp(\beta \hbar
\omega(\mathbf{k},\nu))-1\right]^2}
\textrm{.}
\end{equation}

The volumetric thermal expansion coefficient can also be calculated from the free energy. The
thermal expansion coefficient is defined as
\begin{equation}
\alpha = \frac{1}{V} \left( \frac{\partial V}{\partial T} \right)_p = \frac{1}{V} \left(
\frac{\partial V}{\partial p} \right)_T \left( \frac{\partial p}{\partial T} \right)_V = \kappa_T
\left( \frac{\partial p}{\partial T} \right)_V
\textrm{,} 
\end{equation}
where $\kappa_T$ is the isothermal compressibility. The pressure is given by
\begin{equation}
p = -\left( \frac{\partial F}{\partial V} \right)_T
\textrm{,}
\end{equation}
which leads to the expression
\begin{equation}
\alpha = \frac{\kappa_T}{V} \sum_{\mathbf{k},\nu} \gamma_{\mathbf{k},\nu} c_{\mathbf{k},\nu}
\textrm{,}
\end{equation}
where $\gamma_{\mathbf{k},\nu}$ are the k-vector dependent Gr\"uneisen parameters
\begin{equation}
\gamma_{\mathbf{k},\nu} = -\frac{V}{\omega(\mathbf{k},\nu)}{\partial \omega(\mathbf{k},\nu)}
{\partial V} = \frac{ \partial \ln \omega(\mathbf{k},\nu)}{\partial \ln V}
\textrm{,}
\end{equation}
which describe the dependence of the phonon frequencies on the lattice volume.  The linear thermal
expansion can be obtained in a similar way and the derivation can be easily extended to
non-isotropic cases. 

We note that through the Gr\"uneisen parameters anharmonic corrections of the potential energy are
involved in the thermal expansion coefficient. The approximation that the vibrational free-energy function depends on the volume of the
crystal through the change of the phonon frequencies described by the first-order approximation
\begin{equation}
\omega(\mathbf{k},\nu, V) = \omega(\mathbf{k},\nu, V_0) + \frac{\partial
\omega(\mathbf{k},\nu)}{\partial V} \Delta V
\end{equation}
is usually referred to as the quasi-harmonic approximation~\cite{dovebook}.

At low temperatures, if most of the anharmonic effects are due to lattice expansion, the
quasi-harmonic approximation can be successfully applied. However, if the average displacement of the
atoms is so large that the potential energy cannot be approximated by quadratic terms anymore, the
approximation fails. In such cases, we can use a classical simulation method such as molecular
dynamics to sample the phase space and calculate observables using these samples. We should note
that this is strictly valid only in case of high temperatures, where $T_{\textrm{MD}} \approx
T_{\textrm{QM}}$.

However, if the anharmonic effects are large even at low temperatures, precise results
can be obtained by methods that treat the quantum character of the nuclei explicitly, for example by
path-integrals\cite{pi01} or explicitly solving the nuclear Schr\"odinger equation\cite{vib01}. 
Path-integral methods have been successfully used to calculate the partition function of
semiconductor crystals\cite{thexp04} and hydrogen impurity in metals\cite{path_integral_gillan_01}.
Explicit solution of the nuclear Schr\"odinger equation is routinely performed in the case of
molecules\cite{vib01} by using the system of eigenfunctions of the harmonic solution to
expand the wavefunction.

\chapter{Results}\label{chapter:results}
\section{Atomic energies}

The total energy in Quantum Mechanics is a global property of the system consisting of $N$ atoms and
depends on $3N-6$ variables, namely, the coordinates of the atoms. However, all interatomic potentials
are based on the assumption that the energy can be written as a sum of atomic or bond energies,
which are local and if appropriate, a long-range electrostatic component. In our work, we intend to
estimate the atomic energies by a regression scheme based directly on quantum mechanical data. If
there were a way to extract atomic energies directly from quantum mechanical calculations, these
could be used in the regression. Firstly, we consider ideas that lead to such atomic energies.

In fact, the existence of atomic energies can be justified by showing that the force acting on an
atom does not change \emph{significantly} if the position of another atom that is \emph{far enough
away} is perturbed. This statement can be formulated as
\begin{equation}
\nabla^n_{\mathbf{x}_j} \nabla_{\mathbf{x}_i} E \rightarrow 0
\textrm{ as }
| \mathbf{x}_j - \mathbf{x}_i | \rightarrow \infty \textrm{, for } \forall n \textrm{,}
\end{equation}
which we refer to as the ``strong locality assumption''.

\subsection{Atomic expectation value of a general operator}

The basic idea in the derivation of atomic properties in Quantum Mechanics is partitioning the total
expectation value of an arbitrary operator by using a suitable atomic basis set. This is a
generalisation of the Mulliken charge partitioning scheme. We consider a system of non interacting
electrons moving in an effective potential $V_{\mathrm{eff}}$, which is the case in DFT. Thus the
expectation value of a general operator $ \hat{O} $ is
\begin{equation}\label{eq:expectation_operator}
\langle O \rangle = \sum_i f_i \langle \psi_i | \hat{O} | \psi_i \rangle
\textrm{,}
\end{equation}
where $ f_i $ is the occupation number of the single-electron orbital $ \psi_i $. 
If $ \psi_i $ is expressed in an atomic basis $\{ \phi_{\alpha} \}$ in the form
\begin{equation}
\psi_i = \sum_{\alpha} N_i^{\alpha} \phi_{\alpha}
\textrm{,}
\end{equation}
we can write equation~\ref{eq:expectation_operator} as
\begin{equation}
\langle O \rangle = \sum _i \sum_{\alpha \beta} f_i N_i^{\alpha} \left(N_i^{\beta} \right)^*
\langle \phi_{\beta} | \hat{O} | \phi_{\beta} \rangle
\textrm{.}
\end{equation}
Introducing the density kernel $\mathbf{K}$ as
\begin{equation}
K^{\alpha \beta} = \sum_i f_i N_i^{\alpha} \left(N_i^{\beta} \right)^*
\end{equation}
and the matrix of operator $\hat{O}$ as
\begin{equation}
O_{\beta \alpha} = \langle \phi_{\beta} | \hat{O} | \phi_{\alpha} \rangle
\end{equation}
we obtain
\begin{equation}
\langle O \rangle = \sum_{\alpha \beta} K^{\alpha \beta} O_{\beta \alpha} = \sum_{\alpha}
(\mathbf{KO})_{\alpha \alpha} = \mathrm{Tr} \left( \mathbf{KO} \right)
\textrm{.}
\end{equation}
Each basis function $ \phi_{\alpha} $ belongs to a certain atom, thus we use the partitioning
\begin{equation}\label{eq:local_operator}
\langle O \rangle_A = \sum_{\alpha \in A} \left( KO \right)_{\alpha \alpha}
\textrm{,}
\end{equation}
which conserves the total value
\begin{equation}
\langle O \rangle = \sum_A \langle O \rangle_A
\textrm{.}
\end{equation}

\subsubsection{Mulliken charges}

The total number of electrons is obtained by setting the operator to $\hat{O} = 1 $:
\begin{equation}
N = \sum_i f_i  \langle \psi_i | \psi_i \rangle = \sum_i f_i 
= \sum_{\alpha,\beta} K^{\alpha \beta} \int \! \phi_\alpha(\mathbf{r}) \phi_\beta^*(\mathbf{r})
\end{equation}
which leads to the well-known expression for the Mulliken-charges
\begin{equation}
N_A = \sum_{\alpha \in A} \left( \mathbf{KS} \right)_{\alpha \alpha}
\textrm{,}
\end{equation}
where the elements of the overlap matrix are defined by
\begin{equation}
S_{\alpha \beta} = \int \! \phi^*_\alpha(\mathbf{r}) \phi_\beta(\mathbf{r})
\end{equation}

\subsection{Atomic energies}\label{sec:atomic_energies}

Substituting the Hamiltonian operator $\hat{H}$ into equation~\ref{eq:local_operator}, we obtain a
possible definition for the atomic energies. In the case of Density Functional Theory, the operators
can be formulated as follows.

The total energy can be written as
\begin{equation}
E[\rho] = T_s + E_\mathrm{H}[\rho] + E_\mathrm{xc}[\rho] + E_{Ze}[\rho] + E_{ZZ}
\textrm{.}
\end{equation}

The independent-particle kinetic energy $ T_s $ is given by
\begin{equation}
T_s = - \frac{1}{2} \sum_i^{\textrm{elec.}} f_i \langle \psi_i | \Delta_i | \psi_i \rangle
\textrm{,}
\end{equation}
thus we need to substitute $ \hat{O} = - \frac{1}{2} \sum_i \Delta_i $ and the matrix elements
$\sum_i\langle\phi_{\alpha} |\Delta_i|\phi_{\beta}\rangle$ have to be calculated to obtain the
atomic kinetic energy.

The Hartree energy is defined by the equation
\begin{equation}
E_\mathrm{H}[\rho] = \frac{1}{2} \int \! \! \int \! \mathrm{d}\mathbf{r} \, \mathrm{d}\mathbf{r}'
\frac{ \rho (\mathbf{r}) \, \rho (\mathbf{r}') }{|\mathbf{r} - \mathbf{r}'|}
\textrm{,}
\end{equation}
which we rewrite as
\begin{equation}
E_\mathrm{H}[\rho] = \frac{1}{2} \int \! \mathrm{d}\mathbf{r} \, \rho (\mathbf{r})
V_\mathrm{H}(\mathbf{r}) =
\frac{1}{2} \sum_i f_i \langle \psi_i |  \hat{V}_H | \psi_i \rangle
\textrm{,}
\end{equation}
where the Hartree-operator can be obtained as
\begin{equation}
V_\mathrm{H}(\mathbf{r}) = \frac{1}{2} \int \! \mathrm{d}\mathbf{r}' \frac{ \rho (\mathbf{r}')
}{|\mathbf{r} - \mathbf{r}'|}
\textrm{.}
\end{equation}

Similarly, the interaction between electrons and nuclei is given by
\begin{equation}
E_{Ze}[\rho] = \int \! \mathrm{d}\mathbf{r} \, \rho (\mathbf{r}) V_\mathrm{ext}
\textrm{,}
\end{equation}
and the exchange-correlation energy is
\begin{equation}
E_\mathrm{xc}[\rho] = \int \! \mathrm{d}\mathbf{r} \, \rho (\mathbf{r}) \epsilon_\mathrm{xc}[
\rho(\mathbf{r}) ]
\textrm{.}
\end{equation}
Hence the operators $ V_\mathrm{ext} $ and $ \epsilon_\mathrm{xc}[ \rho(\mathbf{r}) ] $ are required
to calculate the matrix elements of the external energy matrix and the exchange-correlation energy
matrix.

\subsection{Atomic multipoles}

In general, the multipole coefficients of an arbitrary charge distribution $ \rho(\mathbf{r}) $ can be
obtained as
\begin{equation}
\mu \, _{\xi_1, \xi_2, \ldots, \xi_l} =
\frac{1}{l!} \int \! \mathrm{d} \mathbf{r} \, x_{\xi_1} x_{\xi_2} \ldots x_{\xi_l} \,
\rho(\mathbf{r})
\textrm{,}
\end{equation}
where $ x_{\xi_1}, x_{\xi_2}, \ldots,x_{\xi_l} $ are the Cartesian coordinates.
This definition can be regarded as an expectation value of the general position operator $ \hat{X} $,
therefore it can be substituted into equation~\ref{eq:local_operator}, to produce the definition of
atomic multipoles:
\begin{equation}\label{eq:atomic_mult}
\mu \, _{A, \, \xi_1, \xi_2, \ldots, \xi_l} = \frac{1}{l!} \sum _{\alpha \in A} \sum _\beta
K^{\alpha \beta}
\int \! \mathrm{d} \mathbf{r} \, x_{\xi_1} x_{\xi_2} \ldots x_{\xi_l} \,
\phi _\alpha (\mathbf{r}) \phi _\beta ^* (\mathbf{r})
\textrm{,}
\end{equation}
where $ x_{\xi_l} $ is measured from atom $ A $.

It is interesting to note that the expression for atomic multipoles in equation~\ref{eq:atomic_mult}
can be obtained by defining the atomic charge density $ \rho_A $ as
\begin{equation}
\rho_A(\mathbf{r}) = \sum _{\alpha \in A} K^{\alpha \beta} \phi _\alpha (\mathbf{r}) \phi _\beta ^*
(\mathbf{r})
\textrm{.}
\end{equation}
This definition of the atomic charge density is consistent with general physical considerations, for
example it gives the total electron density when summed for all atoms:
\begin{equation}
\rho(\mathbf{r}) = \sum _A \rho_A (\mathbf{r})
\textrm{.}
\end{equation}

\subsection{Atomic energies from ONETEP}\label{sec:onetep_atomic}

ONETEP \cite{Skylaris05}, the order-N electronic total energy package is a numerical implementation
of Density Functional Theory. Unlike usual implementations of DFT, the computational resources
required for the calculation of the energy of a particular atomic system scales linearly with the number
of electrons, which makes it exceptionally efficient in investigations of large systems.
However, in our work we exploited another feature of ONETEP, namely, that it uses local basis functions.

\subsubsection{Wannier functions}
The electronic structure of periodic crystalline solids is usually represented by Bloch orbitals $
\psi_{n \mathbf{k}} $, where $n$ and $\mathbf{k}$ are quantum numbers of the band and crystal
momentum, respectively.  The Bloch states are eigenfunctions of the Hamiltonian of the crystal,
obeying the same periodicity.  Because of the fact that they are usually highly delocalised, it is
often difficult to deduce local properties from Bloch orbitals, for instance, bonding between atoms
or atomic charges.

An equivalent representation of the electronic structure is provided by Wannier functions\cite{wannier01},
which are connected to the Bloch orbitals via a unitary transformation.
Denoting the Wannier functions of band $n$ of cell $ \mathbf{R} $ by $ w_n(\mathbf{r}-\mathbf{R}) $,
we express the transformation as follows:
\begin{equation}
w_n(\mathbf{r}-\mathbf{R}) = \frac{V}{8 \pi^3} \int \! \mathrm{d} \mathbf{k} \, e^{-i \mathbf{k} \mathbf{R}} 
\, \psi_{n\mathbf{k}}(\mathbf{r})
\textrm{.}
\end{equation}
The back transformation is given by
\begin{equation}\label{eq:wannier}
\psi_{n\mathbf{k}}(\mathbf{r}) = \sum_{\mathbf{R}}  e^{i \mathbf{k} \mathbf{R}}
w_n(\mathbf{r}-\mathbf{R})
\textrm{,}
\end{equation}
where the sum is performed over all the unit cells in the crystal.

The Wannier functions obtained in equation~\ref{eq:wannier} are not unique, because it is possible
to mix the Bloch states of different band numbers by a unitary matrix $\mathbf{U}^{(\mathbf{k})}$.
The resulting Wannier functions are also a complete representation of the electronic structure,
although their localisation features are different:
\begin{equation}\label{eq:localised_wannier}
w_n(\mathbf{r}-\mathbf{R}) = \frac{V}{8 \pi^3} \int \! \mathrm{d} \mathbf{k} \, e^{-i \mathbf{k}
\mathbf{R}}
\left( \sum_m U_{mn}^{(\mathbf{k})} \psi_{m\mathbf{k}}(\mathbf{r}) \right)
\textrm{.}
\end{equation}
Since both transformations in \ref{eq:wannier} and \ref{eq:localised_wannier} are unitary, and the
original Bloch states are orthogonal, the resulting Wannier functions are also orthogonal.

\subsubsection{Nonorthogonal generalised Wannier functions}

The matrix $\mathbf{U}$ can be optimised in such a way that the resulting Wannier functions are maximally
localised, as described in \cite{wannier01}. However, orthogonality and localisation are two
competing properties, and more localised Wannier functions can be obtained if the orthogonality
constraint is removed.

The linear combination of the Bloch orbitals of different bands can be performed by using a
non-unitary matrix, resulting in nonorthogonal Wannier functions\cite{onetep01} $ \phi_{\alpha
\mathbf{R}} $:
\begin{equation}
\phi_{\alpha \mathbf{R}}(\mathbf{r}) = \frac{V}{8 \pi^3} \int \! \mathrm{d} \mathbf{k} \, e^{-i
\mathbf{k} \mathbf{R}}
\left( \sum_m M_{m\alpha}^{(\mathbf{k})} \psi_{m\mathbf{k}}(\mathbf{r}) \right)
\textrm{.}
\end{equation}
In ONETEP, Wannier functions are constrained in a localisation sphere centred on atoms, i.e.
$\phi_{\alpha \mathbf{R}} \equiv 0$ outside the localisation sphere, providing an atomic basis set.
The radius of the localisation sphere is set by considering the electronic structure of the system
or it can be increased until convergence of the physical properties is achieved. The nonorthogonal
Wannier functions are optimised during the electronic structure calculation, hence they represent
the ``best possible'' atomic basis functions of a particular system. In our studies of the atomic
properties, we used these Wannier functions as the atomic basis set for calculating atomic
properties with our definition for these properties given in equation~\ref{eq:local_operator}.

\subsection{Locality investigations}\label{sec:locality}

In order to use the atomic energies obtained from quantum mechanical calculations as the target data
of our regression scheme we have to ensure that the atomic energies are local. We tested the degree
of this locality through the variation of the local energy caused by the perturbation of atoms
outside a spatial cutoff. If the atomic energies are local, they can be regarded purely as functions
of the local atomic environment and can be fitted by the Gaussian Process method.

The basic idea for testing the degree of the locality is that we generate a number of
configurations, where the nearest neighbours of a certain atom were held fixed, while the positions of other
atoms were allowed to vary. We calculated the local energy of the atom whose neighbourhood was fixed
for each of these configurations and compared them. We then repeated this process for different
neighbourhood configurations.

\begin{figure}[htb]
  \begin{center}
  \includegraphics[width=30mm]{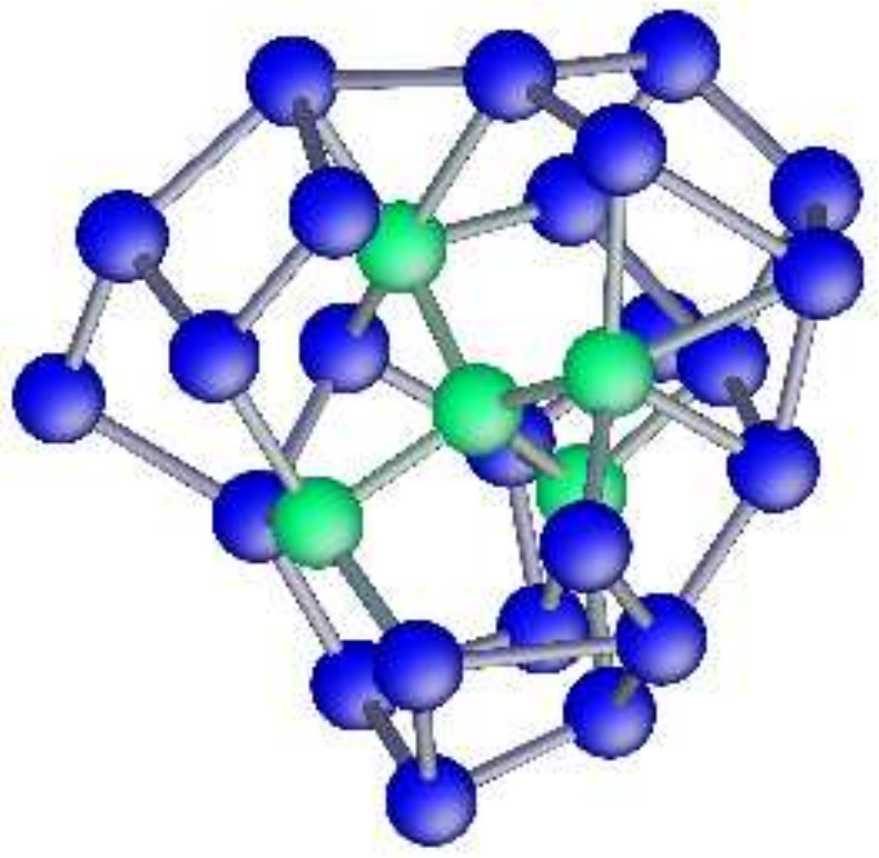}
  \end{center}
  \caption{An example configuration of the examined Si clusters. The atoms which were fixed during
the molecular dynamics simulation are shown in green.}\label{fig:clust}
\end{figure}

As a test system, we used clusters of 29--71 silicon atoms. The configurations were generated by molecular
dynamics simulation at 3000~K, where the forces were obtained from the Stillinger-Weber
potential\cite{sw01}. We performed the electronic structure calculations of the different clusters
using ONETEP, and we also used ONETEP to determine the atomic energies, as described in
section~\ref{sec:atomic_energies}. A typical cluster is shown in figure~\ref{fig:clust}.

We examined the components of the atomic energies which depend principally on the electron density
of the central atom. We calculated the average variation of the atomic kinetic, nonlocal and
exchange-correlation energies and also, the total atomic energy corrected for the long-range
interactions. The atomic energy was calculated as
\begin{equation}
E_i = E^\mathrm{kin}_i + E^\mathrm{nonloc}_i + E^\mathrm{xc}_i + E^{ee}_i + E^{Ze}_i -
\frac{1}{2} \sum_j ^\mathrm{atoms} \hat{L}_i \hat{L}_j \frac{1}{r_{ij}}
\textrm{,}
\end{equation}
where the operator $\hat{L}$ can be written as
\begin{equation}
\hat{L}_i = q_i + \mathbf{p}_i \cdot \nabla_i + \mathbf{Q}_i : \nabla_i
\nabla_i + \cdots \textrm{.} 
\end{equation}

\begin{figure}[htb]
  \begin{center}
  \includegraphics[width=6cm]{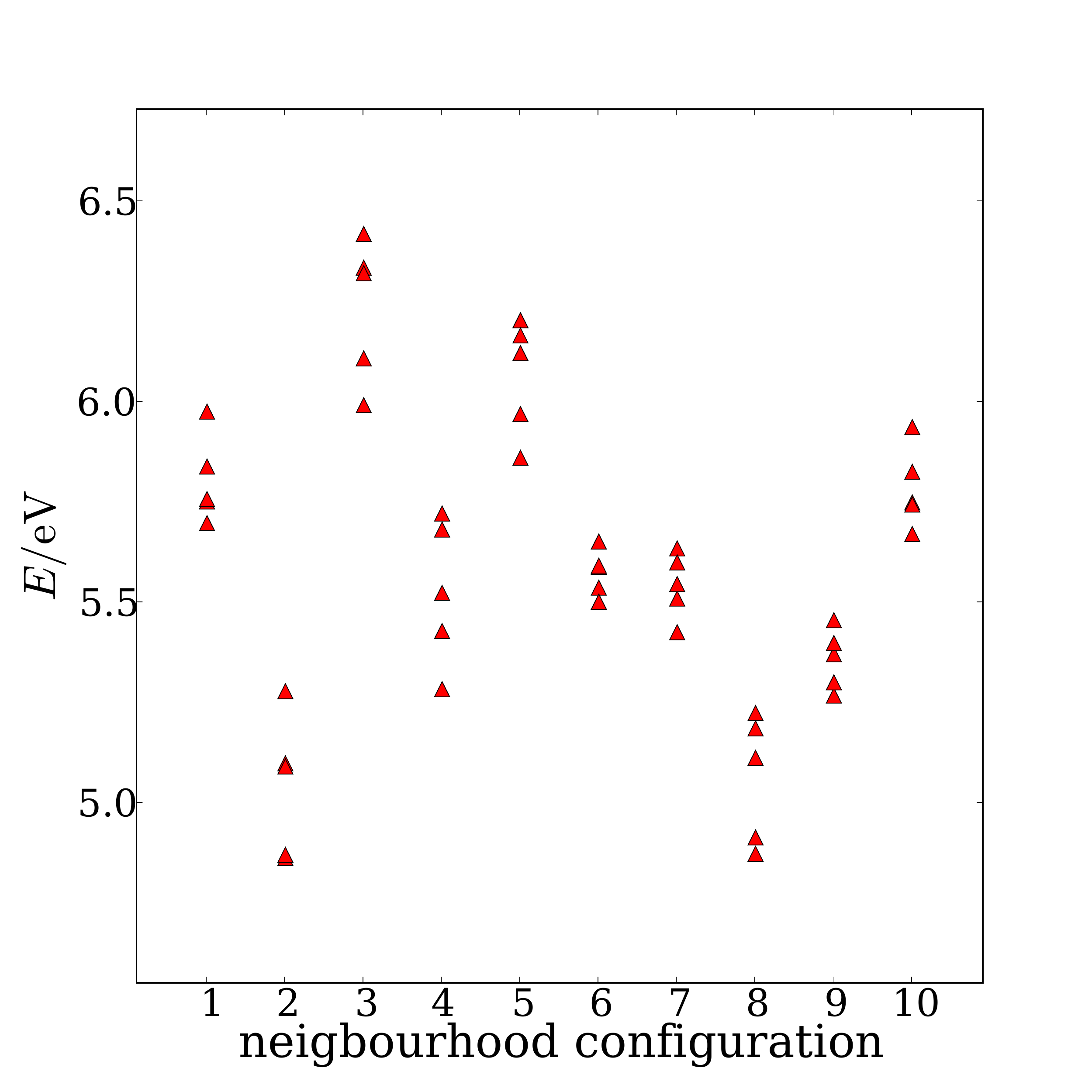}
  \end{center}
  \caption{Atomic energies $E = E^\mathrm{kin}_i + E^\mathrm{nonloc}_i + E^\mathrm{xc}_i$ of silicon
clusters for different neighbourhoods.}\label{fig:kinxc}
\end{figure}

\begin{figure}[hbt]
  \begin{center}
  \includegraphics[width=6cm]{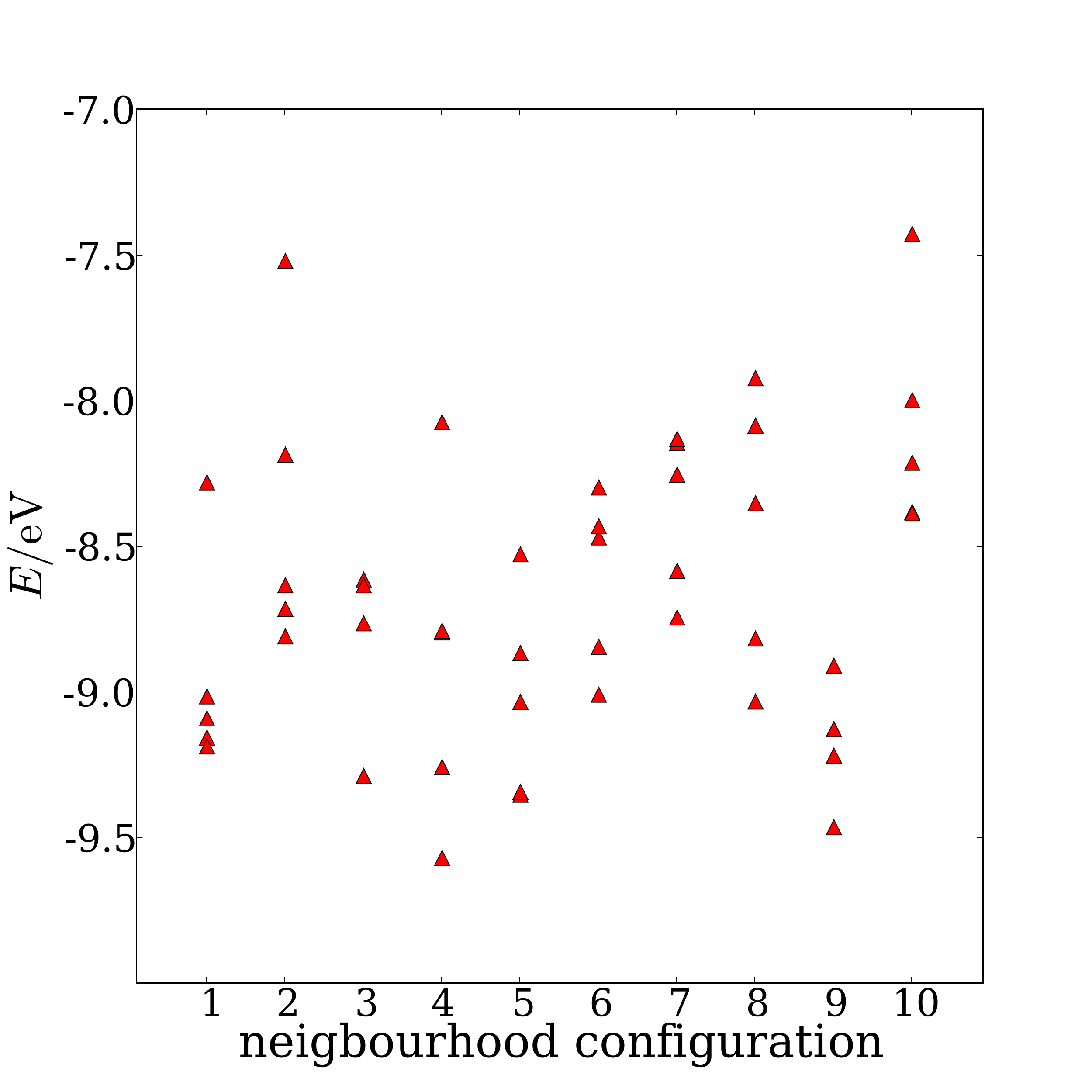}
  \end{center}
  \caption{Atomic energies of silicon clusters corrected for Coulomb contributions: $E = E_i -
\frac{1}{2} \sum_j ^\mathrm{atoms} \hat{L}_i \hat{L}_j \frac{1}{r_{ij}} $ for different
neighbourhood configurations.}\label{fig:enmult}
\end{figure}

The variations of the sum of kinetic, nonlocal and exchange-correlation terms are depicted in figure
\ref{fig:kinxc}, while figure \ref{fig:enmult} shows the spread of atomic energies corrected for
long-range interactions.
Ideally, variations in the atomic energies should be within $0.1$~eV for each neighbourhood as this
is usually reckoned to be the standard DFT error. It is obvious that our results do not fit into
this range. These results are not satisfactory and indicate that either the atomic energies depend
on more neighbours, or that the atomic energies calculated by this particular method are not local.
However, we found when using our final implementation of Gaussian Process (described in
section~\ref{sec:gap_formula}) an explicit definition of local energies is not necessary, as the
Gaussian Process infers these from total energies and forces. We shall discuss the inferred atomic
energies in section~\ref{sec:gap_atomic_energies}.

\section{Gaussian Approximation Potentials}

We have implemented the Gaussian Process to infer atomic energies from total energies and atomic
forces. Gaussian Processes belong to the family of non-linear, non-parametric regression methods,
i.e. not having fixed functional forms. The atomic environments are represented by the four dimensional
bispectrum, which is invariant to permutation of neighbouring atoms and the global rotation of the
environment.  In order to demonstrate the power of this new tool,
we built potentials for a few technologically important materials and we examined how closely the
fitted potential energy surface is to the original, quantum mechanical one. At this stage of the
work, most of the configurations we used for the training were close to the crystalline structure of
the material, hence the use of the current potentials is limited to crystalline phases. 
However, to show the ability of our potential to describe mode widely varying configurations, in the case
of carbon we built a potential that could describe the $sp^2-sp^3$ transition of the carbon atoms,
the $(111)$ surface of diamond and a simple point defect.

Our aim is twofold. On the one hand, we would like to generate potentials for general use, which can
be extended, if needed. On the other hand, there are applications where ``disposable'' force fields
are sufficient. For example, when simulating a crack or defects in a crystalline material, only a
restricted part of the potential energy surface is accessible. In these cases, a purpose-built
potential can be used, which can be generated more rapidly.

\subsection[GAP for diamond, silicon and germanium]{Gaussian Approximation Potentials for simple
semiconductors: diamond, silicon and germanium}\label{sec:GAP_dia}

Our first application of the Gaussian Approximation Potentials was a set of potentials for simple
semiconductors. We calculated the total energies and forces of a number of configurations, which
were generated by randomly displacing atoms in the perfect diamond structure. We included 8-atom
and 64-atom supercells at different lattice constants and we perturbed the lattice vectors in some
cases. The atoms were displaced at most by 0.3~\AA.

The parameters of our representation are the spatial cutoff and the resolution of the bispectrum.
We set the former to 3.7~\AA, 4.8~{\AA} and 5.0~{\AA} for carbon, silicon and germanium,
respectively. The resolution of the bispectral representation can be changed by varying a single
parameter, the maximum order $J_{\textrm{max}}$ of the spherical harmonics coefficients we use when
constructing the bispectrum. We used $J_{\textrm{max}} = 5$ in all cases. During the sparsification,
we chose 300 atomic neighbourhoods in all cases. Due to the method of generating these configurations
all the neighbourhoods were similar, thus we decided to select the set of atomic environments for
the sparsification randomly.

The electronic structure calculations were performed using CASTEP\cite{castep01}. We used the local
density approximation for carbon and the PBE generalised gradient approximation for silicon and
germanium. The electronic Brillouin zone was sampled by using a Monkhorst-Pack k-point grid, with a
k-point spacing of at most 1.7~$\textrm{\AA}^{-1}$.  The plane-wave cutoff was set to 350~eV, 300~eV
and 300~eV for C, Si and Ge, respectively, and the total energies were extrapolated for infinite
plane-wave cutoff. Ultrasoft pseudopotentials were used with 4 valence electrons for all ions.

\begin{figure}
\begin{center}
\includegraphics[width=6.5cm]{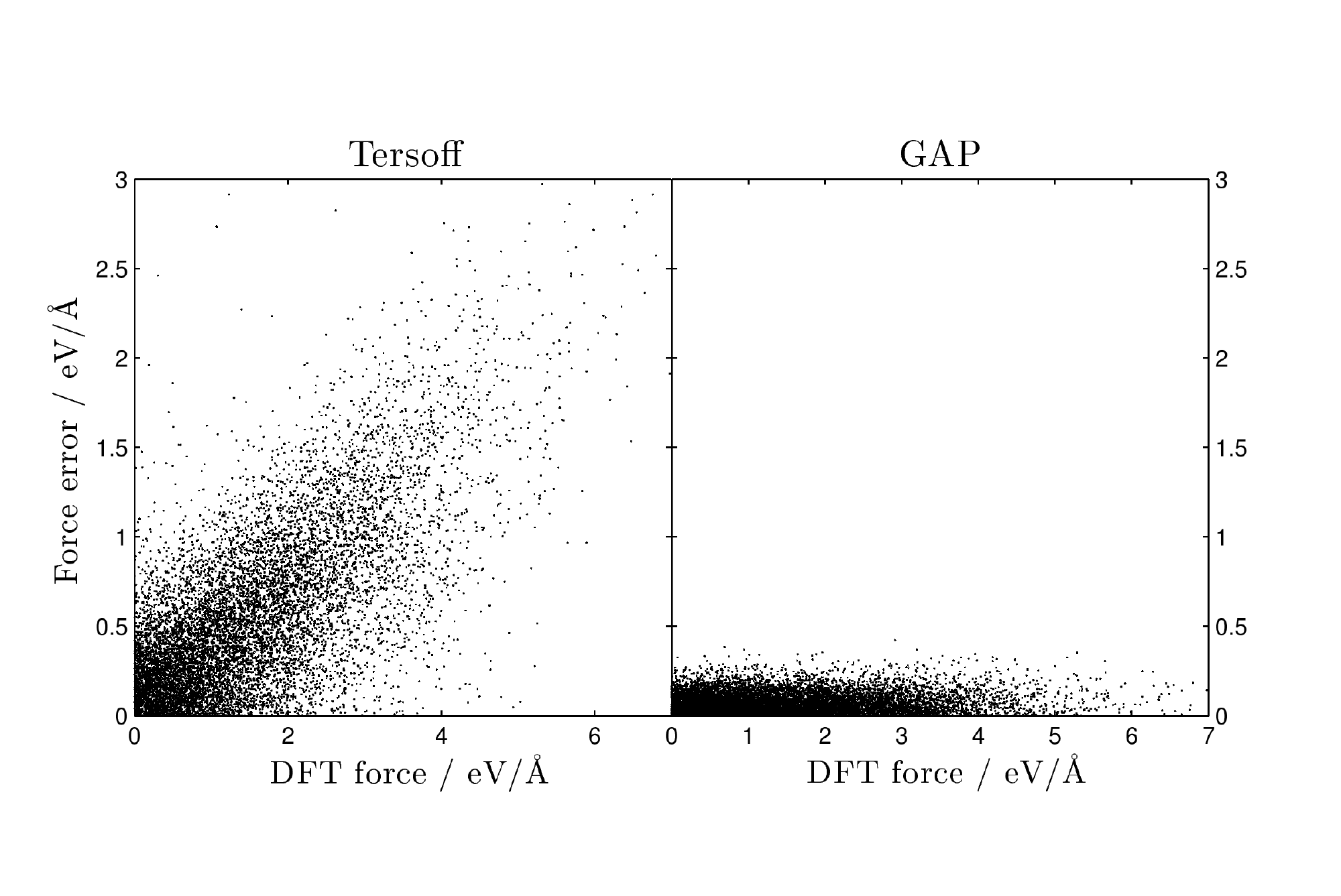}
\includegraphics[width=6.5cm]{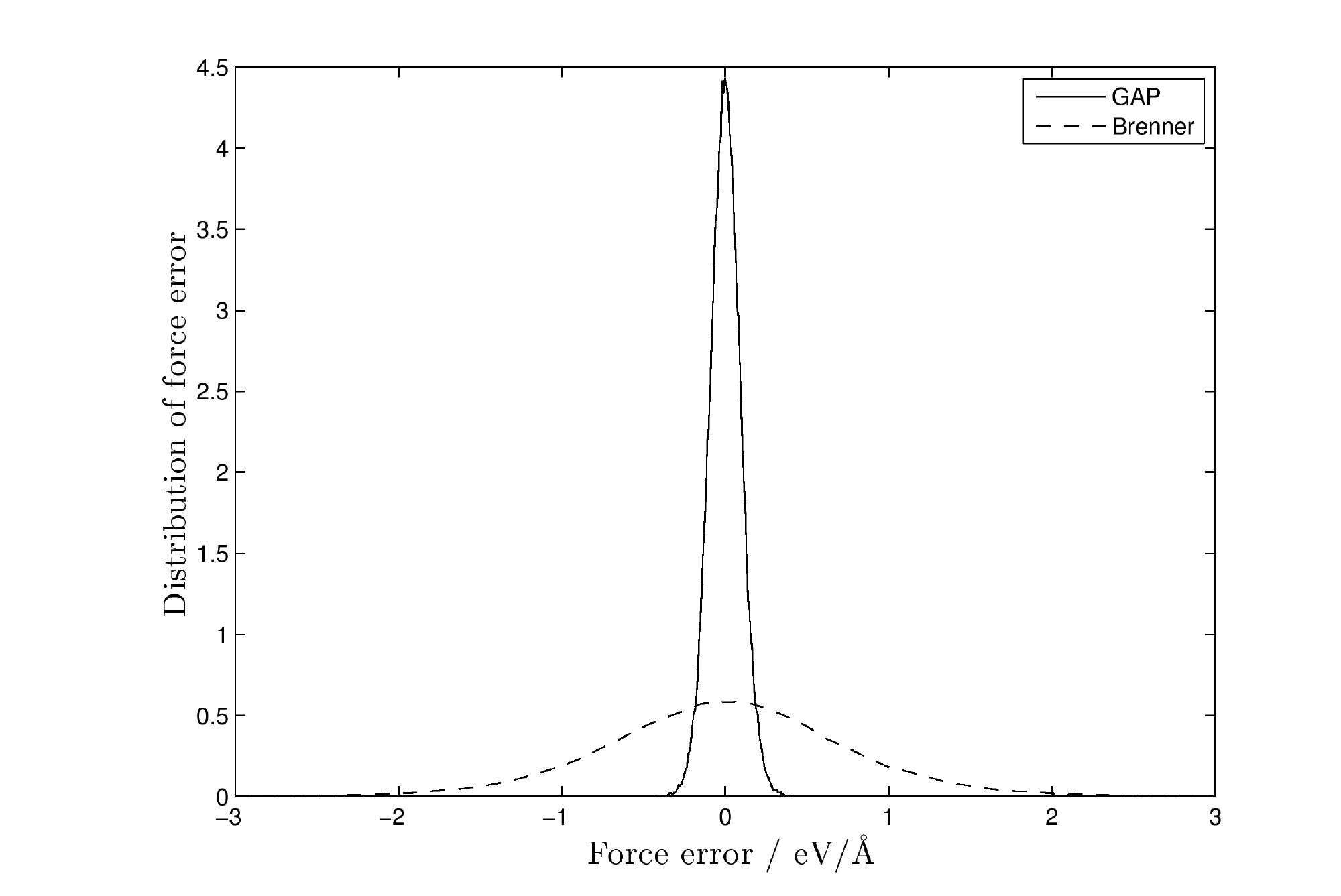}
\end{center}
\caption{\label{fig:corr_GAP_brenner} Force errors compared to DFT forces for GAP and the Brenner
potential in diamond. The left panel shows the force errors at different DFT forces. On the right panel, the
distribution of the force errors is shown.}
\end{figure}

In figure~\ref{fig:corr_GAP_brenner} we show the performance of GAP, compared to the
state-of-the-art interatomic potential, the Brenner potential\cite{brenner03}.
The set of configurations used for testing was obtained from a long ab initio molecular dynamics run
of a 64-atom supercell at 1000~K. The absolute values of the components of the difference between
the predicted and the DFT forces are shown as a function of the DFT force components and the
distribution of these differences is also displayed. The force and energy evaluation with the
Gaussian Approximation Potential for diamond, in the current implementation, is about 4000 times
faster than Density Functional Theory in the case of a 216-atom supercell.

\begin{figure}
\begin{center}
\includegraphics[width=6.5cm]{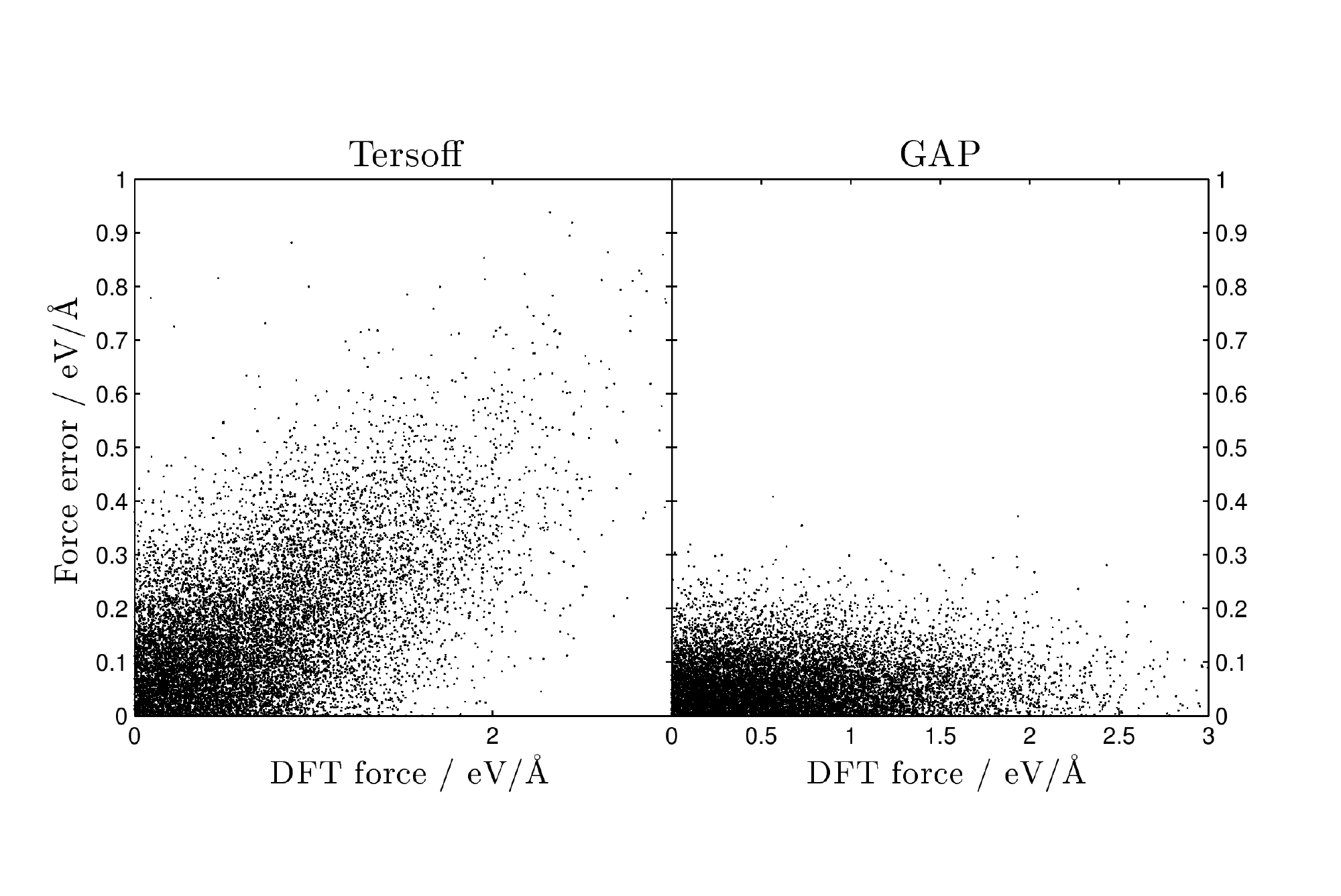}
\includegraphics[width=6.5cm]{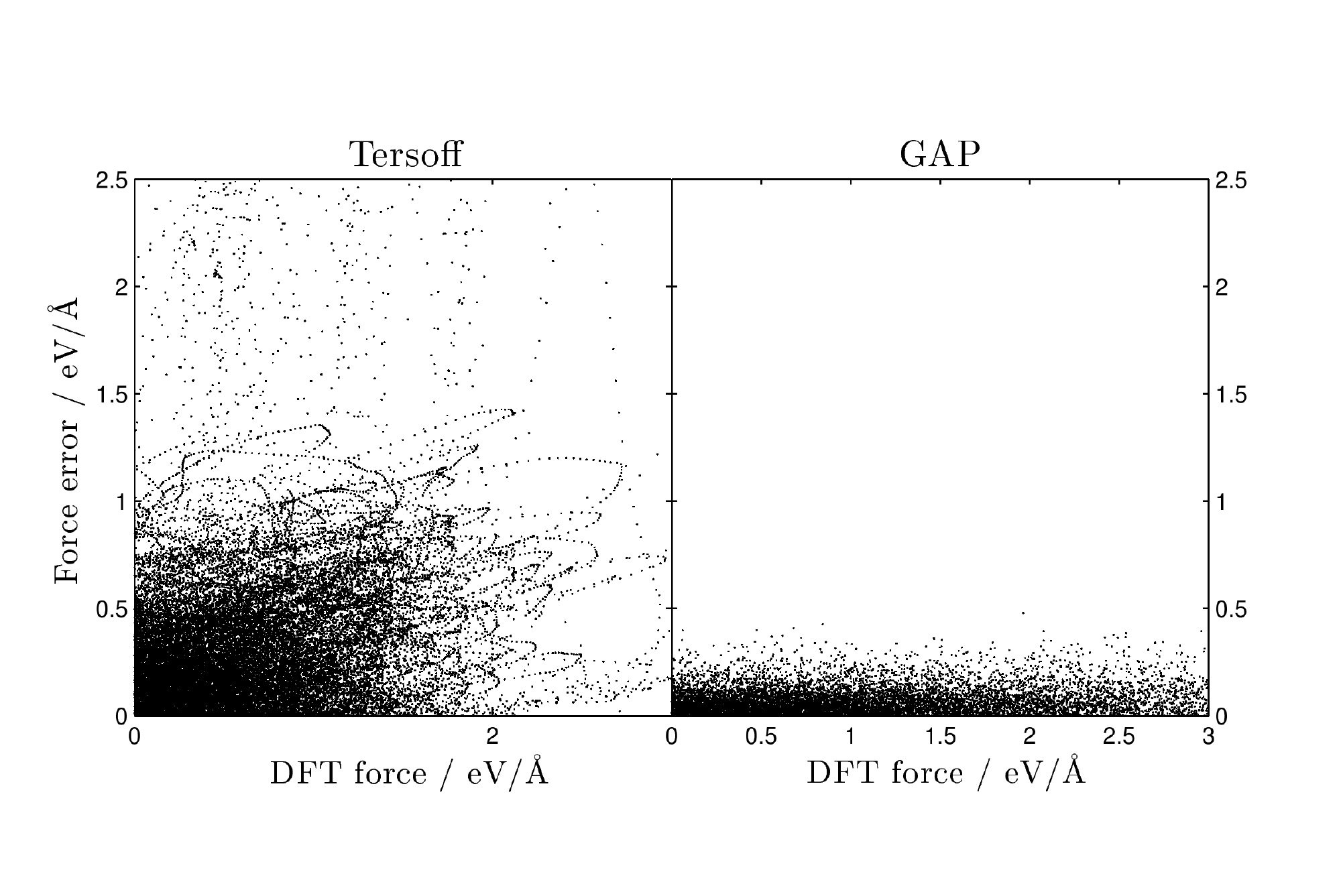}
\end{center}
\caption{\label{fig:corr_Si} Force errors compared to DFT forces for GAP and the Tersoff
potential. The silicon potentials are shown in the left panel and the germanium potentials in the
right.}
\end{figure}

We show in figure~\ref{fig:corr_Si} the results for our potentials which were developed to model the
two other group IV semiconductors, silicon and germanium, compared to the Tersoff potential.

The strict localisation of the atomic energies places a limit on the accuracy
with which the PES can be approximated. If we consider an atom whose environment \emph{inside}
$r_{\textrm{cut}}$ is fixed, but the position of other atoms are allowed to vary, the forces on this
atom will still show a variation, depending on its environment \emph{outside} the cutoff. An
estimate of this theoretical limit can be obtained by calculating the force on an atom inside a fixed
environment in various configurations. For carbon atoms in the diamond structure with
$r_{\textrm{cut}}$ = 3.7~{\AA} this error estimate is 0.1~eV/\AA.

\subsection{Parameters of GAP}

In diamond, we carried out the GAP training process using different parameters to determine the
accuracy of the representation. We truncated the spherical harmonics expansion in
equation~\ref{eq:4d_exp} at $J_{\textrm{max}}$, which therefore represents the resolution of the
bispectrum. Employing more spherical harmonics coefficients requires more computational resources,
partially because of the increased number of operations needed for the calculation of the bispectrum
and partially because there are more invariant elements, which affects the calculation of the
covariances in equation~\ref{eq:cov_mat_element}.
\begin{figure}
\begin{center}
\includegraphics[width=10cm]{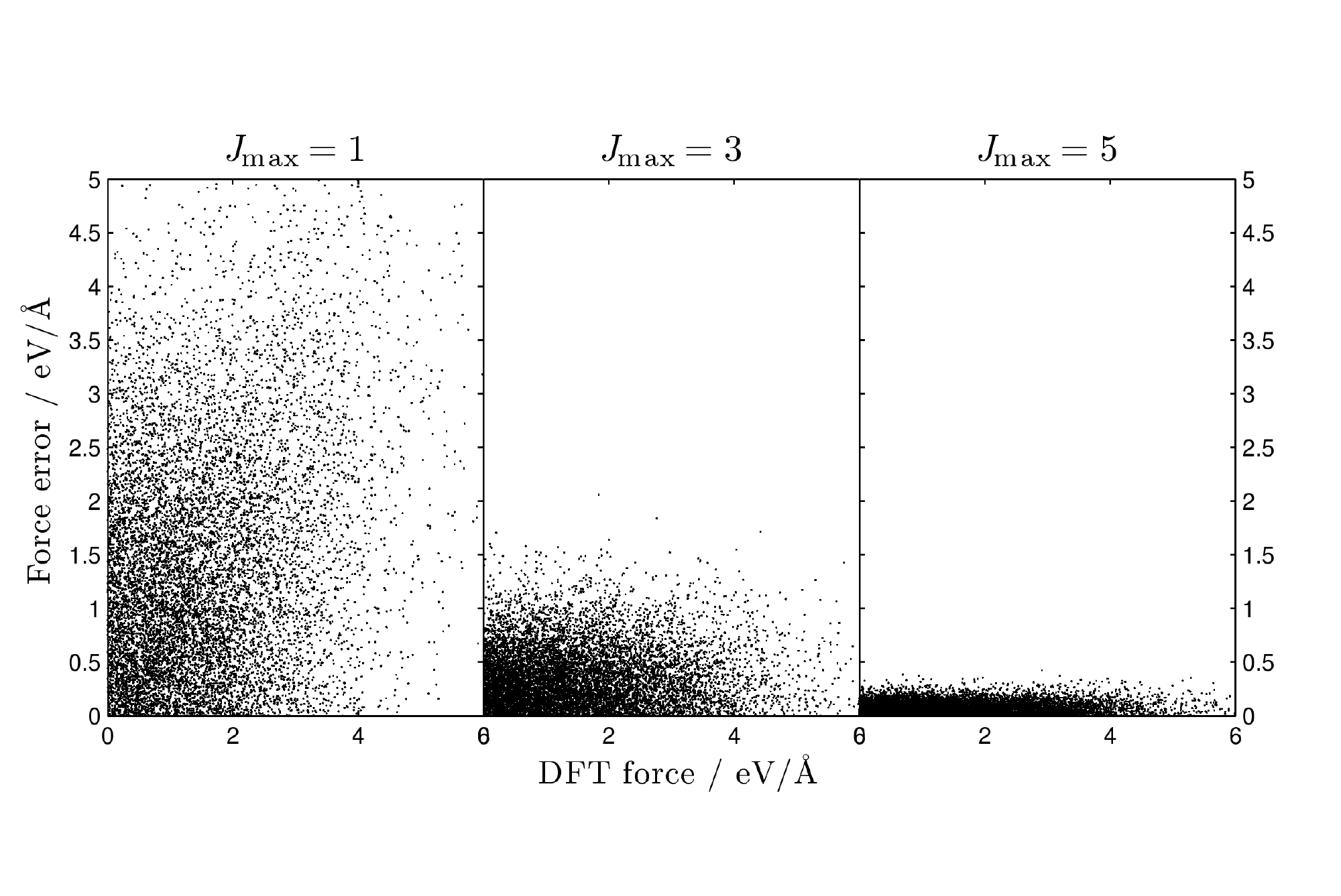}
\end{center}
\caption{\label{fig:force_GAP_J} Force correlation of GAP models for diamond with different resolution of
representation. The number of invariants were 4, 23 and 69 for $J_{\textrm{max}}=$1, 3 and 5,
respectively.}
\end{figure}
Figure~\ref{fig:force_GAP_J} shows the force error of three different GAP models. The cutoffs of all
three models were 3.7~\AA, but the spherical harmonics expansion was truncated at the first, the
third and the fifth channel, respectively. We chose $J_{\textrm{max}}=5$ for our model, as in this
case the standard deviation of the force errors reached the theoretical limit of 0.1~eV/{\AA}
associated with the spatial cutoff.

\begin{figure}
\begin{center}
\includegraphics[width=10cm]{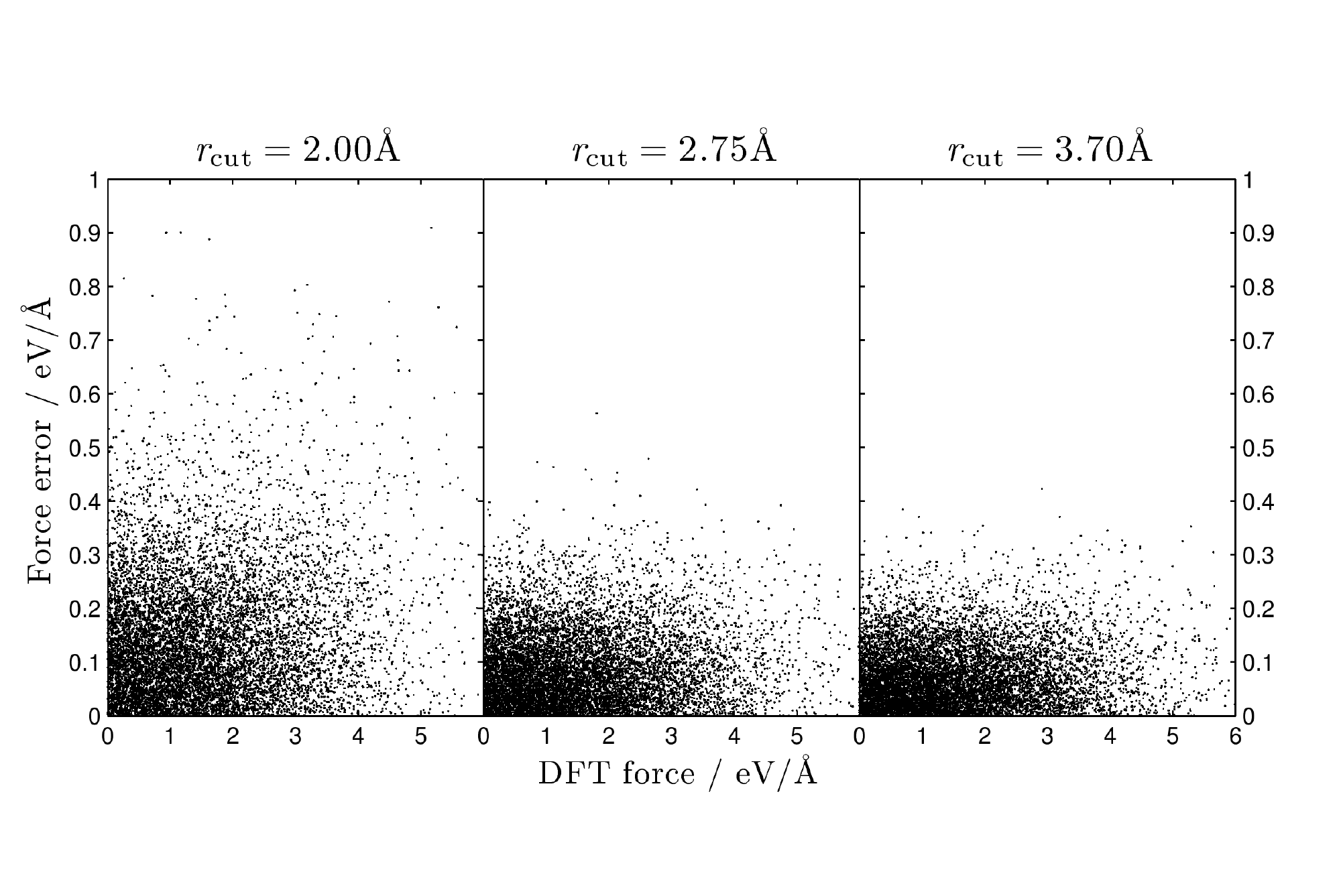}
\end{center}
\caption{\label{fig:force_GAP_rcut} Force correlation of GAP models for diamond with different
spatial cutoffs.}
\end{figure}

Figure~\ref{fig:force_GAP_rcut} shows the force errors of three Gaussian Approximation Potential
models for diamond with cutoffs of 2.0~\AA, 2.75~{\AA} and 3.7~\AA. The difference between the
latter two models is
negligible. However, the elastic moduli calculated from the model with $r_{\textrm{cut}}=2.75$~{\AA}
did not match the elastic moduli of the ab initio model and so we chose $r_{\textrm{cut}}=3.7$~{\AA}
for our final GAP potential.

\subsection{Phonon spectra}

\begin{figure}[hbt]
\begin{center}
\includegraphics[width=12cm]{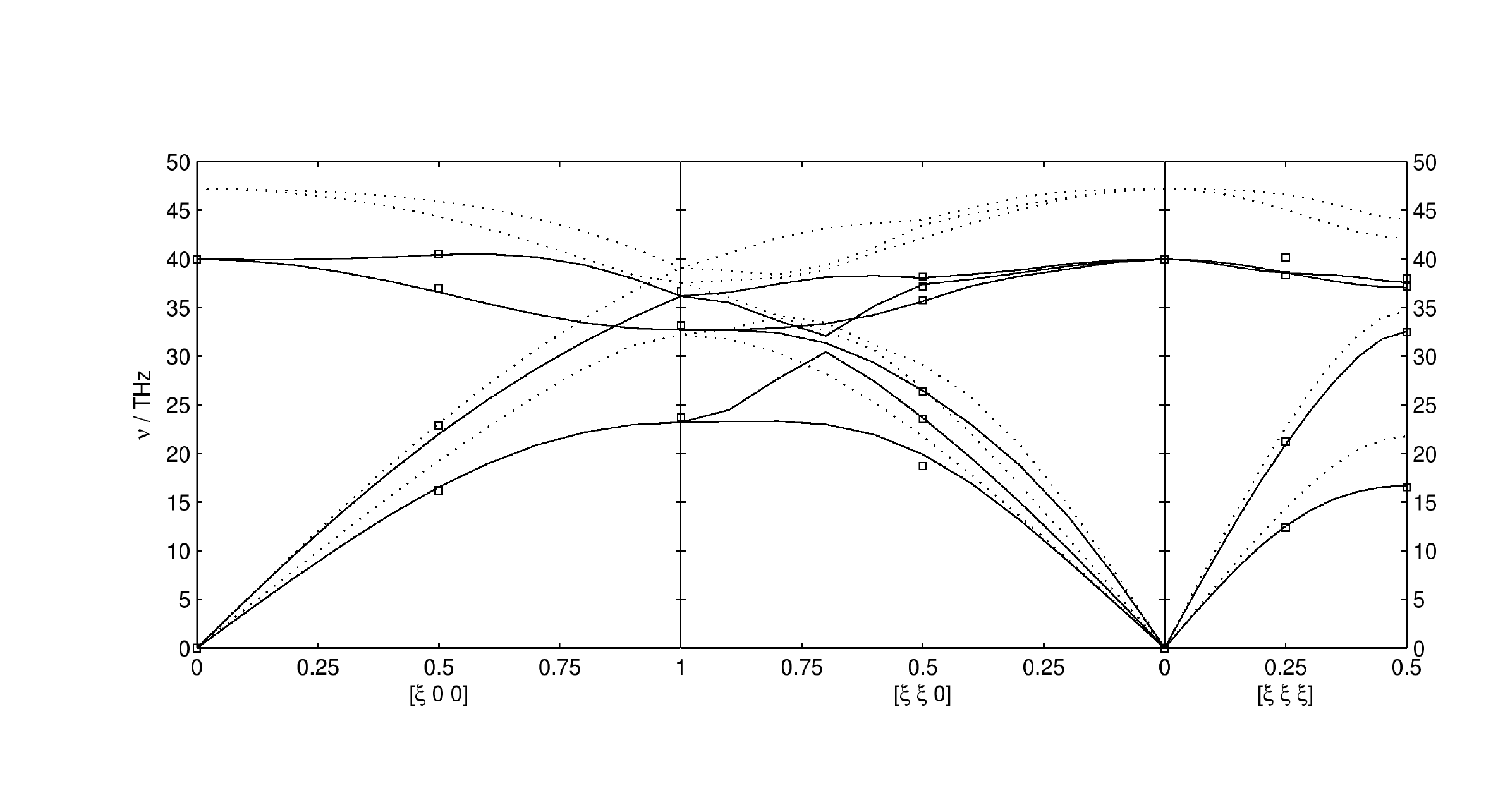}
\end{center}
\caption{\label{fig:dia_phonon} Phonon dispersion of diamond calculated by GAP (solid lines), the
Brenner potential (dotted lines) and LDA-DFT (open squares).}
\end{figure}

\begin{figure}
\begin{center}
\includegraphics[width=12cm]{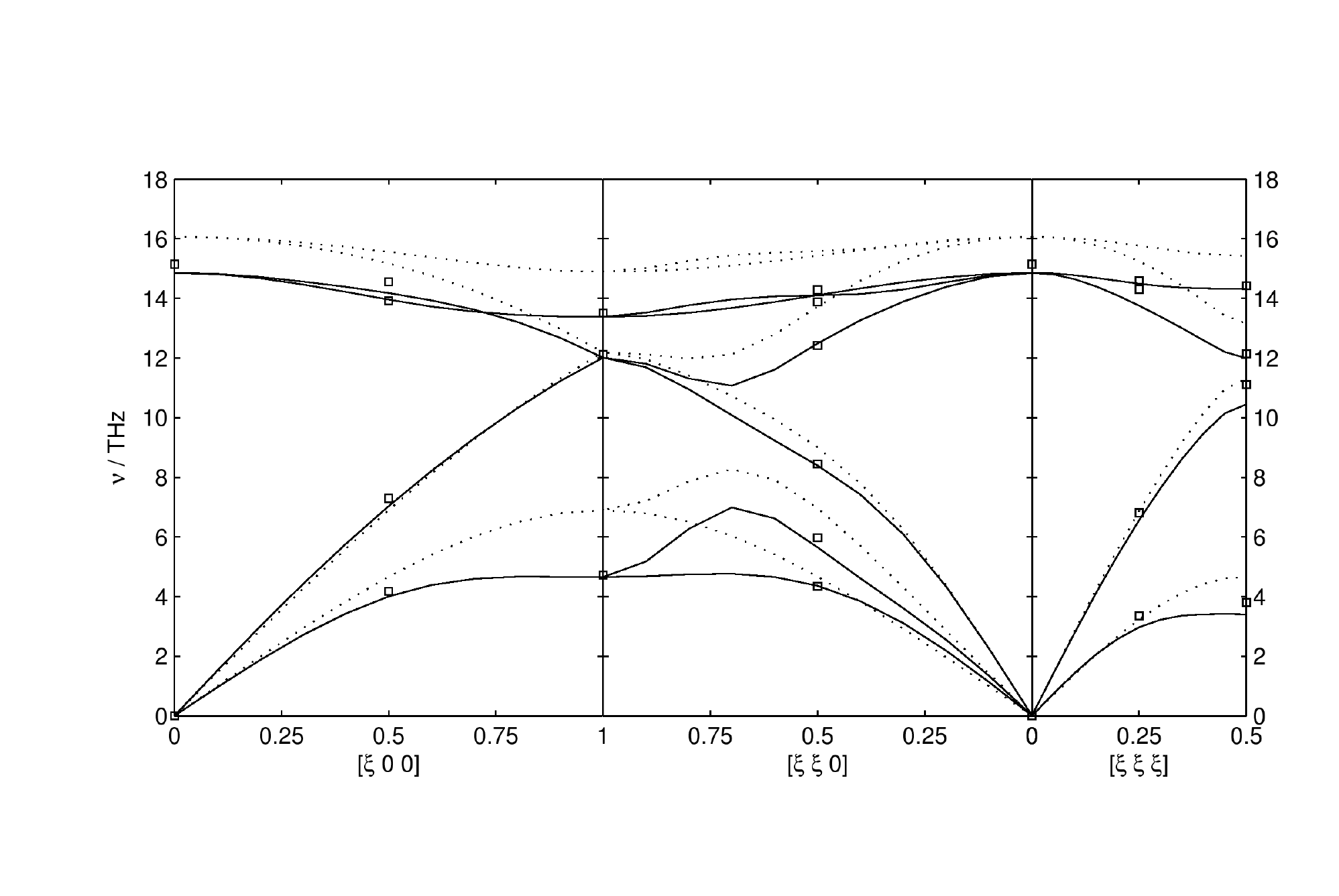}
\caption{\label{fig:si_phonon} Phonon dispersion of silicon calculated by GAP (solid lines), the
Tersoff potential (dotted lines) and PBE-DFT (open squares).}
\end{center}
\end{figure}

\begin{figure}
\begin{center}
\includegraphics[width=12cm]{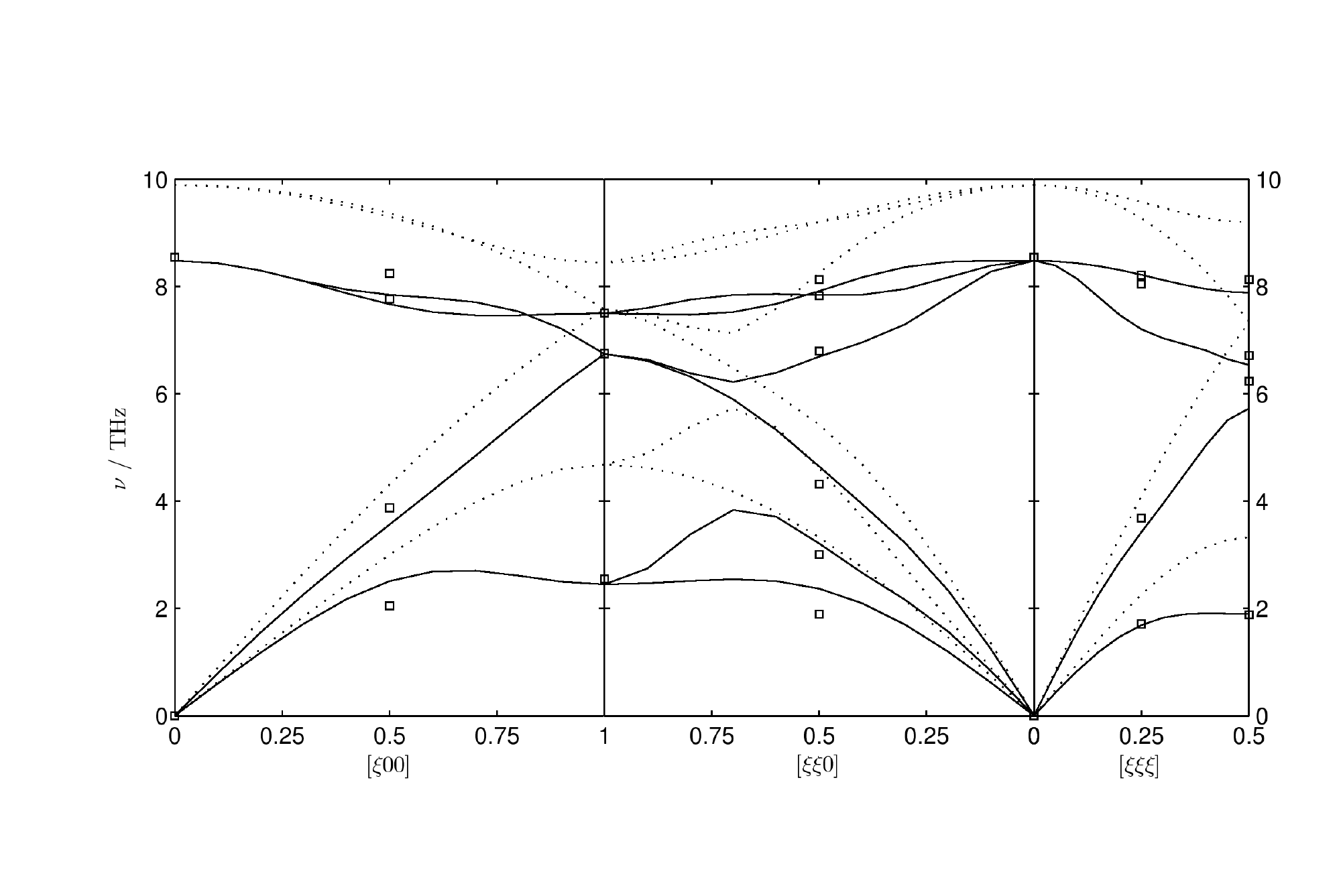}
\end{center}
\caption{\label{fig:ge_phonon} Phonon dispersion of germanium calculated by GAP (solid lines), the
Tersoff potential (dotted lines) and PBE-DFT (open squares).}
\end{figure}

\begin{figure}[hbt]
\begin{center}
\includegraphics[width=12cm]{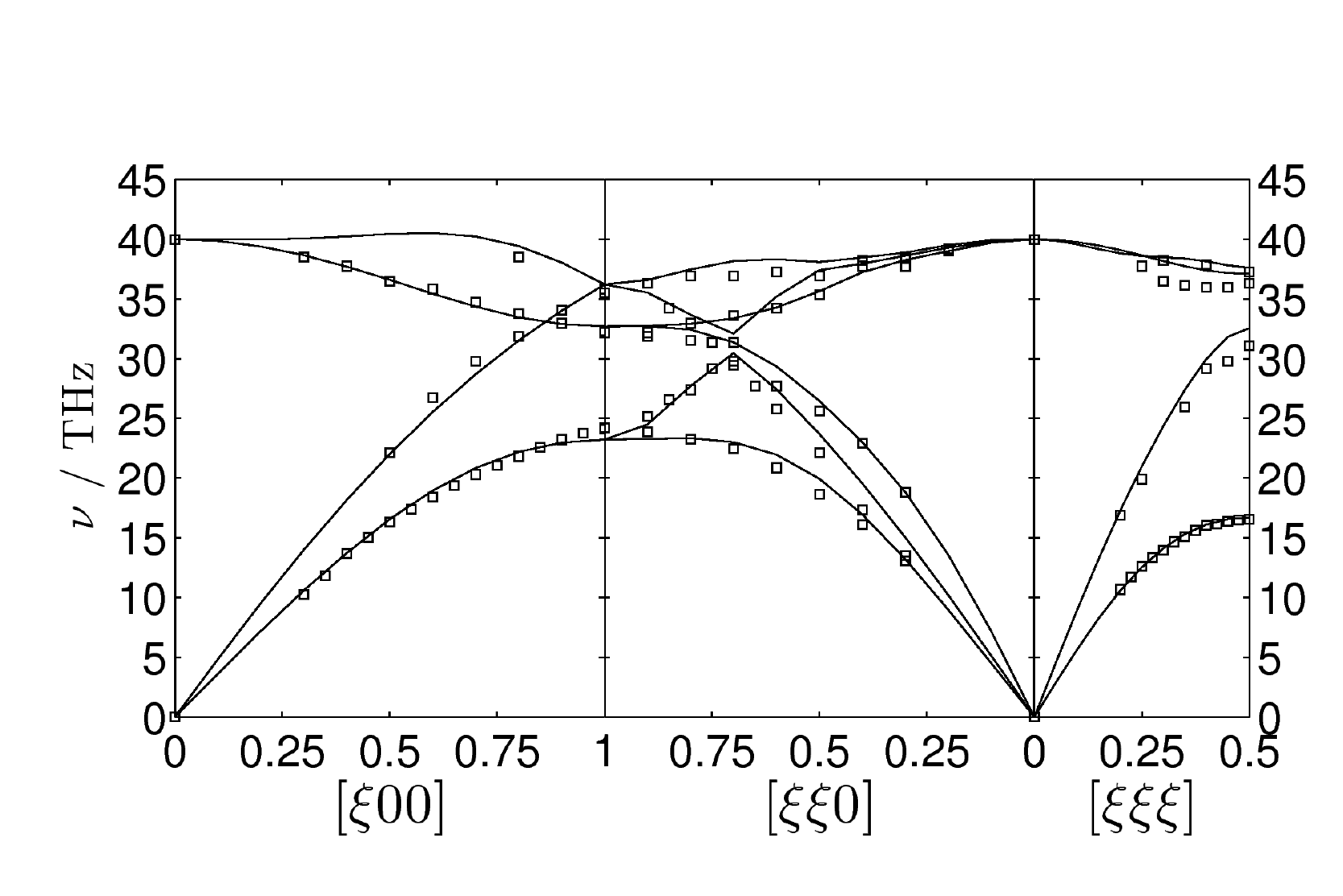}
\end{center}
\caption{\label{fig:dia_phonon_exp_GAP} Phonon dispersion of diamond calculated by GAP (solid lines)
and experimental data points\cite{phonon12,phonon_dia_exp01} (open squares).}
\end{figure}

The force error correlation is already a good indicator of how well our potential fits the original
potential energy surface. In addition, we determined the accuracy of a few other properties. The
phonon dispersion curves represent the curvature of the potential energy surface around the lowest
energy state. We calculated the phonon spectrum by the finite difference method using GAP. The
force-constant matrix of the model was calculated by the numerical differentiation of the forces,
and the phonon spectrum was obtained as the eigenvalues of the Fourier-transform of the
force-constant matrix. The parameters of the GAP potentials are given in
table~\ref{tab:gap_parameter}.
\begin{table}
\begin{center}
\begin{tabular}{cccc}
\hline
\hline
\\[-2.0ex]
   & C & Si & Ge \\[0.5ex]
\hline
\\[-2.0ex]
$r_{\textrm{cut}} / \textrm{\AA}$ & 3.7 & 4.8 & 5.0 \\
$J_{\textrm{max}}$                & 5   & 5   & 5   \\
\hline
\hline
\end{tabular}
\end{center}
\caption{\label{tab:gap_parameter} Parameters of the used GAP potentials.}
\end{table}
We compared the phonon values at a few points in the
Brillouin zone with the ab initio values and the analytic potentials. These results are shown in
figures~\ref{fig:dia_phonon}, \ref{fig:si_phonon} and \ref{fig:ge_phonon} for diamond, silicon and
germanium, respectively.
The GAP models show excellent accuracy at zero temperature over most of the Brillouin zone, with a
slight deviation for optical modes in the $(111)$ direction. The agreement of the phonon spectrum
of GAP with the phonon spectrum of Density Functional Theory suggests that any quantity that can be derived
from the vibrational free-energy, such as the constant-volume heat capacity, at low temperatures
will also show good agreement.
We found excellent agreement between the phonon frequencies calculated by the GAP potential for
diamond and the dispersion curves measured by inelastic neutron
scattering\cite{phonon12,phonon_dia_exp01}, shown in figure~\ref{fig:dia_phonon_exp_GAP}.

We also calculated the elastic constants of our models and these are compared to Density Functional
Theory and existing interatomic potentials in table~\ref{tab:elastic}. We note that to our current
knowledge, no existing analytic potential could reproduce all of the elastic constants of these materials
with an error of only a few percents.

\begin{table}
\begin{center}
\begin{tabular}{lrrr}
&\multispan{3}\hfill C\hfill\\
& DFT & GAP &  Brenner \\
\hline
$C_{11}$   & 1118 & 1081 & 1061 \\
$C_{12}$    & 151 &  157 &  133 \\
$C_{44}^0$  & 610 &  608 &  736 \\
$C_{44}$    & 603 &  601 &  717 \\
\hline
\hline
\vspace{0.0cm}\\
&\multispan{3}\hfill Si\hfill\\
&DFT&GAP&Tersoff\\
\hline
$C_{11}$   & 154& 152& 143 \\
$C_{12}$   &  56&  59&  75 \\
$C_{44}^0$ & 100& 101& 119 \\
$C_{44}$   &  75&  69&  69 \\
\hline
\hline

\vspace{0.0cm}\\
&\multispan{3}\hfill Ge\hfill\\
&DFT&GAP&Tersoff\\
\hline
$C_{11}$   & 108& 114& 138 \\
$C_{12}$   &  38&  35&  44 \\
$C_{44}^0$ &  75&  75&  93 \\
$C_{44}$   &  58&  54&  66 \\
\hline
\hline
\end{tabular}

\end{center}
\caption{\label{tab:elastic} Table of elastic constants, in units of GPa.}
\end{table}

\subsection{Anharmonic effects}

\begin{figure}
\begin{center}
\includegraphics[width=12cm]{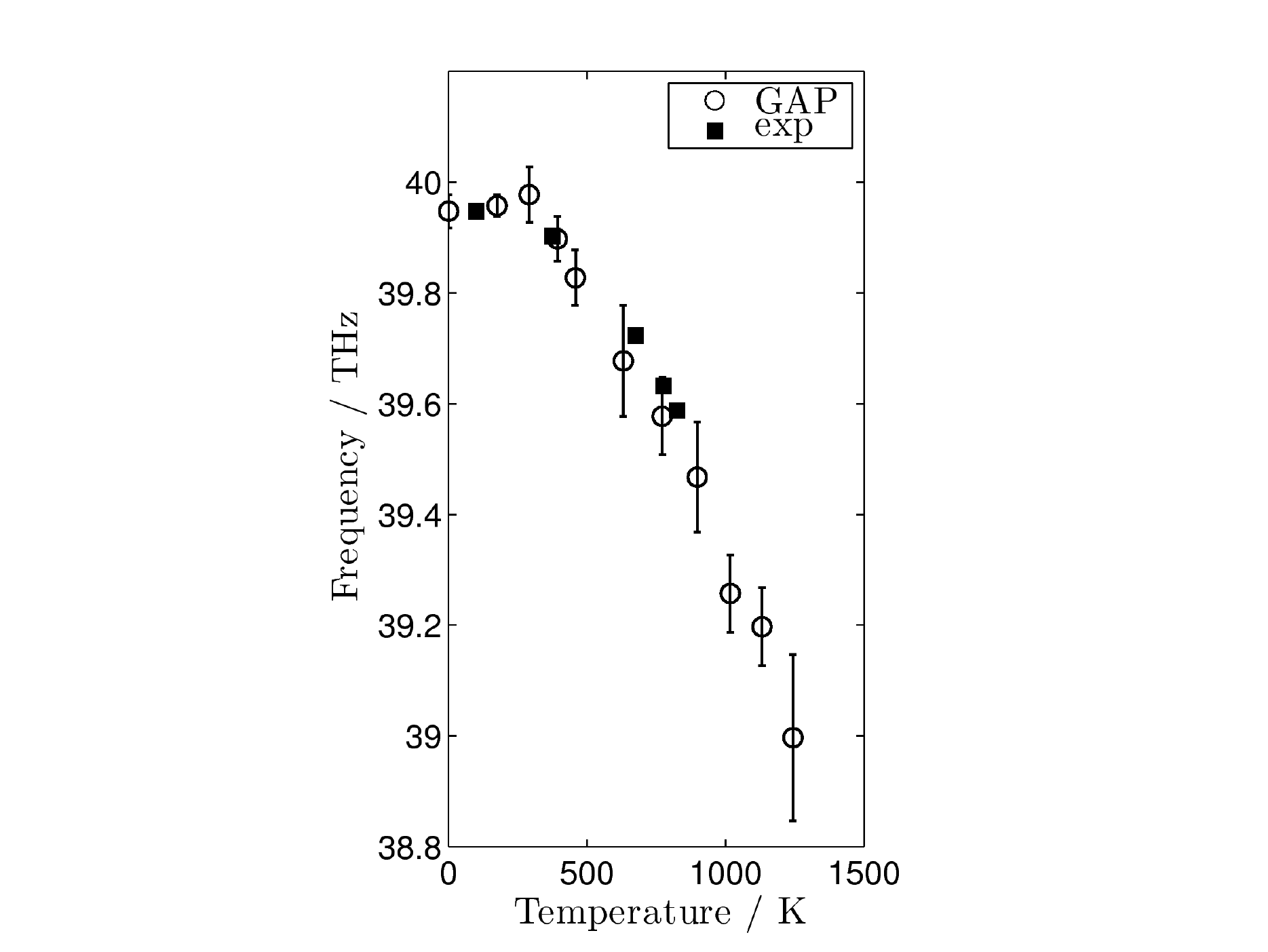}
\caption{\label{fig:phonon_t} Temperature dependence of the optical phonon at the $\Gamma$ point in
diamond.}
\end{center}
\end{figure}

In order to demonstrate the accuracy of the potential energy surface described by GAP outside the
harmonic regime, we calculated the temperature dependence of the optical phonon mode of the $\Gamma$
point in diamond. In fact, the low temperature variation of this quantity has been calculated using
Density Functional Perturbation Theory by Lang et al.\cite{phonon18}. The ab initio calculations
show excellent agreement with experimental values determined by Liu et al.\cite{phonon09}. We
calculated this optical phonon frequency using a molecular dynamics approach. We first performed a
series of constant-pressure molecular dynamics simulations for  a 250-atom supercell at different
temperatures in order to determine the equilibrium lattice constant as a function of temperature.
Then, for each temperature, we used the appropriate lattice constant to run a long microcanonical
simulation, from which we calculated the position-position correlation function. We selected the
phonon modes by projecting the displacements according to the appropriate wavevector. From the
Fourier-transform of the autocorrelation function, we obtained the phonon frequencies by fitting
Lorentzians on the peaks. We present our results in figure~\ref{fig:phonon_t}, where our values for
the phonon frequencies were shifted to match the experimental value at 0~K.

\begin{figure}
\begin{center}
\includegraphics[width=10cm]{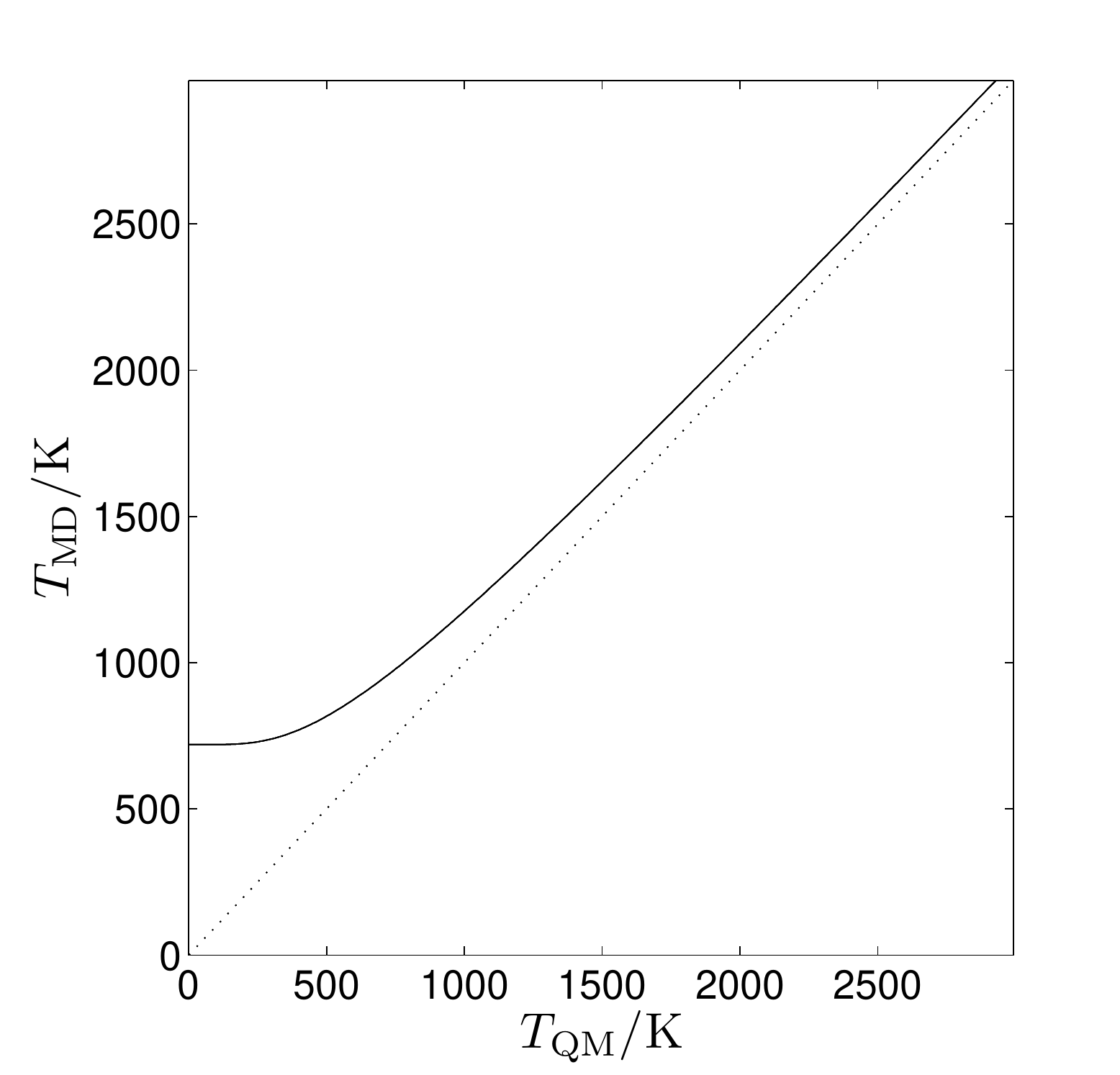}
\caption{\label{fig:tmd2tqm} Temperature of the quantum system described by GAP whose energy is
equal to the average kinetic energy of the classical system, as a function of the temperature of
the classical system. The dotted line is the identity function $f(x)\equiv x$, 
and is merely shown to provide a guide to the eye.}
\end{center}
\end{figure}

We note that even at 0~K there are anharmonic effects present due to the zero-point motion of the
nuclei. We accounted for the quantum nature of the nuclei by rescaling the temperature of the
molecular dynamics runs, by determining the temperature of the quantum system described by the same
phonon density of states whose energy is equal to the mean kinetic energy of the classical molecular
dynamics runs. The scaling function for the GAP model is shown in figure~\ref{fig:tmd2tqm}.

We are aware that at low temperatures this approximation is rather crude, and the correct way of
taking the quantum effects into account would be solving the Schr\"odinger equation for the nuclear
motion. However, we note that the anharmonic correction calculated by Lang et al. by Density
Functional Perturbation Theory\cite{phonon18} and our value show good agreement.
\begin{table}
\begin{center}
\begin{tabular}{cc}
\hline
\hline
\\[-2.0ex]
   & $\Delta \nu_{\textrm{anharmonic}} / \textrm{THz}$ \\[0.5ex]
\hline
LDA & 0.95 \\
GAP & 0.93 \\
\hline
\hline
\end{tabular}
\end{center}
\caption{\label{tab:anharmonic_gamma} Anharmonic shift of the $\Gamma$ phonon frequency in diamond.}
\end{table}

\subsection{Thermal expansion of diamond}

\begin{figure}
\begin{center}
\includegraphics[width=12cm]{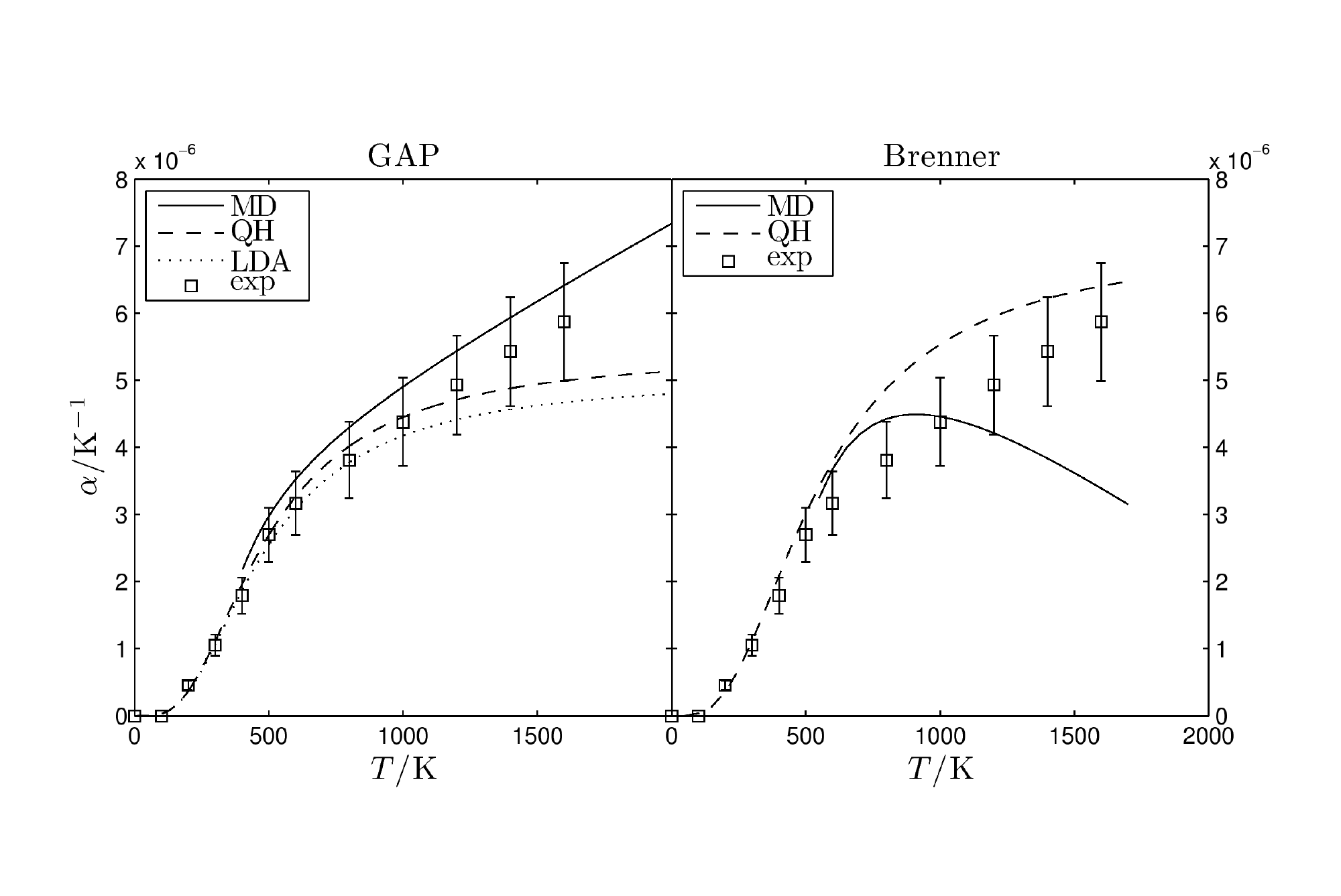}
\end{center}
\caption{\label{fig:thexp} Temperature dependence of the thermal expansion coefficient.}
\end{figure}

Another phenomenon that occurs as a result of the anharmonicity of the potential energy surface is
thermal expansion. The temperature dependence of the thermal expansion coefficient calculated from
first principles using the quasi-harmonic approximation is remarkably close to the experimental
value at low temperatures. However, at larger temperatures the quasi-harmonic approximation is less
valid, because other anharmonic effects, which cannot be modelled assuming first-order dependence of
the phonon frequencies on the lattice constant, are more significant. This effect can be calculated
exactly by solving the nuclear Schr\"odinger equation for the nuclear motion, or by classical
molecular dynamics simulation.  Herrero and Ram\'irez used a path-integral Monte Carlo method to
calculate the thermal expansion of diamond modelled by the Tersoff potential\cite{thexp04}. We determined the
thermal expansion by calculating the equilibrium lattice constant by running a series
of constant-pressure molecular dynamics simulations at different temperatures. We fitted the
analytic function
\begin{equation}
a(T) = c_1 T + c_2 T^2 + c_{-1} T^{-1} + c_0
\end{equation}
to the lattice constants, and then calculated the thermal expansion using the definition
\begin{equation}
\alpha(T) = \frac{1}{a(T)}\left(\frac{\mathrm{d}a}{\mathrm{d}T}\right)_T
\textrm{.}
\end{equation}
The same analytic function was used by Skinner to obtain the thermal expansion coefficient from the
experimental lattice constants~\cite{thexp03}. Our results are shown in figure~\ref{fig:thexp},
together with the experimental values~\cite{thexp03} and values calculated by LDA and GAP using the
quasiharmonic approach. The results obtained by using the Brenner potential is shown in the right
panel of figure~\ref{fig:thexp}. It can be seen that the thermal expansion is extremely well
predicted using GAP in molecular dynamics simulations.

The GAP results for the thermal expansion coefficients obtained from the quasiharmonic approximation show
excellent agreement with the LDA values. This verifies that the potential energy surface
represented by the GAP model is, in fact, close to the ab initio potential energy surface, even
outside the harmonic regime. In the case of Density Functional Theory, the molecular dynamics
simulation would be computationally expensive, because a large supercell has to be used to minimise
finite-size effects. However, with GAP, these calculations can be easily performed and the thermal
expansion coefficients obtained match the experimental values well, even at high temperatures.

\section{Towards a general carbon potential}

The ultimate aim of our research is to create potentials for general use. In the case of carbon,
describing the diamond phase is certainly not sufficient. Although we still have to add many more training
configurations to complete a general carbon potential, we demonstrate the capabilities of the GAP
scheme by extending the scope of the diamond potential described in the previous section to include
graphite, surfaces and vacancies.

We generated a set of randomised graphite configurations in a similar fashion to the diamond
training configurations. We randomised the atomic positions of the carbon atoms in 54- and
48-atom supercells of rhombohedral and hexagonal graphite and we also considered a number of uniaxially
compressed supercells. The training configurations also included diamond configurations with a vacancy
and $(111)$ surfaces, in particular, configurations of the unreconstructed $(111)$ surface and the
$2 \times 1$ Pandey-reconstruction were included in the training set.

\begin{figure}
\begin{center}
\includegraphics[width=10cm]{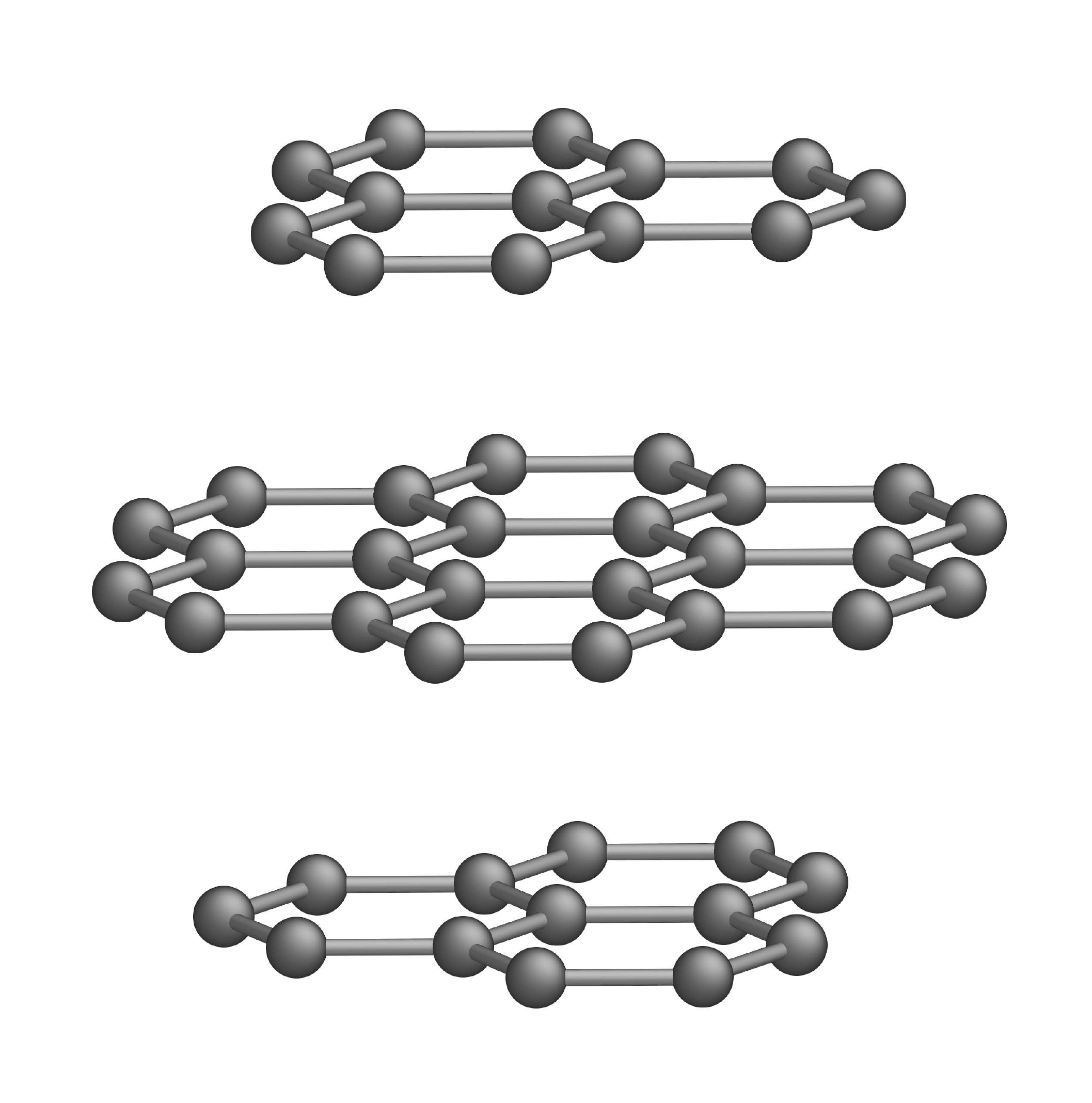}
\end{center}
\caption{\label{fig:rhombo_gra} Rhombohedral graphite.}
\end{figure}

\begin{figure}
\begin{center}
\includegraphics[width=10cm]{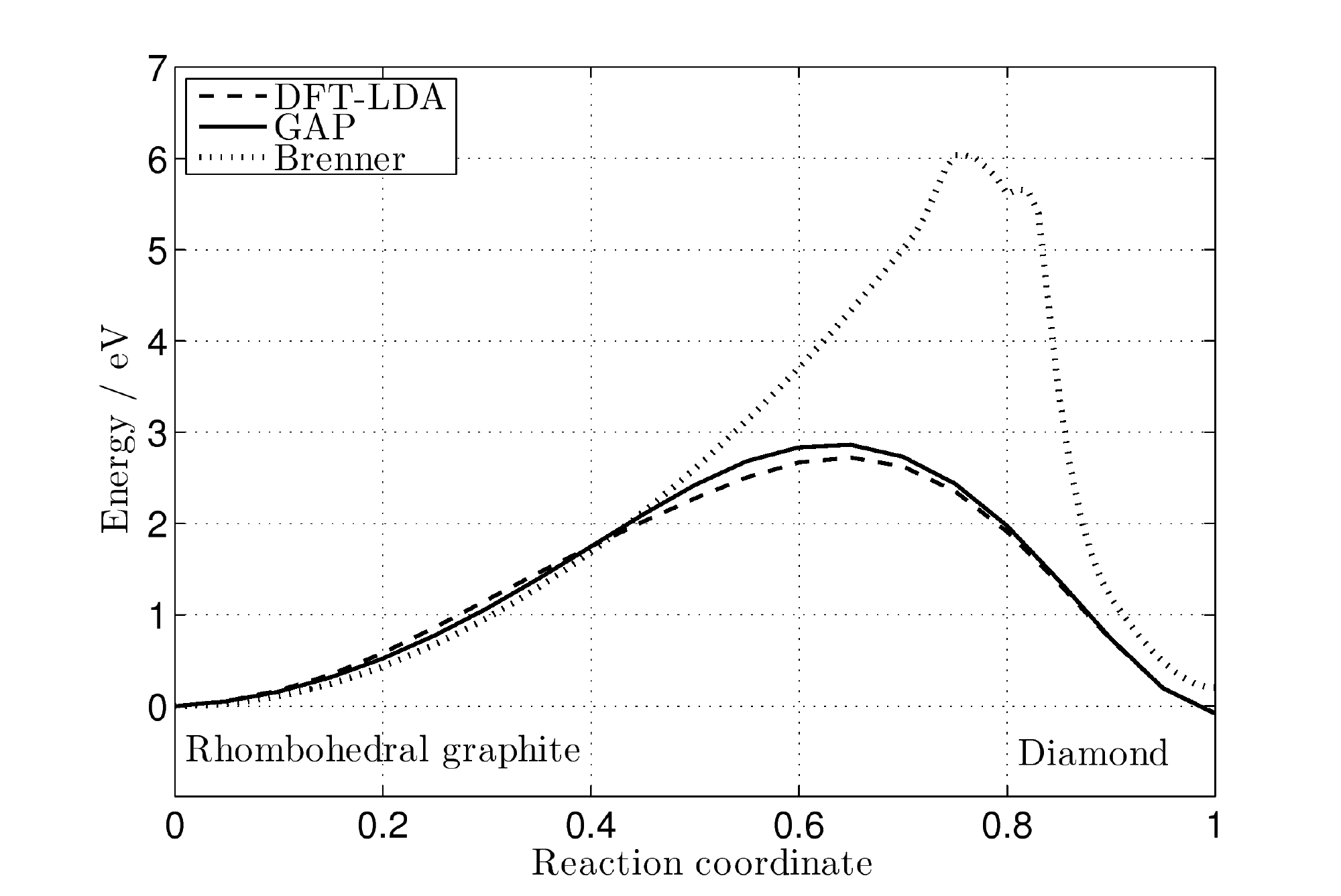}
\end{center}
\caption{\label{fig:dia2gra} The energetics of the linear transition path from rhombohedral graphite
to diamond calculated by DFT, GAP and the Brenner potential.}
\end{figure}

We tested how accurately the resulting GAP potential reproduces the rhombohedral graphite-diamond transition. Fahy et al.
described a simple reaction coordinate that transforms the 8-atom unit cell of rhombohedral graphite
(figure~\ref{fig:rhombo_gra}) to the cubic unit cell of diamond.
In figure~\ref{fig:dia2gra} we show the energies of the intermediate configurations between
rhombohedral graphite and diamond calculated using GAP, DFT and the Brenner potential. The lattice vectors $\mathbf{a}_{\alpha}$ and the atomic coordinates
$\mathbf{r}_i$ of these configurations were generated by
\begin{align}
\mathbf{a}_{\alpha} &= (1-x)\mathbf{a}_{\alpha}^{\textrm{graphite}} +
x\mathbf{a}_{\alpha}^{\textrm{diamond}} \textrm{, where } \alpha=1,2,3 \\
\mathbf{r}_i &= (1-x)\mathbf{r}_i^{\textrm{graphite}} +
x\mathbf{r}_i^{\textrm{diamond}} \textrm{, where } i = 1,\ldots,8
\textrm{.}
\end{align}
The reaction coordinate $x$ corresponds to graphite at $x=0$ and to diamond at $x=1$. It can be seen
that the Brenner potential cannot describe the change in the bonding of the carbon atoms, whereas
the GAP potential reproduces the quantum mechanical barrier accurately.

\begin{figure}
\begin{center}
\includegraphics[width=10cm]{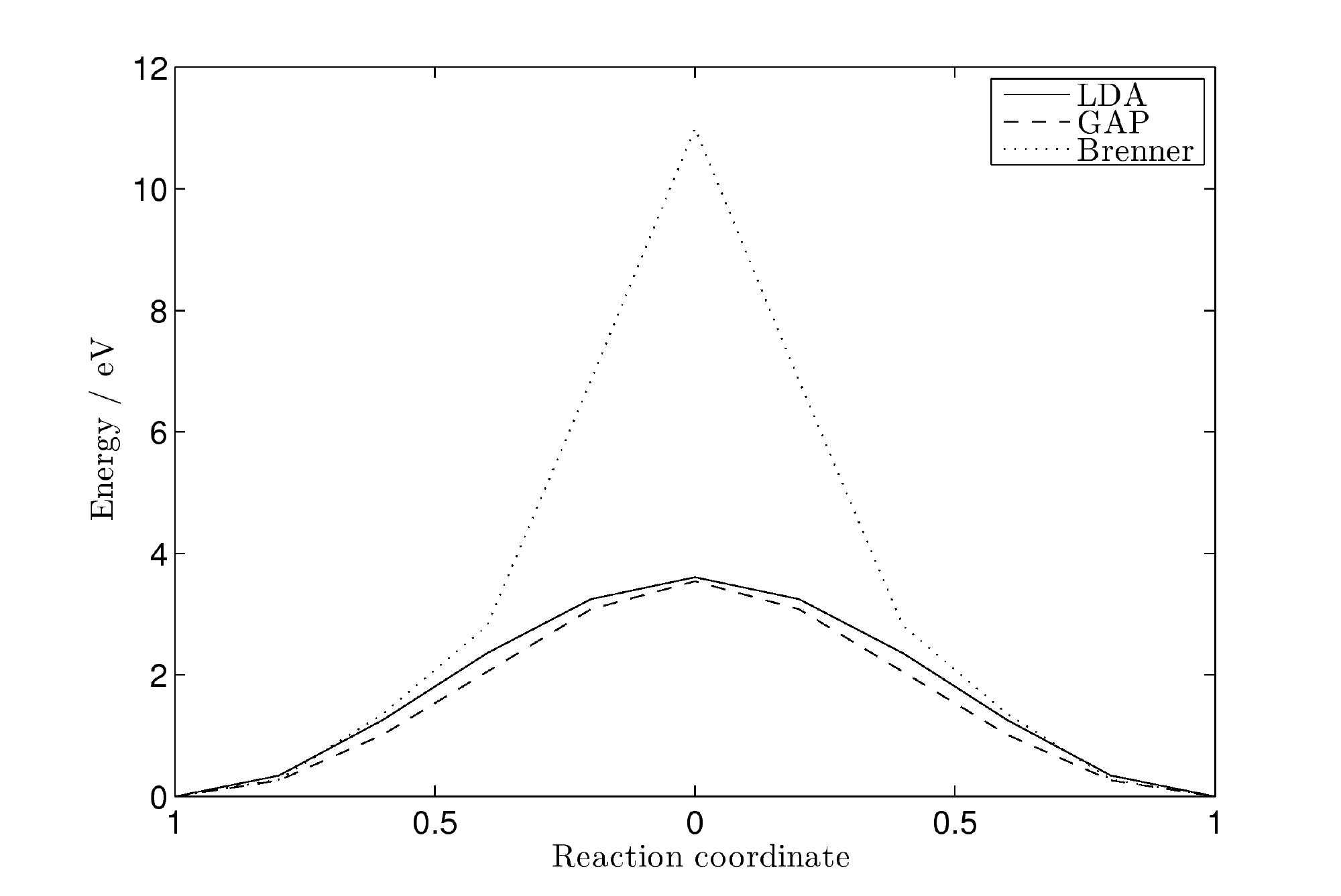}
\end{center}
\caption{\label{fig:vac} Energy along a vacancy migration path in diamond by DFT, GAP and the Brenner potential.}
\end{figure}

We also calculated the energetics of the vacancy migration in a similar fashion, i.e. along a linear
path between two configurations, where the vacancies are at two neighbouring lattice sites. Our
results are shown in figure~\ref{fig:vac}. The GAP model predicts the same the energies as the
Density Functional Theory, whereas the Brenner potential overestimates the energy barrier of the
migration.

Our results for the surface energies of the diamond $(111)$ surface are presented in
table~\ref{tab:surf} again showing very good agreement between GAP predictions and LDA results.
\begin{table}
\begin{center}
\begin{tabular}{ccccc}
\hline
\hline
\\[-2.0ex]
   & LDA-DFT & GAP & Brenner & Tersoff \\[0.5ex]
\hline
unreconstructed & 6.42 &  6.36 &  4.46  & 2.85  \\
$2\times1$      & 4.23 &  4.40 &  3.42  & 4.77  \\
\hline
\hline
\end{tabular}
\end{center}
\caption{\label{tab:surf} Surface energies in the units of $\mathrm{J/m^2}$ of the unreconstructed
and $2\times1$ Pandey-reconstructed surface of the $(111)$ diamond surface.}
\end{table}

\section{Gaussian Approximation Potential for iron}

The Gaussian Approximation Potential scheme is not limited to simple semiconductors. We
demonstrate this by applying the scheme to a metallic system, namely the body-centred cubic (bcc) phase of iron. We included configurations in
the training set where the lattice vectors of the 1-atom primitive cell were randomised and where
the positions of the atoms in 8 and 16-atom supercells were also randomised. These configurations
were represented by 50 sparse points in the training set for the GAP potential. The spatial cutoff
for the GAP potential was 4.0~{\AA} and we used the spherical
harmonics coefficients for the bispectrum up to $J_{\textrm{max}} = 6$.

\begin{figure}
\begin{center}
\includegraphics[width=10cm]{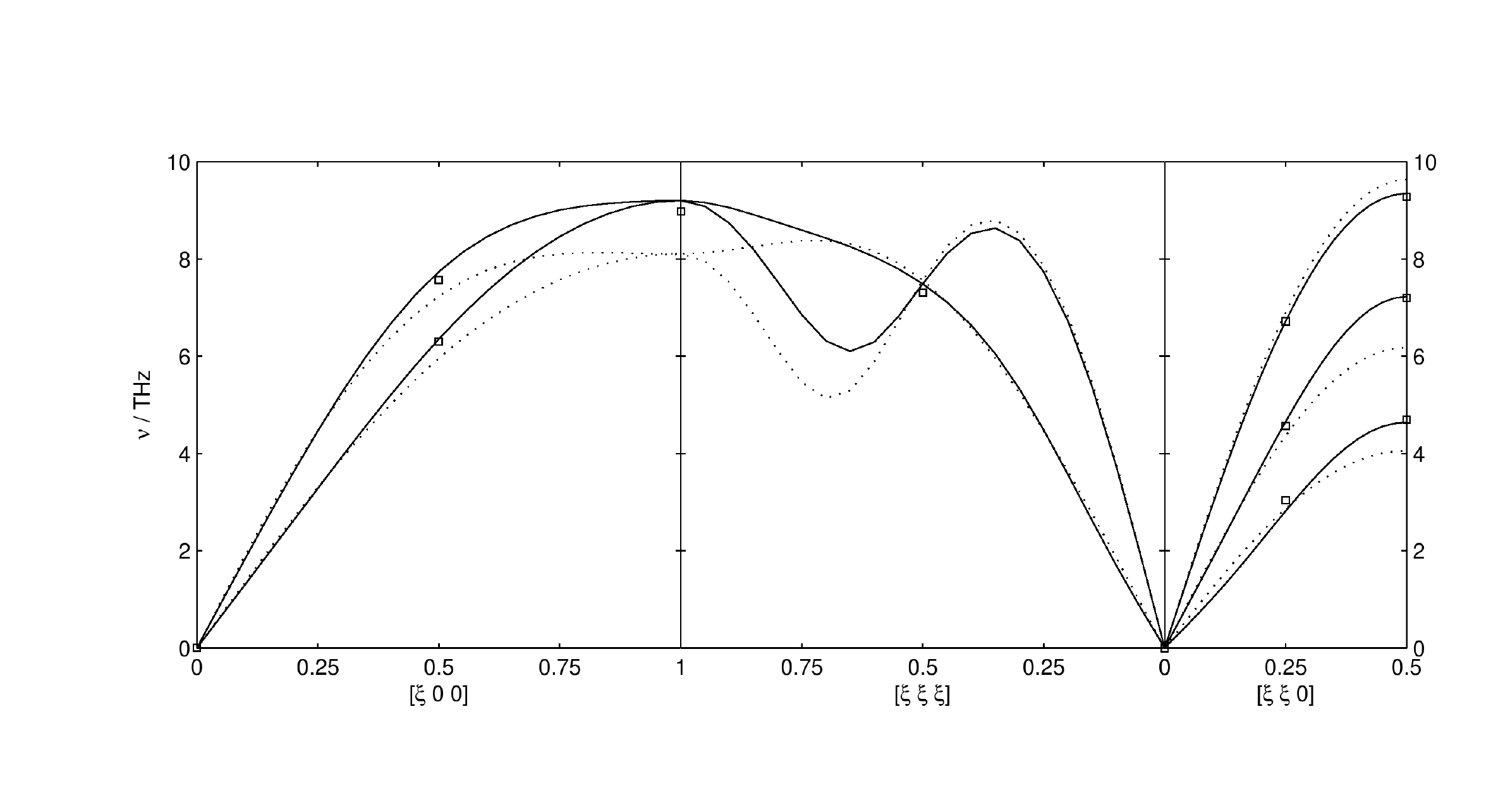}
\end{center}
\caption{\label{fig:fe_phonon} Phonon spectrum of iron using the GAP potential (solid lines), the
Finnis-Sinclair potential (dotted lines) and PBE-DFT (open squares).}
\end{figure}

\begin{figure}
\begin{center}
\includegraphics[width=10cm]{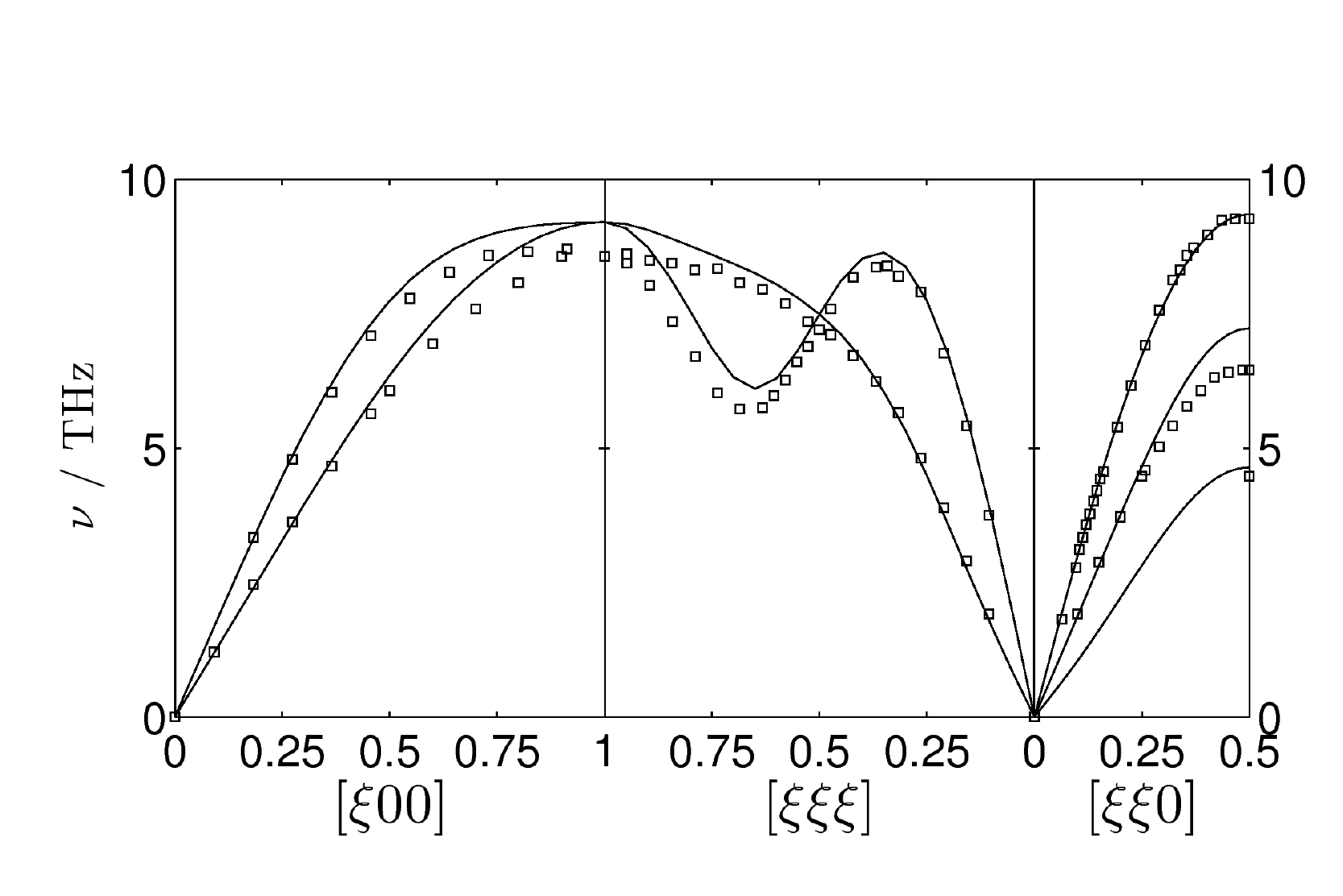}
\end{center}
\caption{\label{fig:fe_phonon_exp} Phonon dispersion of iron using the GAP potential (solid lines)
and experimental values (open squares)\cite{phonon_Fe_exp_01}.}
\end{figure}

We checked the accuracy of our potential by calculating the phonon spectrum along the high symmetry
directions and comparing the phonon frequencies at a few k-points
with Density Functional Theory. These spectra, together with those generated by the
Finnis-Sinclair potential are shown in figure~\ref{fig:fe_phonon}.
In figure~\ref{fig:fe_phonon_exp} we compared the phonon frequencies calculated by the GAP potential
to the experimental values obtained by the
neutron-inelastic-scattering technique\cite{phonon_Fe_exp_01}. The main features of the phonon
dispersion relation, for example, the crossing of the two branches along the $[\xi,\xi,\xi]$
direction, are reproduced by the GAP potential. The errors in the frequencies can be attributed to
our Density Functional Theory calculations.

The elastic moduli calculated with our model, the Finnis-Sinclair potential\cite{FS01} and Density Functional
Theory are given in table~\ref{tab:fe_elastic}. The elastic properties and the phonon dispersion
relations described by the GAP model show excellent agreement with the values calculated by Density
Functional Theory.
\begin{table}
\begin{center}
\begin{tabular}{cccc}
\hline
\hline
\\[-2.0ex]
   & PBE-DFT & GAP & Finnis-Sinclair \\[0.5ex]
\hline
$C_{11}$  & 236 &  222 & 245 \\
$C_{12}$  & 160 &  156 & 138 \\
$C_{44}$  & 117 &  111 & 122 \\
\hline
\hline
\end{tabular}
\end{center}
\caption{\label{tab:fe_elastic} Elastic moduli of iron in units of GPa calculated using different models.}
\end{table}

\section{Gaussian Approximation Potential for gallium nitride}

So far our tests of the Gaussian
Approximation Potentials were limited to single-species systems, but the framework can be extended
to multispecies systems. Here we report our first attempt to model such a system, the cubic phase of
gallium nitride. Gallium nitride (GaN) is a two-component semiconductor with a wurtzite or
zinc-blende structure. There is a charge transfer between the two species.

As in our previous work, the configurations for fitting the GAP model were generated by randomising
the lattice vectors of the primitive cell and randomly displacing atoms in larger supercells.  Owing
to the charge transfer, we need to include the long-range Coulomb-interaction in our model. We
decided to use the charges obtained from the population analyses of the ground state electronic
structure of a number of atomic configurations. Due to the fact that these configurations are
similar, the fluctuation of the atomic charges was not significant, hence we chose to use a simple,
fixed charge model with $-1\, e$ charge on the nitrogen atoms and $1\,e$ charge on the gallium
atoms. We calculated the electrostatic forces and energies for each training configuration by the
standard Ewald-technique\cite{ewald03} and subtracted these from the forces and energies obtained
from the Density Functional Theory calculations. We regarded the remaining forces and energies as
the short-range contribution of the atomic energies, and these were used for the regression to
determine the GAP potential.  The cutoff of the GAP potential was chosen to be 3.5~{\AA},
$J_{\textrm{max}}=5$ and we sparsified the training configurations using 300 sparse points.

\begin{figure}
\begin{center}
\includegraphics[width=10cm]{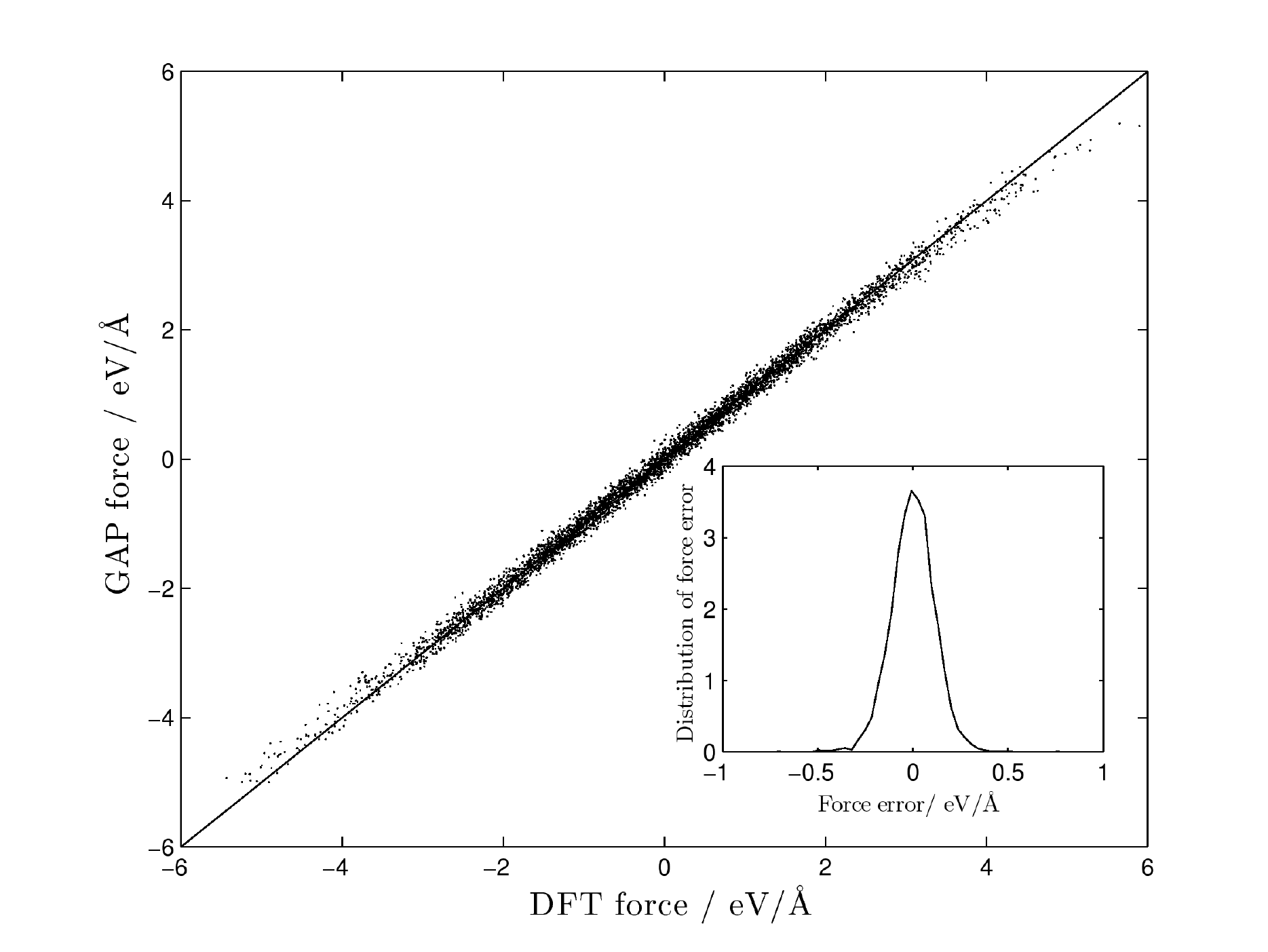}
\end{center}
\caption{\label{fig:corr_gan} Force components in GaN predicted by GAP vs. DFT forces. The diagonal
line is the $f(x)\equiv x$ function, which represents the perfect correlation. The inset depicts the
distribution of the difference between the force components.}
\end{figure}

\begin{figure}
\begin{center}
\includegraphics[width=10cm]{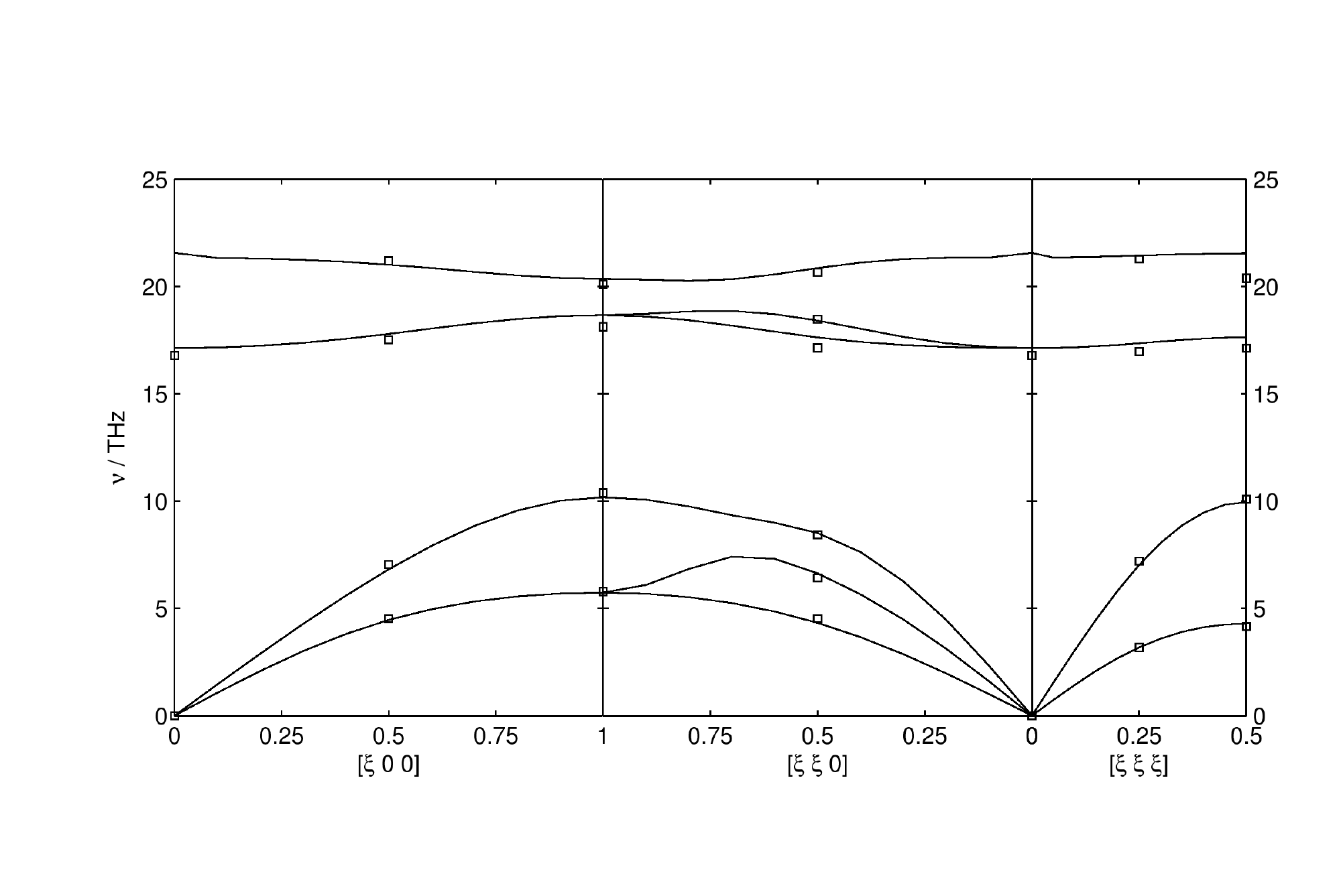}
\end{center}
\caption{\label{fig:gan_phonon} Phonon spectrum of GaN, calculated by GAP (solid lines) and PBE-DFT
(open squares).}
\end{figure}

We checked the correlation of the predicted forces of the resulting GAP potential with the ab initio
forces, and the results are shown in figure~\ref{fig:corr_gan}. We used 64-atom configurations where
the atoms were randomly displaced by similar amounts to the training configurations. 
The phonon spectrum calculated by GAP is shown in figure~\ref{fig:gan_phonon} and the elastic moduli
are listed in table~\ref{tab:gan_elastic}. 

\begin{table}
\begin{center}
\begin{tabular}{ccc}
\hline
\hline
\\[-2.0ex]
   & PBE-DFT & GAP \\[0.5ex]
\hline
$C_{11}$  & 265 & 262  \\
$C_{12}$  & 133 & 136  \\
$C_{44}$  & 153 & 142  \\
\hline
\hline
\end{tabular}
\end{center}
\caption{\label{tab:gan_elastic} Elastic moduli of GaN in units of GPa calculated using PBE-DFT and
GAP.}
\end{table}
Even this simple GAP model for gallium nitride shows remarkable accuracy in these tests, which we take as evidence that we
can adapt GAP to multispecies systems. However, in the case of very different neighbourhood
configurations we will probably have to include variable charges, and we will possibly have to consider the
contributions of multipole interactions in the long-range part of the potential. This is the subject
of future research.

\section{Atomic energies from GAP}\label{sec:gap_atomic_energies}

In section~\ref{sec:atomic_energies} we investigated a possible definition of atomic energies based
on localised atomic basis sets. According
to our results in section~\ref{sec:locality}, however, those atomic energies could not be used in our potential generation scheme
because they showed a large variation between numerically identical local environments.
Instead, we employed some extensions of the Gaussian Process regression method---learning from
derivatives, use of linear combination of function values and sparsification---, which make
the explicit definition of atomic energies unnecessary. Nonetheless, we found it striking that
an alternative possible definition of the quantum mechanical atomic energies, i.e. the ones inferred by the
Gaussian Approximation Potentials appeared to be successful. In other words, using these atomic energies we can obtain the
most commensurate forces and total energies for a given spatial cutoff, therefore these atomic
energies are optimal in this sense.

\begin{figure}[htb]
  \begin{center}
  \includegraphics[width=10cm]{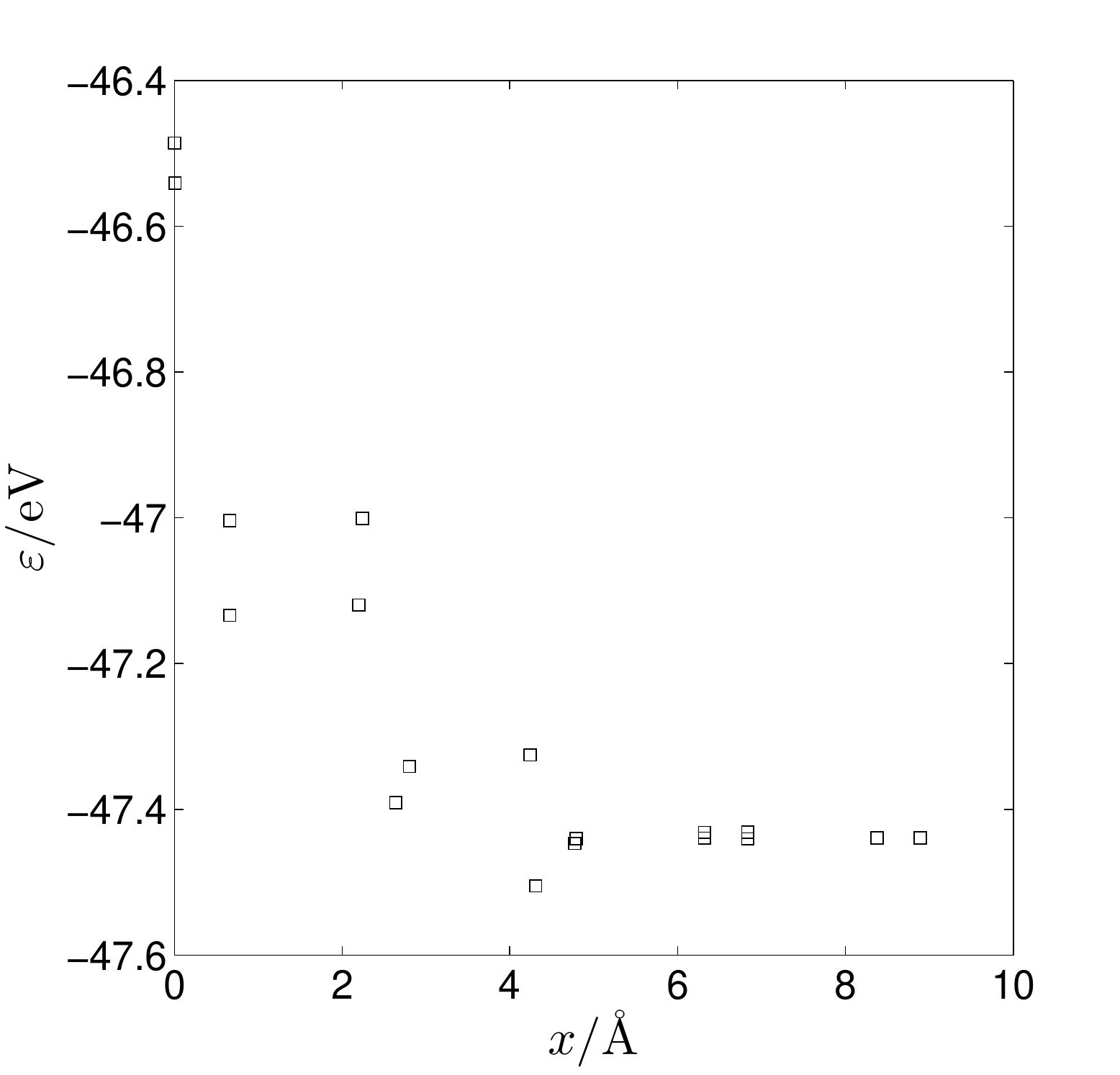}
  \end{center}
  \caption{Atomic energies of carbon atoms in a slab of diamond with two $(111)$ surfaces as a
function of the distance of the atom from the surface.}\label{fig:evsx}
\end{figure}

We show two examples which demonstrate that the atomic energies predicted by GAP are consistent with
physical considerations. In the first application, we calculated the atomic energies of the atoms in
a 96-atom slab of diamond, which had two $(111)$ surfaces. The training configurations were
generated by scaling the lattice vectors and  positions of the atoms of the minimised configuration
by a constant factor and randomising the atomic positions, and each of these steps was started from
a previous one. This means that in 20 steps, we created a series of samples between the minimised
structure and a completely randomised, gas-like configuration. We calculated the total energy and
the forces of the configurations by DFTB\cite{dftb01}, and used these to train a GAP model. The cutoff
of the model was 2.75~{\AA} and the atomic environments were represented by 100 sparse teaching
points. We used this model only to determine the atomic energies in the original slab. The atomic
energies of the carbon atoms as a function of their distance from the surface are plotted in
figure~\ref{fig:evsx}.
It can been seen that the atomic energy is higher at the surface and then gradually reaches the bulk
value towards the middle of the slab.

We also calculated the atomic energies defined by GAP in a gallium-nitride crystal where permutational defects were
present. We created two configurations which contained such defects. The first one was generated by
swapping the positions of a gallium and nitrogen atom in a 96-atom wurtize-type supercell, and then
we swapped the positions of another pair to generate the second configuration. We calculated the
total energies and forces of the two configurations by Density Functional Theory and used this data
to train a very simple GAP potential.  The cutoff of the model was 3.5~{\AA} and we used six sparse
point to represent the atomic environments. We used this model to calculate the atomic energies in
the same two configurations.  Certainly, the resulting potential is not a good representation of the
quantum mechanical potential energy surface, but it still detects the defects and predicts higher atomic
energies for the misplaced atoms.
\begin{figure}[htb]
  \begin{center}
  \includegraphics[height=3.5cm]{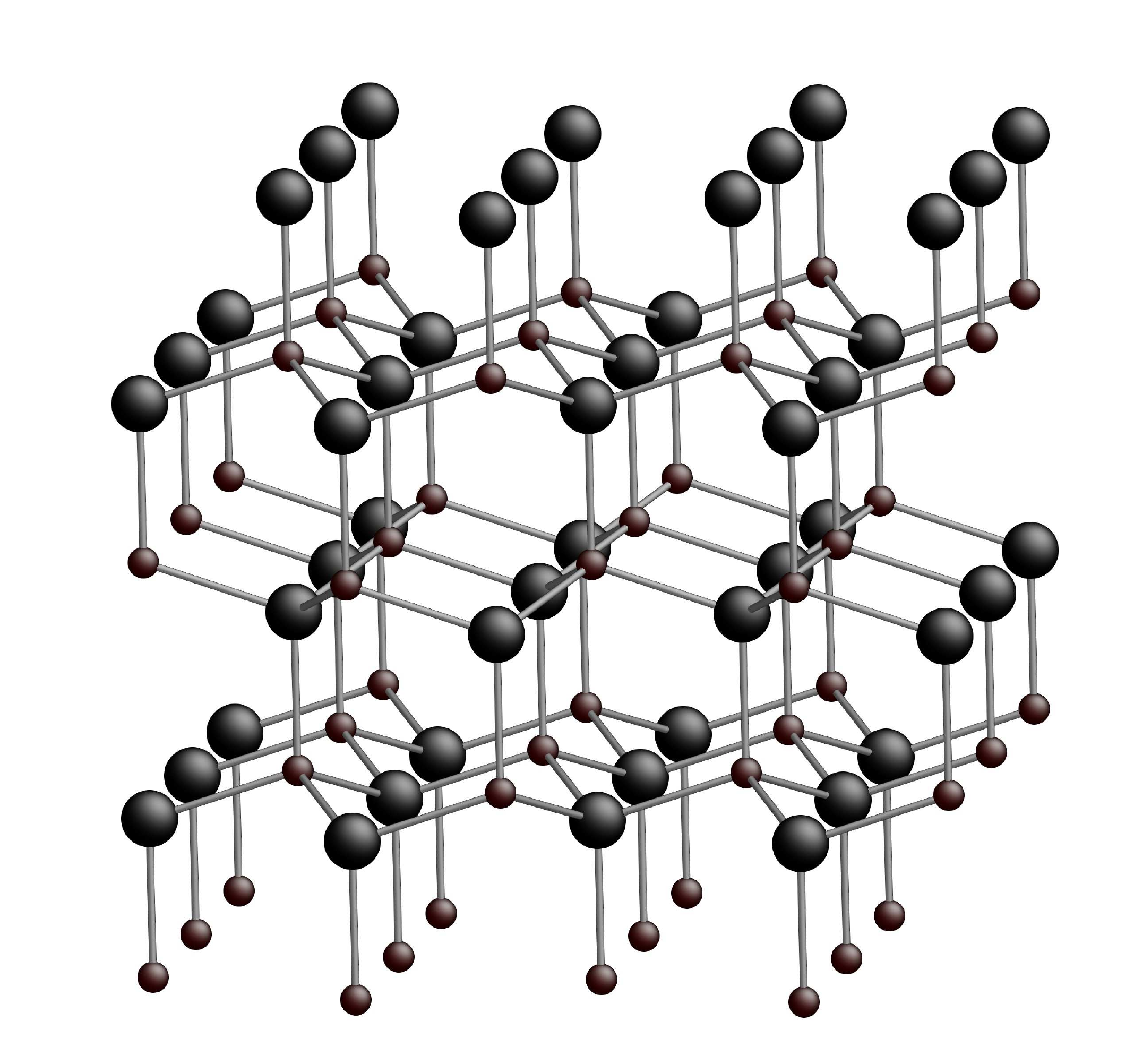}
  \includegraphics[height=3.5cm]{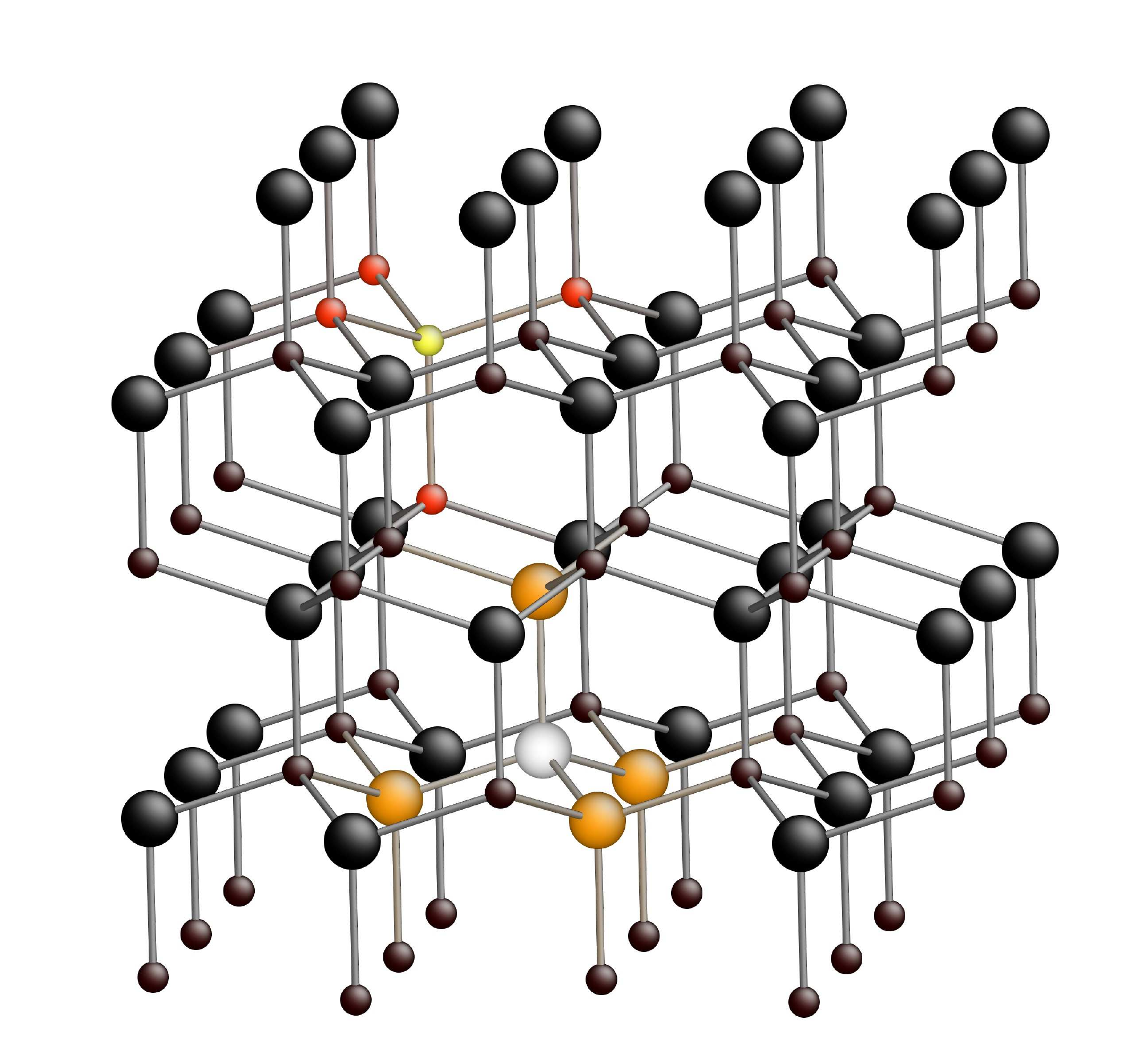}
  \includegraphics[height=3.5cm]{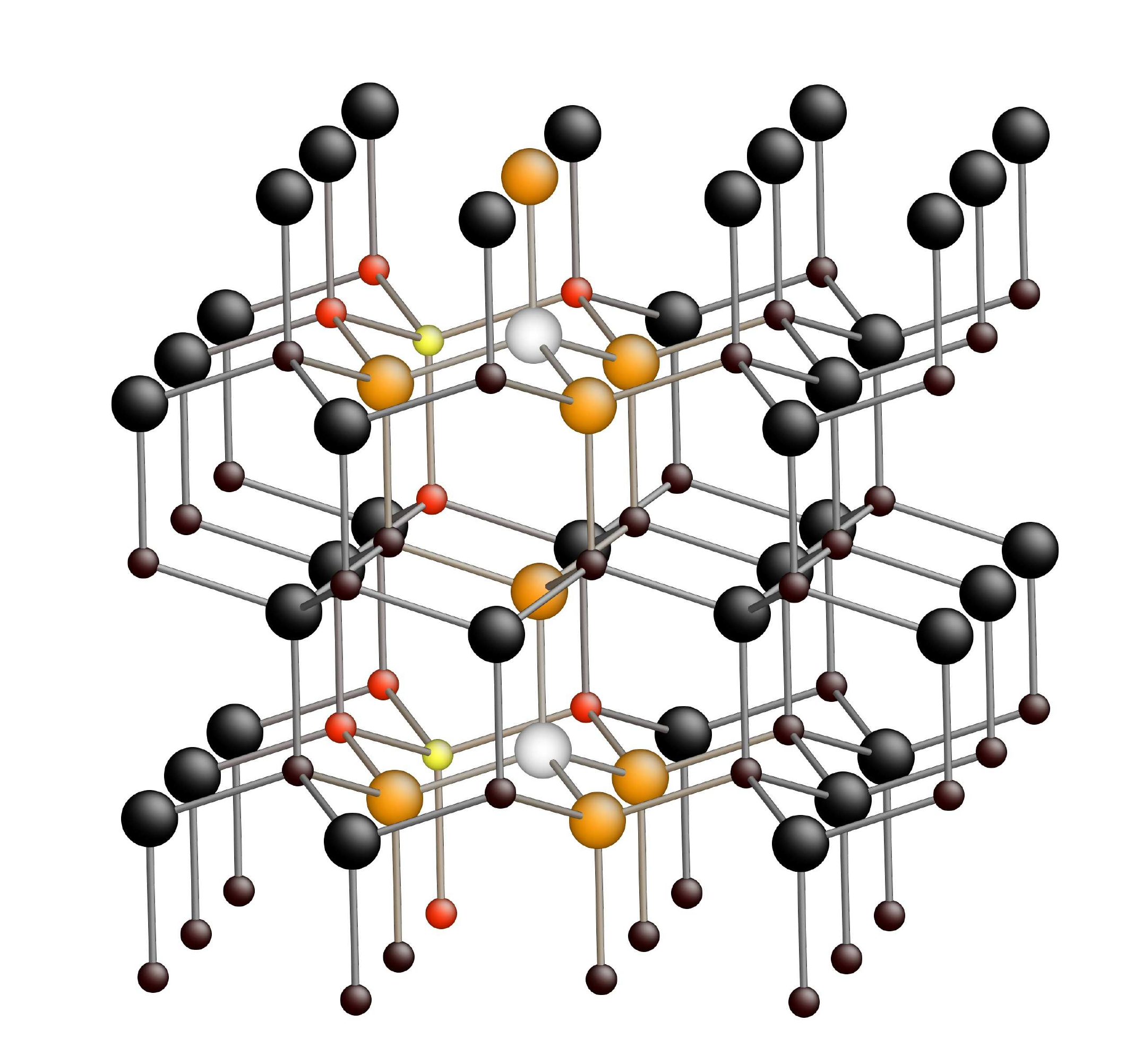}
  \includegraphics[height=3.5cm]{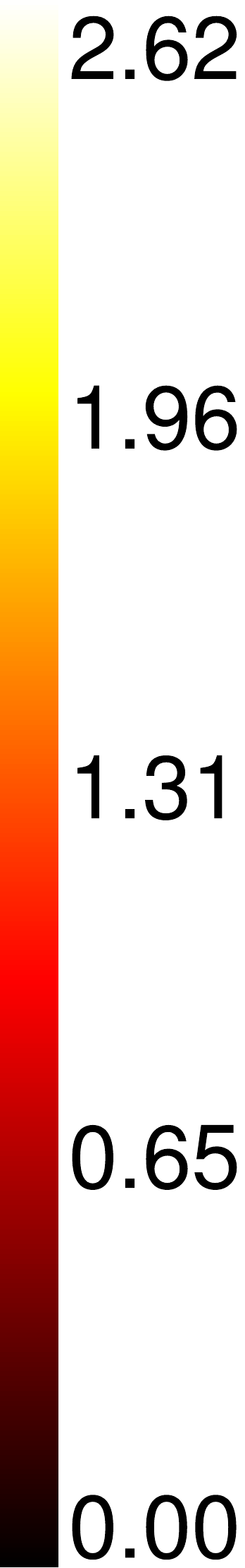}
  \end{center}
  \caption{The atomic energies in gallium-nitride crystals. We show the perfect wurtzite structure
on the left, a crystal containing a single defect (a gallium and a nitrogen atom swapped) is on the
middle and a crystal containing two defects (a further gallium-nitrogen pair swapped) is on the
right. The smaller spheres represent the nitrogen atoms, the larger ones represent the gallium
atoms. The coloured bar on the right shows the energy associated with the colour shades, in
eV.}\label{fig:gan_perm}
\end{figure}

Figure~\ref{fig:gan_perm} shows the configurations with the defects and the perfect lattice. The
colouring of atoms represent their atomic energies. It can be seen that the atomic energies of the atoms
forming the defect and surrounding it are higher.

In random structure search applications\cite{pickard_silane_01} GAP can be directly employed to detect
permutational defects. If there are more than one species present in the structure, the structure
search can result in many similar lattices, none of which are perfect, because of the large number
of permutations of different species. GAP models, which are generated on the fly, can be used to
suggest swaps of atoms between the local minima already found, which can then result in lower energy
structure. Using GAP as an auxiliary tool in such structure searches can possibly achieve a
significant speedup in searching for the global energy minimum.

\section{Performance of Gaussian Approximation Potentials}\label{sec:gap_performance}

The total computational cost of Gaussian Approximation Potentials consists of two terms. The first
term, which is a fixed cost, includes the computation of the ab initio forces and energies of the
reference calculations and the generation of the potential. The time required to generate the
potential scales linearly with the number of atomic environments in the reference configurations and
the number of sparse configurations. In our applications, performing the DFT calculations typically took
100 CPU hours while the generation of a GAP potential was about a CPU hour.

Even for small systems, GAP potentials in our current implementation are order of magnitudes faster
than Density Functional Theory, but significantly---about a hundred times---more expensive than analytical potentials.
Calculation of the energies and forces requires about 0.01~s for every atom on a single CPU core.
For comparison, a timestep of a 216-atom simulation cell takes about 190~s per atom on a single core
by CASTEP, which corresponds to 20,000-fold speedup. The same calculation for iron would take a
million times longer by CASTEP.

\chapter{Conclusion and further work}\label{chapter:conclusion}

During my doctoral studies, I implemented a novel, general approach to building interatomic
potentials, which we call Gaussian Approximation Potentials. Our potentials are designed to
reproduce the
quantum mechanical potential energy surface (PES) as closely as possible, while being significantly faster
than quantum mechanical methods. To achieve this, we used the concept of Gaussian Process from
Inference Theory and the bispectral representation of atomic enviroments, which we derived and
adapted using the Group Theory of rotational groups.

I tested the GAP models on a range of simple materials, based on data obtained from Density Functional
Theory. I built interatomic potentials for the diamond lattices of the group IV semiconductors and
I performed rigorous tests to evaluate the accuracy of the potential energy surface. These tests showed
that the GAP models reproduce the quantum mechanical results in the harmonic regime, i.e. phonon
spectra, elastic properties very well. In the case of diamond, I calculated properties which are
determined by the anharmonic nature of the PES, such as the temperature dependence of the optical phonon
frequency at the $\Gamma$ point and the temperature dependence of the thermal expansion
coefficient. Our GAP potential reproduced the values given by Denstity Functional Theory and experiments.

These potentials constituted our initial tests of the scheme, and represented only a small part of the PES. In the case of
carbon, I extended the GAP model to describe graphite, the diamond $(111)$ surface and vacancies in the
diamond lattice. I found that the new GAP potential described the rhombohedral graphite-diamond transition, the
surface energies and the vacancy migration remarkably well.

To show that our scheme is not limited to describing monoatomic semiconductors, I generated a potential for
bcc iron, a metal, and for gallium nitride, an ionic semiconductor. Our preliminary tests, which
were the comparison of the phonon dispersion and the elastic moduli with Density Functional Theory
values, demonstrate that GAP models can easily be built for different kinds of materials. I also
suggest that the Gaussian Approximation Potentials can be generated on the fly and used as
auxiliary tools for example, in structure search applications.

\section{Further work}

In my thesis I presented preliminary tests and validation of our potential generation scheme.
In the future, we intend to build models and perform large scale simulations on a wide range of
materials. The first step will be to create a general carbon potential, which can describe amorphous and
liquid carbon at a wide range of pressures and temperatures as well as defects and surfaces. We are
also planning to create ``disposable'' potentials, which can be used, for instance, in the case of
crack simulations. These do not have to be able to describe the high-temperature behaviour of the
materials, as only a restricted part of the configurational space is accessible under the conditions of
the simulation.  The description of electrostatics will be soon implemented, with charges and
polarisabilities which depend on the local environment and the electric field. This will allow us to
simulate more complex systems, for example silica or water and our ultimate aim is to build
interatomic potentials---force fields---for biological compounds. None of these potentials have to
be based on Density Functional Theory, for instance it might be necessary to use more accurate solutions of
the electronic Schr\"odinger equation.  Finally, using GAP as a post-processing tool to determine
atomic energies derived from on Quantum Mechanics is also a future direction of our research, for example,
in structure searches.

\appendix
\appendixpage
\addtocontents{toc}{\protect\setcounter{tocdepth}{-1}}
\chapter{Woodbury matrix identity}

The likelihood function in equation~\ref{eq:sparse_likelihood} is used during the sparsification
procedure in order to optimise the hyperparameters and the sparse points. At first sight, it seems
that the inverse of an $N \times N$ matrix has to be calculated, the computational cost of which
would scale as $N^3$.  However, by using the \emph{matrix inversion lemma}, also known as the
Woodbury matrix identity, the computational cost scales only with $NM^2$ if $N>>M$.
If we want to find the inverse of a matrix, which can be written in the form $\mathbf{Z} +
\mathbf{UWV}^T$, the Woodbury matrix identity states that
\begin{equation}
(\mathbf{Z} + \mathbf{UWV}^T)^{-1} = \mathbf{Z}^{-1} - \mathbf{Z}^{-1}\mathbf{U}(\mathbf{W}^{-1} +
\mathbf{V}^T \mathbf{Z}^{-1} \mathbf{U})^{-1} \mathbf{V}^T \mathbf{Z}^{-1}
\textrm{.}
\end{equation}
In our case, $\mathbf{Z}$ is an $N \times N$ diagonal matrix, hence its inverse is trivial, and
$\mathbf{W}^{-1}$ is $M \times M$. The order of the operations can arranged such that none of them
requires more than $NM^2$ floating point operations:
\begin{multline}
\mathbf{t}^T (\mathbf{C}_{NM} \mathbf{C}_M^{-1} \mathbf{C}_{MN} + \mathbf{\Upsilon})^{-1} \mathbf{t} = \\
\mathbf{t}^T \mathbf{\Upsilon}^{-1} \mathbf{t} 
- (\mathbf{t}^T \mathbf{\Upsilon}^{-1}) \mathbf{C}_{NM} (\mathbf{C}_M^{-1} + \mathbf{C}_{MN}
\mathbf{\Upsilon}^{-1} \mathbf{C}_{NM})^{-1} \mathbf{C}_{MN} (\mathbf{\Upsilon}^{-1} \mathbf{t})
\textrm{,}
\end{multline}
where $\mathbf{\Upsilon} = \mathbf{\Lambda}+ \sigma^2 \mathbf{I}$.
In the evaluation of the second term in equation~\ref{eq:sparse_likelihood} we used the \emph{matrix
determinant lemma}, which is analogous to the inversion formula:
\begin{equation}
\det (\mathbf{Z} + \mathbf{UWV}^T) = \det (\mathbf{W}^{-1} +
\mathbf{V}^T \mathbf{Z}^{-1} \mathbf{U}) \det (\mathbf{W}) \det (\mathbf{Z})
\textrm{.}
\end{equation}
In our implementation, the determinants are calculated together with the inverses, without any
computational overhead.

We also note that at certain values of the hyperparameters the matrix $\mathbf{C}_M$ is ill
conditioned. In the original Gaussian Process, the covariance matrix $\mathbf{Q}$ can also be ill
conditioned, but by adding the diagonal matrix $\sigma_{\nu}^2\mathbf{I}$ this problem is
eliminated, except for very small values of the $\sigma_{\nu}$ parameters. Snelson
suggested\cite{snelson_private} that a small diagonal matrix $\xi^2\mathbf{I}$ should be added to
$\mathbf{C}_M$ to improve the condition number of the matrix. This small ``jitter'' factor can be
regarded as the internal error of the sparsification.

\chapter{Spherical harmonics}

\section{Four-dimensional spherical harmonics}
The spherical harmonics in three dimensions are the angular part of the solution of the Laplace
equation
\begin{equation}
\left( \frac{\partial^2}{\partial x^2} + \frac{\partial^2}{\partial y^2} + \frac{\partial^2}{\partial
z^2}  \right) f = 0
\textrm{.}
\end{equation}
This concept can be generalised to higher dimensions. In our case, we need the solutions of the four
dimensional Laplace equation
\begin{equation}
\left( \frac{\partial^2}{\partial x^2} + \frac{\partial^2}{\partial y^2} + \frac{\partial^2}{\partial
z^2} + \frac{\partial^2}{\partial z_0^2} \right) f = 0
\textrm{,}
\end{equation}
which can be written in the form of the three-dimensional rotation matrices, the Wigner D-functions.

The definition of the elements of the rotational matrices is
\begin{equation}
D^{(l)}_{mm'}(R) = \langle Y_{lm} | \hat{R} | Y_{lm'} \rangle
\textrm{,}
\end{equation}
where the rotation $\hat{R}$ is defined by three rotational angles. The rotational operator is 
usually described as three successive rotations
\begin{itemize}
\item rotation about the $z$ axis by angle $\alpha$,
\item rotation about the new $y'$ axis by angle $\beta$,
\item rotation about the new $z'$ axis by angle $\gamma$,
\end{itemize}
where $\alpha$, $\beta$ and $\gamma$ are called the Euler-angles.
The Wigner D-functions are usually formulated as the function of these three angles and denoted as
$D^J_{MM'}(\alpha,\beta,\gamma)$. However, in some cases the rotation can be described more
conveniently in terms of $\omega$, $\theta$ and $\phi$, where the rotation is treated as a single
rotation through angle $\omega$ about the axis $\mathbf{n}(\theta,\phi)$. The vector $\mathbf{n}$ is
determined by the polar angles $\theta$ and $\phi$.

The rotational matrices in the form $U^J_{MM'}(\omega,\theta,\phi)$, where the four dimensional polar
angles are $2\theta_0 \equiv \omega$, $\theta$ and $\phi$ are the four dimensional spherical
harmonics.

The matrix elements can be constructed as
\begin{equation}
U^J_{MM'}(\theta_0,\theta,\phi) = \left\{
\begin{array}{c}
(-iv)^{2J} \left(\frac{u}{-iv}\right)^{M+M'} e^{-i(M-M')\phi} \times \\
\sum_s \frac{\sqrt{(J+M)!(J-M)!(J+M')!(J-M')!}}{s!(s+M+M')!(J-M-s)!(J-M'-s)!} (1-v^{-2})^s,\\
M+M' \ge 0 \\
\hspace{10pt} \\
(-iv)^{2J} \left(\frac{u^*}{-iv}\right)^{-M-M'} e^{-i(M-M')\phi} \times \\
\sum_s \frac{\sqrt{(J+M)!(J-M)!(J+M')!(J-M')!}}{s!(s-M-M')!(J+M-s)!(J+M'-s)!} (1-v^{-2})^s,\\
M+M' \le 0 
\end{array} \right. \textrm{,}
\end{equation}
where
\begin{align}
v &= \sin \theta_0 \sin \theta \\
u &= \cos \theta_0 - i \sin \theta_0 \cos \theta
\textrm{.}
\end{align}

In our application, each time an entire set of $U^J_{MM'}$ has to be calculated, thus the use of
recursion relation is computationally more efficient. The recursion relations are
\begin{equation}
\begin{split}
U^J_{MM'}(\theta_0,\theta,\phi) &= \sqrt{ \frac{J-M}{J-M'} } u^*
U^{J-\frac{1}{2}}_{M+\frac{1}{2}M'+\frac{1}{2}}(\theta_0,\omega,\theta) \\
& - i \sqrt{ \frac{J+M}{J-M'} } v e^{i \phi}
U^{J-\frac{1}{2}}_{M-\frac{1}{2}M'+\frac{1}{2}}(\theta_0,\omega,\theta) \\
& \textrm{for } M' \ne J
\end{split}
\end{equation}
and
\begin{equation}
\begin{split}
U^J_{MM'}(\theta_0,\theta,\phi) &= \sqrt{ \frac{J+M}{J+M'} } u
U^{J-\frac{1}{2}}_{M-\frac{1}{2}M'-\frac{1}{2}}(\theta_0,\omega,\theta) \\
& - i \sqrt{ \frac{J-M}{J+M'} } v e^{i \phi} 
U^{J-\frac{1}{2}}_{M+\frac{1}{2}M'-\frac{1}{2}}(\theta_0,\omega,\theta) \\
& \textrm{for } M' \ne -J
\end{split}
\textrm{.}
\end{equation}
The actual implementation does not involve the explicit calculation of the polar angles, we
calculate the spherical harmonics in term of the Cartesian coordinates $x$, $y$, $z$ and $z_0$. The
first two four-dimensional spherical harmonics are
\begin{equation}
U^0_{00} = 1
\end{equation}
and
\begin{align}
U^{\frac{1}{2}}_{ \pm \frac{1}{2} \pm \frac{1}{2}} &= \frac{1}{\sqrt{2}} \frac{z_0 \pm i z}{r} \\
U^{\frac{1}{2}}_{ \pm \frac{1}{2} \mp \frac{1}{2}} &= -\frac{i}{\sqrt{2}} \frac{x \mp i y}{r}
\textrm{,}
\end{align}
which are indeed analogous to their three-dimensional counterparts.

\section{Clebsch-Gordan coefficients}

We used the following formula to compute the Clebsch-Gordan coefficients:

\begin{multline}
C^{c \gamma}_{a \alpha b \beta} = \delta_{\gamma,\alpha+\beta} \Delta(abc) \\
\times \sqrt{(a+\alpha)! (a-\alpha)! (b+\beta)! (b-\beta)! (c+\gamma)! (c-\gamma)! (2c+1)} \\
\times \sum_z \frac{(-1)^z}{z! (a+b-c-z)! (a-\alpha-z)! (b+\beta-z)!(c-b+\alpha+z)! (c-a-\beta+z)!}
\textrm{,}
\end{multline}
where $\Delta$-symbol is
\begin{equation}
\Delta(abc) = \sqrt{ \frac{ (a+b-c)! (a-b+c)! (-a+b+c)!}{(a+b+c+1)!} }
\textrm{.}
\end{equation}

\bibliography{thesis}
\bibliographystyle{h-physrev3}

\end{document}